\documentclass[10pt,journal,compsoc]{IEEEtran}
\ifCLASSOPTIONcompsoc
  \usepackage[nocompress]{cite}
\else
  \usepackage{cite}
\fi

\usepackage[switch]{lineno}  %
\usepackage{booktabs}
\usepackage{subfigure}
\usepackage{amssymb}

\usepackage{paralist}
\usepackage{xspace}
\usepackage{mathrsfs}
\usepackage{color}
\usepackage{algpseudocode}
\usepackage{amsmath}
\usepackage{multirow}
\usepackage{listings}
\usepackage{algpseudocode}
\usepackage{multirow}
\usepackage{bm}
\usepackage{amsthm}
\usepackage[ruled,vlined]{algorithm2e} 
\usepackage{graphicx}

\def\ie{\textit{i.e.}\xspace}
\def\etal{\textit{et al.}\xspace}
\def\etc{\textit{etc.}\xspace}

\def\eg{\textit{e.g.}\xspace}
\def\iid{\textit{i.i.d.}\xspace}
\def\st{\xspace\textbf{s.t.}\xspace}

\newcommand{\CUTXY}[1]{{}}
\newcommand{\linote}[1]{{\small $\langle${\textbf{\textcolor{red}{#1}}}$\rangle$ \normalsize}}

\newcommand{\prob}[1]{{\textbf{Pr}\left(#1\right)}}

\newcommand{\defeq}{\stackrel{\text{def}}{=}}

\newtheorem{claim}{Claim}

\newtheorem{theorem}{Theorem}

\newtheorem{definition}{Definition}

\newtheorem{assumption}{Assumption}
\newtheorem{lemma}[theorem]{Lemma}

\begin{document}
%\title{Revenue-Competitive Truthful Online Mechanism for Trading \\ Time-Sensitive Valued {Data-Copyright}}
%\title{Revenue-Competitive Truthful Online Trading of Time-Sensitive Valued Digital Product}
\title{Competitive Online Truthful Time-Sensitive-Valued Data Auction}
%\title{Revenue-Competitive Truthful Online Mechanism for Trading Time-Sensitive Valued {Digital Product}}
\author{
\IEEEauthorblockN{
Shuangshuang~Xue,~\IEEEmembership{Student Member,~IEEE,}
Xiang-Yang~Li,~\IEEEmembership{Fellow,~IEEE,} }\\
%Lan~Zhang,~\IEEEmembership{Member,~IEEE,}
%and~Hu~Ding,~\IEEEmembership{Member,~IEEE}}\\	
}

\IEEEtitleabstractindextext{
\begin{abstract}
%A number of data trading platforms have emerged with the explosion of big data based technologies and applications. 
Digital product, such as original data and data-copyright, as an irreplaceable form of data products, can be transferred by the owner to potential users
  to maximize its potential usage while protecting the traded data from being resold. 
 Moreover, the  value of the digital product and data typically changes over time in real data application scenarios  such as monitoring and recommendation systems,
  \ie, its value is time-sensitive.
In this work, we investigate online mechanisms for trading time-sensitive valued digital product and data-copyright.
We adopt a continuous function $d(t)$ to represent the data value fluctuation over time $t$. 
Our objective is to design an \emph{online} trading mechanism achieving  \emph{truthfulness} and \emph{revenue-competitiveness}
  with respect to the offline widely-celebrated Vickrey auction. 
However, designing such \emph{single-round online}  mechanisms (\ie, the copyright can only be sold to one unique buyer) is much more challenging 
  than that in a multi-round scenario due to the short of opportunities to learn the trading information, 
  or than that in a non-time-sensitive scenario as each user's valuation changes over time.  
We first prove several lower bounds on the  revenue competitive ratios of  individual-rational mechanisms under various assumptions, such as
  a lower bound of $\Omega(n)$ when there is  an arbitrarily unknown function $d(t)$, and 
  a lower bound of $\Omega(\frac{\log n}{\log \log n})$ when the known function $d(t)$ is non-increasing.
We then propose several online truthful auction mechanisms for various adversarial models, such as
 a randomized observe-then-select mechanism $\mathcal{M}_R$ and prove that it is \textit{truthful} and $\Theta(\log^2 n)$-competitive
  when $d(t)$ is known and non-increasing and the number of users  $n_c$ in each \emph{discount-class} is of same order. 
With same assumption, we show that a simpler mechanism $\mathcal{M}_1$  is \textit{truthful} and $\Theta(\log n)$-competitive.
Furthermore, without any restrictions on the sizes of the discount-classes, we  prove that the  mechanism $\mathcal{M}_1$ could have a competitive ratio
 as ad as  $\Omega(n^2)$.
Then we present an effective  truthful weighted-selection mechanism $\mathcal{M'}_W$  by relaxing the assumptions on the sizes of the discount-classes.
We prove that it achieves a competitive ratio $\Theta(n\log n)$ for any known non-decreasing discount function $d(t)$, and $n_c \ge 2$.
When the optimum expected revenue can be estimated within a constant factor, we propose a truthful online posted-price 
 mechanism that achieves a constant competitive ratio.
%We also present several reserved-price mechanisms when the valuations of users are drawn from some given known distribution and analyze their performances.
We conduct  extensive numerical evaluations to evaluate the practical performances of proposed mechanisms
 and our results demonstrate that our mechanisms perform very well in most cases.
 
\end{abstract}

}

\maketitle

\section{Introduction}
\label{sec:intro}

%%%%%%% introduction section, written by Xiang-Yang Li

Big data and AI techniques have demonstrated remarkable capabilities in many areas~\cite{lecun2015deep},
such as noise monitoring~\cite{noisetube}, traffic analysis~\cite{D2019A},
visual object recognition~\cite{donahue2014decaf},  recommendations on e-commerce websites~\cite{zhang2019deep}.
Big data, as a key ingredient to unlock the power of artificial intelligence, %next generation AI technologies,
has become a strategic resource for economic and social development similar to the land, oil and capital~\cite{fernandez2020data}. 
Powerful deep learning models, such as speech recognition~\cite{hinton2012deep}, deep face recognition~\cite{deng2019arcface}, and deepfake forensics~\cite{li2020celeb},
 heavily rely on large amounts of specific training data, \eg, sensor data, picture, video, and so on.
Thus, for data consumers, effective approaches for data acquisition are urgently in need.
For example, one can adopt some crowdsourcing techniques to collect the needed data,
 or buy the required data from  data owners who are willing to share their data for profits, or buy the machine learning services without actually buying the data directly.

To facilitate the data exchange between data owners and data-based service providers, 
 emerging works are exploring various approaches to build data trading markets~\cite{liang2018survey,li2018can}.
Designing effective data trading markets has attracted a great amount of attentions recently~\cite{liang2018survey,li2018can,yu2017data,li2018can,pei2020data},
 and a number of data trading platforms  have emerged, such as
  DataExchange, Datacoup,  DATATANG,  GBDEx, Datacoup, Qlik, CitizenMe Factual, XOR Data Exchange.
% DataExchange\footnote{https://dataexchange.io/},  Datacoup\footnote{http://datacoup.com/}, 
% DATATANG\footnote{http://www.datatang.com/},  GBDEx\footnote{http://www.gbdex.com/website/},
% Datacoup, Qlik, CitizenMe Factual\footnote{https://www.factual.com/}, XOR Data Exchange.
In a typical trading scenario, data providers sell their data to the platform from where consumers can discover and purchase the data.
There are two main procedures for users of the data platform: data purchasing and data selling. 

Data trading is significantly different from trading of traditional commodities~\cite{li2018can}. 
There are many technical challenges and concerns in  designing effective data trading platforms, such as 
 trustable data rights determination and management~\cite{reichman1997intellectual,subramanya2006digital,rosenblatt2002digital}, 
 effective data quality evaluation~\cite{peng2015pay,pipino2002data,cai2015challenges,li2019todqa}, data privacy~\cite{zhang2018crowdbuy,torra2017data,isaak2018user,qian2018towards} especially unstructured data privacy, 
 privacy-preserving computing and learning~\cite{du2020patronus,zhang2018cloak,taeho-pda,li2017multi}, 
 strategy-proof trading mechanism design~\cite{cao2017data,an2019truthful},
 data tracing~\cite{tracing-L0WXCWL19,wang2018data}, 
 law and rule enforcement~\cite{dolvzan2017trading,veale2018algorithms},  
  accountability~\cite{jung2017accounttrade,jung2018accounttrade},  
  and intelligent data applications.
Data pricing, a critical fundamental mechanism for data markets, has become a hot topic both in industry and academia~\cite{pei2020data}.
Designing data pricing mechanisms for revenue (approximately) maximization, or social welfare maximization has attracted a great amount of attentions~\cite{pei2020data,sen2013survey,yu2017data,zheng2017online,tian2018toward,mao2019pricing,dhangwatnotai2015revenue,hart2017approximate}.  
%In this work, we focus on designing effective and efficient trading and pricing mechanisms.

In general, revenue maximization considers the aspects of cost minimization in data collection and profit maximization in data selling. 
Some existing works try to minimize the cost in the process of data purchasing via crowdsourcing~\cite{zhao2014crowdsource,zheng2017trading,han2017online,gong2018incentivizing,peng2015pay}. 
In this work, we focus on designing revenue-competitive truthful online mechanisms for trading data related digital products.
In other words, the seller of the data (which could be the data owner or the data trading platform) selects some buyers from a group of potential buyers
 to maximize the total payment paid by these selected buyers.
Previous works (\eg,~\cite{niu2019making,gong2018incentivizing,yu2017data,zheng2017online}) proposed various data pricing models based on privacy, quality and query-version.
As a supplement of existing work, we here investigate online mechanisms for trading raw data (or dataset\footnote{We use the ``data" or ``dataset" interchangeably to denote one  unit of the trading data or the data product such as an AI model.}) or some form of digital products such as software or machine-learning models by jointly considering the data copy-sensitiveness and time-sensitiveness:
\begin{itemize}
\item \textbf{Copy-sensitiveness}: The more scarce resource the more valuable it is. 
Specifically, being the only owner of some critical data (or some AI models, or some Apps) might benefit the owner to dominate a specific market. 
In this work, we study the scenario that all buyers are competitive to obtain the data-copyright or exclusive usage of some data which is universally unique. 
In other words, the specific data or the data product will be sold to only one carefully selected user to maximize the expected revenue of the seller.
\item \textbf{Time-sensitiveness}: In real data application scenarios, typically the perceived data value will change over time. In some cases such as monitoring~\cite{2020A} and recommendation systems~\cite{chen2018sequential}, the consumers prefer near-term data to improve  the prediction or classification accuracy. Thus, the data value may decrease over time. In other cases such as the time-series data~\cite{jung2016pda,niu2019making}, the value may increase over time since the data gradually becomes more accurate and complete.  
Besides, the variation of data value may be more complicated due to some specific considerations, \eg, increasing for some time and then decreasing. 
In general, we use a continuous "discount" function $d(t)$ 
\footnote{In this paper, we use the term ``discount" though the function $d(t)$ is not necessarily monotone decreasing as a function of $t$.}
to denote the data value fluctuation over time $t$. 
When a buyer $i$ has an \emph{initial} valuation $v_i$ for the data at time $0$, then the valuation becomes $v_i \cdot d(t)$ at time $t \ge 0$.
In this work, we assume that such a value discount function $d(t)$ is known to the buyers and also the seller.
\end{itemize}

In this work, our objective is to design strategy-proof mechanisms that can incentivize buyers to truthfully arrive in time and
 report his true valuation $v_i$ or discounted valuation $v_i d(t)$,  while approximately maximize the revenue and/or social efficiency.
The combined feature of copy-sensitiveness and time-sensitiveness increases the difficulty of mechanism design for revenue maximization in \emph{online data pricing}. 
Unfortunately, existing online mechanisms cannot be applied directly here.
In this work,  we focus on trading time-sensitive valued digital product (such as data or data-copyright, \etc) and 
our objective is to design an online mechanism with the following desirable properties: 
Individual Rationality, Incentive Compatibility (both time-IC and value-IC), Consumer Sovereignty, Competitiveness, and Computational Efficiency.
A mechanism is called truthful (or strategy-proof) if it satisfies both  Individual Rationality (IR, non-negative utility),  Incentive Compatible(value-bidding and time-arrival) and Consumer Sovereignty(CS, a chance to win).  We focus on designing truthful auction mechanisms with approximately optimum revenue.

When selling the data to potential buyers for  revenue maximization, typically there are two different approaches.
One is the \emph{posted price mechanism}, where
  the seller posts a public price for a single data and the buyers take-it-or-leave-it.
  \ie, the buyer decides to buy if its discounted valuation $v_i d(t)$ is at least this posted price. 
Previous works~\cite{xu2016dynamic, zheng2017online,mao2018online} usually model it as a multi-armed bandit (MAB) optimization.
However, the MAB-based approaches are all multi-round-play games, 
 while the problem studied here is to  consider the single-round trading for a specific digital good  or data copyright. 
The other is \emph{auction  mechanism} where each buyer reports a possible bid when arrives, 
 and the seller determines whether to sell the product to this buyer and at what price.
One typical auction mechanism is  \emph{online auction mechanism} when users arrive online.
Then when a user $i$ arrives, the user needs report a value representing his reserving price $r_i$ (called bid) 
 for his target data; and then the seller will \emph{immediately} decide whether to sell the data or not and the decision is irrevocable. 
Typically, the seller also needs to immediately determine the price the buyer needs to pay online when the user arrives.
In certain situations, the seller is allowed to postpone his decision on the amount of payment that the buyer needs to pay.

The problem studied here is similar to the online secretary problem, and its variations such as discounted secretary problem.
Without considering the time-sensitiveness,
 Hajiaghayi \etal~\cite{hajiaghayi2004adaptive} have presented a constant-competitive truthful online auction of single item.
However, when considering the discount function $d(t)$, each bidder's true valuation fluctuates overtime and thus the existing mechanism may result in an arbitrarily large competitive ratio with respect to offline Vickrey for revenue. 
When considering discounting valuations, by using  a two-stage sampling-accepting process,
 Wu \etal~\cite{wu2014strategy} presented a strategy-proof online auction in the multi-round-play setting, \ie, each data can be sold up to $g\geq 1$ copies in each time slot $t$.
 Unfortunately, such mechanisms cannot be directly applied to our setting where only one buyer will get the data.
The single-round online trading task studied in this work
is much more challenging due to the short of opportunities to learn the trading information. 

To the best of our knowledge, this work is the first to study the single-round online mechanism in the time-sensitive setting. 
Observe that even when all buyers are truthful in advance, all online algorithms have competitive ratio at least $\Omega(\frac{\log n}{\log \log n})$~\cite{babaioff2009secretary}. 
Generally, for all truthful online mechanisms, 
 we will first show  a lower bound of $\Omega(n)$ on the revenue competitive ratio when we have an arbitrary unknown discount function $d(t)$ (\S\ref{without_any_prior_knowledge}),    
 and a lower bound $\Omega(\frac{\log n}{\log\log n})$ for non-increasing discount functions. 
We design a sequence of online truthful auction mechanisms by the following approaches: 
 1) partition the discount function into different classes such that the discount values in each discount-class is within a constant $B>1$ of each other, \ie,
  the discount value in class $c$ is in the range $[B^{-c}, B^{-c+1}]$ (assuming that the maximum discount value is $1$);
  2) then run some mechanism on users from one carefully selected discount-class (in this sub-procedure, we will ignore the discount values of users as they are within a constant factor
   of each other).
Then we design an online \textit{ truthful} and $O(\log^2 n)$-competitive mechanism (called $\mathcal{M}_R$) when there is an \emph{arbitrary known} non-increasing
 discount function $d(t)$ (\S\ref{alg_unknown_random} and \S\ref{analysis_M_R}) and the number of users per discount-class is within a constant factor of each other.
 We also present  a simpler truthful mechanism $M_1$ with $O(\log n)$-competitive ratio.
Furthermore, without any restrictions on the sizes of the discount-classes, we  prove that the  mechanism $\mathcal{M}_1$ could have a competitive ratio
 as bad as $\Omega(n^2)$.
The design of online approximate algorithm, due to its worst-case nature, can be quite pessimistic when the input instance at hand is far from worst-case. 
Then we  propose an heuristic weighted selection mechanisms $\mathcal{M}_W$ (see \S\ref{alg_unknown_weighted}) and 
 a modified weighted selection mechanism $M'_W$ (see \S\ref{modified-weight}).
We show that the modified weighted selection mechanism $M'_W$ achieves a competitive ratio $O(n \log n)$ 
 for any known non-decreasing function $d(t)$, and by only assuming that each discount class has at least 2 users, \ie, $n_c \ge 2$.
When the optimum revenue can be estimated within a constant factor, we then design a mechanism $M_Z$ with a constant-competitive ratio. 
We also present several reserved-price mechanisms when the valuations of users are drawn from some given known distribution.
All mechanisms has linear time computation complexity.
We conduct extensive simulations to study the performance of proposed mechanisms,
especially the competitive ratios and the impact of different parameters.
Our numerical results  show that all mechanisms $\mathcal{M}_R$, $M_W$, and $M'_W$ do  achieve better performance 
 than its theoretical worst-case results in terms of revenue, and the performance of $\mathcal{M}_W$ is about ten times better than $\mathcal{M}_R$ in our evaluation (\S\ref{eva}).

The rest of the paper is organized as follows. 
In Section~\ref{sec:related}, we briefly review related work in the literature. 
In Section~\ref{sec:model}, we introduce the model of online auction with time discounting values, and recall important solution concepts used in this paper. 
We first analyze some lower bound of the competitive ratios of  truthful mechanisms in several adversary models in Section~\ref{sec:bound} for a various adversary models.
In Section~\ref{sec:alg}, we first assume that the adversary is arrival-sequence adaptive online, \ie, the valuations are not selected by the adversary and the adversary
 can only determine the arrival sequence of users.
We present a sequence of strategy-proof online auction mechanisms and analyze the computational and economic properties
 of these mechanisms. 
% In Section~\ref{alg_special_case}, we present several mechanisms when the valuations of users are drawn from a  known distribution.
In Section~\ref{sec:eva}, we report the extensive numerical evaluations of  performances of our proposed mechanisms compared with some related mechanisms proposed in the literature.
We conclude the paper in Section~\ref{sec:conclusion} with the discussion of some future research questions.

\section{Related work}
\label{sec:related}
%%%% related work section

%\subsection{Pricing Time-sensitive Goods}
%Revenue maximization in online auction.
%Competitive analysis.
%Auctions on Digital Products with Unlimited Supplies

%In the literature of revenue maximization mechanism design, existing works can be divided into two categories. %based on the assumptions about the agents' private valuations.   
%One is \textbf{Bayesian setting} in which the agents' private valuations are \iid drawn from a prior known distribution~\cite{myerson1981optimal}.
%The other category is about \textbf{prior-free} (\ie, without known distribution) approximations to the optimal Algorithm~\cite{goldberg2006competitive}.
%Auction is one of the dynamic selling mechanisms that yields good results when it is infeasible to do the market research to determine the best fixed price.

In the era of big data, data pricing has attracted considerable attentions and a series of data pricing strategies from different perspectives have been proposed. 
Niu \etal~\cite{niu2019making} studied trading time-series data by  compensating the data owners for their privacy losses, 
Yu \etal~\cite{yu2017data} proposed data pricing strategies based on data quality.
Zheng~\etal ~\cite{zheng2017online} investigated online pricing mechanisms for version-based data trading.
In this work, we focus on designing the online mechanism for trading time-sensitive valued data-copyright.

%\paragraph{Online Mechanism:} 
A closely related problem is to choose the maximum from a sequence. 
This well known problem is also called the secretary problem or optimal stopping problem.
Previous works~\cite{rasmussen1975choosing, babaioff2009secretary} have studied how to select an element with a discount function.
However, these approaches are not designed in the scenario where the participating agents could be selfish, thus, these methods cannot guarantee truthfulness
when the valuations of bidders are private information.
The design of profit-maximizing truthful online auctions for digital goods, or information, goods such as electronic books, software, and digital copies of music have been extensively explored in academic area.
Goldberg~\cite{goldberg2000competitive} is the first to introduce the notion of competitive auctions. 
They studied a class of offline single-round, sealed-bid  auctions for items in unlimited supply, and proposed truthful, constant-competitive auctions via random sampling.
Bar-Yossef \etal~\cite{bar2002incentive} studied the competitive incentive-compatible online auctions for unlimited digital goods.
When considering limited supply, Hajiaghayi \etal~\cite{hajiaghayi2004adaptive} presented constant-competitive truthful online mechanisms 
with respect to the offline Vickrey for revenue, using  a two-stage sampling-accepting process. 
In addition, a number of  previous works~\cite{blum2004online,blum2005near,zheng2017online,xu2017dynamic} explored online pricing mechanisms based on the multi-armed bandit problem. 
However, these mechanisms did not take time-sensitive valuations into consideration.

Since data has some new peculiar properties, \ie, time-sensitive valuations, previous works can not be applied any more. 
When we take the time-sensitive valuations into consideration, the problem become more challenging.
Some recent pricing models started to focus on goods whose valuations are time sensitive. %In other words, bidders' valuations are changing with time.
Lavi \etal~\cite{lavi2015online} studied the allocation of multiple identical items that they ``expire'' 
 at different times and proposed online auction mechanisms to approximately maximize the social welfare.
Wu \etal~\cite{wu2014strategy} presented a strategy-proof online auction with discounting valuations, but 
 they assume the discounting functions are known to the seller, and 
 their objective is to maximize social welfare instead of revenue.
They assume that the items can be sold up to $g$ highest bidding agents in each time slot $t$.
Xu \etal~\cite{xu2016dynamic} focused on pricing private personal data via multi-armed bandit approach with time-variant rewards.
Mao \etal~\cite{mao2018online} designed an online pricing mechanism based on the feature of discounting valuations with unlimited number of copies.
Romano~\cite{romano2020online} proposed posted-pricing mechanisms with unknown time-discounted valuations drawn from some distributions.
However, when we consider selling only one copy to some carefully selected buyers,
 the problem model is totally different and it becomes more challenging due to the short of opportunities of learning future information.

\section{Trading Model and Problem Definition}
\label{sec:model}

%%%%% \section{Model and Problem Definition} \label{sec:model}

\begin{figure}[ht!]
\begin{center}
	{\includegraphics[width=0.8\linewidth,trim={0.1cm 0.1cm 0.1cm 0.1cm}, clip]{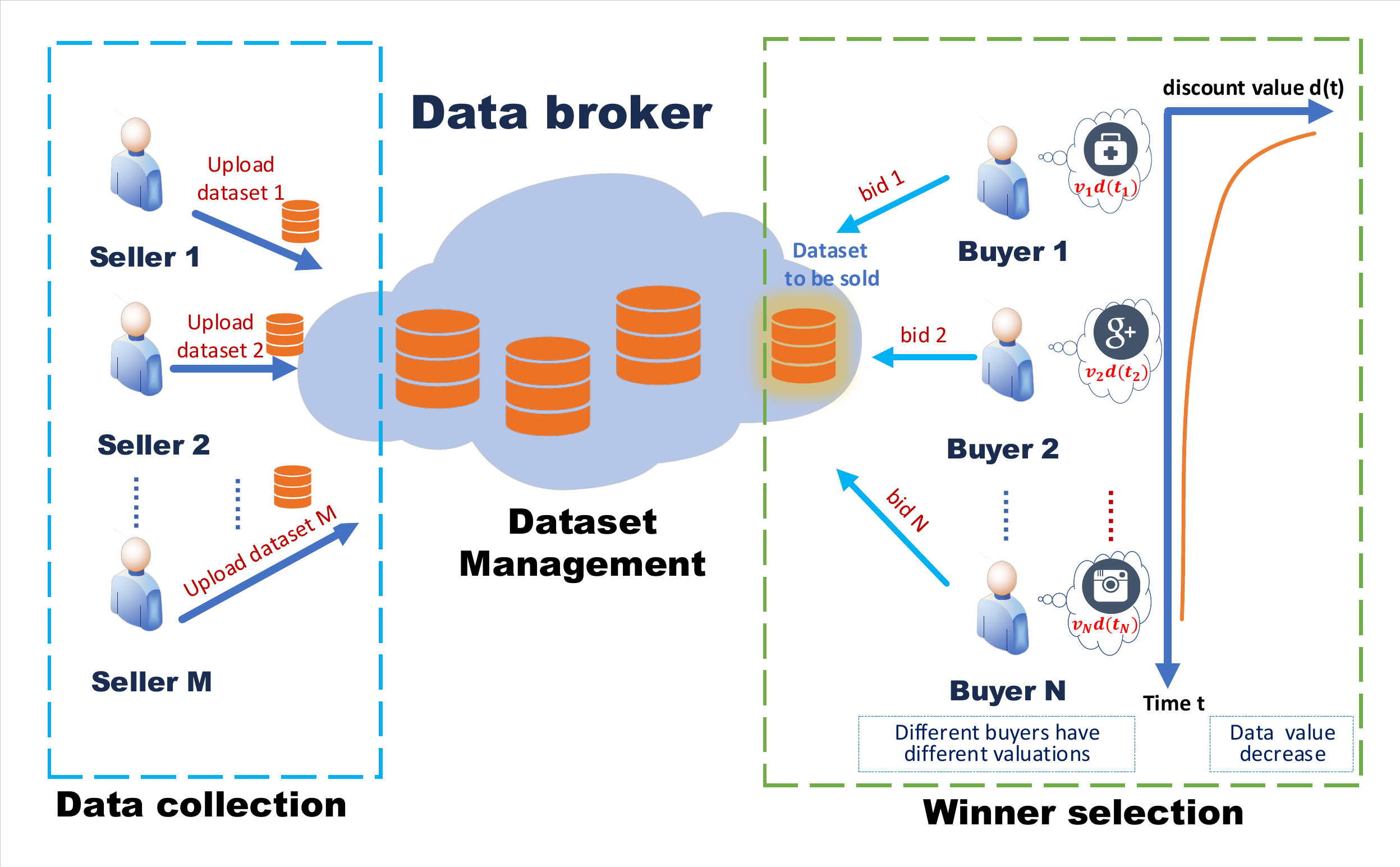}}
	\vspace{-0.15in}	
	\caption{System model for data trading.}
	\label{fig:system}
\end{center}
\vspace{-0.15in}
\end{figure}

In this section, we present the  system models of trading data related product such as data or data-copyright with time discounting valuations, 
 and introduce some related definitions and  concepts from algorithmic mechanism design.
There are three parties in our model: data \textit{owner}, data \textit{seller} and data \textit{consumer (\ie, buyer)},
see Figure~\ref{fig:system} for illustration.
The seller collects data from the owners and consumers buy some data from the seller to facilitate their data-driven services.
Potentially all these transactions are conducted on a data sharing/trading platform.

\textbf{Data broker:} 
Suppose the data broker has a set of datasets $\mathcal{D}$ and each dataset contains a set of organized data items after some process. 
 \eg, data cleaning and labeling.
Each dataset is associated with a description\footnote{The description may include the data content, time of generation, quality, volume, \etc. Methods of data exhibition is another challenging but important issue in data trading market.} for exhibition, and the data buyers decide which data set to buy according to the description.
For each dataset $\mathbb{D}\in\mathcal{D}$, the broker needs to take into account of three kinds of cost: 
 \textit{initial} cost  in data collection and cleaning; \textit{maintaining} cost in the procedures of data storage, organization, management,
   \etc; \textit{trading} cost including the cost for copy and others. 
We use  $C_{i}^{s}, C_{m}^s, C_{t}^s$ to denote dataset $\mathbb{D}_{s}$'s  initial, maintaining and trading cost respectively. 
For the sake of simplicity, we assume the total cost is a fixed and known constant in our system.
The objective of the data broker is to sell the data to potential buyers to gain maximum profit, or to maximize the social welfare.

\textbf{Data value:} It is very complicated to estimate the data value in practice. 
For example, different data consumers often have different evaluations on the same dataset and thus report different prices.
Data broker can estimate dataset's value by data quality, data function, market survey or other types of data value.	
Each potential data buyer $i$ has his own data evaluation $v_i$ according to data application scenario.
A data buyer  may estimate data value by current market situation and future forecast.
For instance, the data value may change following a \textit{non-increasing} function of time $d(t)$ in reality. 
Consequently, it is reasonable to assume that the broker will \textit{stop} providing the dataset $\mathbb{D}$ at some time $ t_e$,
when the dataset's market value is lower than the cost for holding it. 
In our model, we assume that $d(t)$ is known in advance to data seller and data buyers. 
Actually, this is a reasonable assumption in practice; for example, we can learn $d(t)$ by sufficient survey in the market. 

\textbf{Data buyers:}
In some trading scenarios, a buyer may cooperate with others to get lower purchasing cost, for example, share digital data (\eg movie, music) with other potential buyers. 
Here we assume there are \textit{no collusions} between buyers/bidders; actually
we assume that the buyers compete with each other and do not want to share their target datasets with others.
For a specific dataset $\mathbb{D}$, the $j$-th arrived buyer $i$, arriving at time $t_j$, has a \textbf{private} initial valuation $v_i$, representing the maximum price he is willing to pay; he has the utility  $v_i \cdot d(t_j)-p(t_j)$, if picked by the mechanism,
where $p(t_j)$ is the payment by buyer $i$ at the time $t_j$. 
All buyers are assumed to be selfish, that is, to maximize the utility, each buyer will manipulate his reported price $r_{j}$ strategically.  
We also assume that the buyers have the full knowledge of the auctioneer's mechanism when they bid.

\subsection{Trading Model}
\label{pro:formulation}

For simplicity, we focus on the selling of one specific dataset, or one specific data product. 
Assume that there are $n$ buyers ${\textbf U}= \{1, 2, \cdots, i, i+1, \cdots, n\} $ for this data who arrive online in a random order and follow some random process, 
 such as a Poisson process with the arrival rate $\lambda$.
To focus on the design of truthful mechanisms, without worrying about the subtle differences caused by the different arrival times from a random arrival process,
 in this work we assume that users's arrival follows a \emph{fixed sequence of time instances}.
 This sequence of time instances are assumed to be pre-determined, which could be generated by some random process in advance.
 For the sake of simple analysis, let ${\textbf T}=\{t_1, t_2, \cdots, t_i, t_{i+1}, \cdots, \}$ be the time sequence when the $i$-th arrived user will arrive at time $t_i$.
In this work, we use  a set $\textbf{v}= \{v_1, v_2, v_3, \cdots, v_{n-1}, v_n \} $ to denote the set of arbitrary given initial valuations of users, \ie,
 the user $i$ has an initial valuation $v_i$.
 An arrival sequence is a mapping $\pi$ from $\textbf T$ to $\textbf U$, \ie, the $j$-th arrived user is $\pi(j)$.
Given an arrival sequence $\pi$
 we use a vector $\pi({\textbf v}) = \vec {\textbf v} = [ v_{\pi(1)}, v_{\pi(2)}, v_{\pi(3)}, \cdots, v_{\pi(n-1)}, v_{\pi(n)} ] $  to denote
   a given sequence of initial valuations of users arrived in order $\pi$.
Let $T=[0,t_e]$ denote the time horizon of the auction, here time $0$ is the starting time of the auction, and $t_e$ is the time when the seller stops selling the data.
Each buyer $i$ has a \textit{private initial} valuation $v_i$ for the data at the time of data production.
The buyers' initial valuations are \textit{arbitrary unknown} without any prior knowledge, and the buyers have the full knowledge of the auctioneer's mechanism when they bid. 
We model the interactive process between the seller and the buyers as an \textbf{online auction}.
%It is reasonable to assume that the seller will {stop} providing the data at some time, when the data's value  is lower than the cost (considering at least
% three  kinds of cost: \textit{initial} cost in data collection, \textit{maintaining} cost in data management and \textit{trading} cost.) 

\textbf{Auction Mechanism Setting}:
In the auction mechanism setting, we assume the following.
We assume that the $j$th-arrived buyer has a user ID $\pi(j)$. Here $\pi$ is a mapping from time $t_j$ to a user with ID $\pi(j)$.
For simplicity of notations, sometimes we assume that $\pi(j)=j$ if no confusion is caused.
When the $j$-th buyer arrives, he submits a price (i.e., bid) $r_{j}$ 
 representing the maximum reported price he is willing to pay. 
For each $j$-th arrived buyer, let $t_j$ denote his arrival time instance and $v_j$ be his \textit{private} initial valuation. 
We say $h_{j}=v_{j}\cdot d(t_j)$ is the \textit{true} maximum value of the $j$-th  buyer (actually with an ID $\pi(j)$) arrived at time $t_j$ for the data product.
If $r_j = h_j$, we say that the user $j$ is truthful.
 In this work, we assume that the discount function $d(t)$ satisfies that $0 < d(t) \le 1$ for any value $t \ge 0$.
 In other words, for any user $i$, the valuation of the product cannot exceed its initial valuation $v_i$.
After receiving the bid, the seller must decide whether to sell  the data or not immediately and the decision is irrevocable. 
Here, we study a real trading scenario that the buyers are \textit{competitive} and there are \textit{no collusions} among them. 
In addition, we assume that all buyers are \textit{rational}:
 each buyer $j$ arrived at time $t_j$ reports his price $r_{j}$ strategically to maximize the utility.  
Then buyer $j$ arrived at time $t_j$ has the \textbf{utility}:
\begin{eqnarray}
\small
\label{for-utility}
u_j=
\begin{cases}
 v_j \cdot d(t_j) - p_j, & \quad \text{if picked as winner and pay $p_j$;}\\
0, & \quad\text{otherwise.}
\end{cases}
\end{eqnarray} 
where $p_{j}$ is the payment by $j$-th arrived buyer  if he is picked by the mechanism at his arrival time $t_j$. 
Let $E[u^j]$ denote the \textbf{expected utility} of user $j$, where $E[\cdot]$ is the expectation over the mechanism's randomizations if it has any.

\textbf{Posted-Price Mechanism Setting}:
In this case, the mechanism decides a posted price $p_t$ for all users arrived starting from $t$ and hereafter.
The seller may dynamically set multiple posted-prices $p_{t_1}$,  $p_{t_2}$, $p_{t_3}$, $\cdots$, $p_{t_m}$.
When a buyer $j$ arrives at a time $t \in [t_i,t_{i+1})$, this buyer $j$ is accepted as the winner iff his reported price $r_j \ge p_{t_i}$.
The winner will be charged a price $ p_{t_i}$.
The mechanism stops if one winner has been determined.
If $g \ge 1$ copies of the data will be sold, the mechanism stops if $g$ copies have been sold or all users have been processed.

\textbf{Revenue-Maximizing Objectives}:
We focus on studying the seller's revenue $R$, i.e.,  the sum of the winning buyers' payments minus the seller's total cost. 
For simplicity, 
we assume the data cost is constant and fixed, and 
we also assume that no two buyers arrive at the same time. 
If multiple buyers satisfy our decision rule at the same time, our mechanism can be modified by selecting the one who reports the highest price.
Formally, we aim to maximize the \textbf{revenue} $R$ of the mechanism
\begin{equation}
\small
\label{equ:objective1}
\begin{aligned}
\begin{cases}
& R= \sum_{j=1}^{n} y_j*p_j\\
& s.t. \quad \sum_{j=1}^n y_j \leq 1, \quad y_j\in\{0,1\}\\
\end{cases}
\end{aligned}
\end{equation}
Here the reported price vector $\mathbf{r}$, allocation vector $\textbf y$, and the payment vector $\textbf p$ are
  initialized with the zero vector $(0, 0,\dots,0)$ and updated online with each arrived buyer.
The mechanism $\mathcal{M}$ computes $(\mathbf{y},\mathbf{p})$ online based on the received report price information $\mathbf{r}$,
\ie, $\mathcal{M}(\mathbf{r}, d(t))=(\mathbf{y},\mathbf{p})$,
Notice that mechanism, composed of the payment rule $p$ and the allocation rule $y$, must be strategy-proof.
\begin{compactenum}
\item
 $\mathbf{y}=(y_1,y_2,\dots,y_n)$ is the \textit{allocation rule},
where $y_j=1$ indicates that the seller accepts the $j$-th arrived buyer as the winner;
$y_j=0$ rejects. 
 In this setting, we only choose one winner, \ie, $\sum_{j=1}^n y_j \leq 1$.
\item
 $\mathbf{p}=(p_1, p_2,\dots, p_n)$ is the \textit{payment rule}, where $p_j$ is the value paid by the $j$-th arrived buyer. 
\end{compactenum}

For the convenience of our understanding, we summarize some important notations in Table~\ref{tab:notationMR}.
\begin{table}[th]
{	\centering
\caption{Notations and abbreviations used in this work.}
\label{tab:notationMR}
\vspace{-0.15in}
\begin{center}
\small
	\begin{tabular}{c| p{0.7\columnwidth}}
	\toprule
	\textit{Notation} & \textit{Description} \\	
	\hline $\vec {\textbf v}$, or $\pi({\textbf v})$ & A vector of the valuations, \ie, a permutation $\pi$ of the initial valuation set $\textbf v$. \\
	\hline $t_{j}$ & The arrival time  of $j$-th arrived user in a given arrival time-sequence. \\	
	\hline $r_{j}$ & The reported price of $j$-th arrived user in a given arrival sequence. \\
	\hline $d_{j}$ & The discount value   of $j$-th arrived user for a given discount function and arrival time-sequence. \\		
	\hline $r_{(j)}(\pi)$ & The $j$-th largest report price in sequence $\pi$, sometimes abbreviated as $r_{(j)}$. \\
	\hline $r_{(j)}(\Pi)$ & The expected value of $j$-th largest report price given the set of arrival sequence $\Pi$. \\
	\hline $v_{\pi(k)} $ & The  valuation of the  $k$-th arrived user in an arrival sequence $\pi$. \\
	\hline $v_{k} $ & The $k$-th largest initial valuation in the valuation set $\textbf V$. \\
	\hline $v_{(k)}^c (\pi) $ & The $k$-th largest initial valuation of users in the class $c$ in the arrival sequence $\pi$. \\
	\hline $d_{max}^c$ & The maximum discount value in class $c$. \\		
	\hline $d_{min}^c$ & The minimum discount value in class $c$. \\
	\hline $\hat c$ & The number of reserved discount classes. \\
	\hline $n_c$ & The number of users in a discount class $c$.\\
	\hline $T_c$ & The time-interval of the discount class $c$.\\
	\hline $I_c$ & The discount values interval of the  class $c$.\\
	\hline $c^{(k)} (\pi)$ &  The class id where the $k$-th largest report price $r_{(k)}$ is from in arrival sequence $\pi$.\\
	\hline $\varrho({\cal M})$ & The competitive ratio of a mechanism $\mathcal{M}$ against a \emph{fully adaptive-online} adversary.\\
	\hline  $\varrho_{d}({\cal M})$ & The competitive ratio of a mechanism $\mathcal{M}$ against a \emph{valuation adaptive-online} adversary.\\
	\hline  $\varrho_{\vec{\textbf v}}({\cal M})$ & The (worst case) competitive ratio of a mechanism $\mathcal{M}$ against a \emph{discount-function adaptive-online} adversary.\\
	\hline $\rho_{\textbf{v},d}({\cal M})$ & The \emph{expected} competitive ratio of a mechanism $\mathcal{M}$ 
 against  an \emph{arrival-sequence adaptive-offline} adversary. \\
	\hline		
	\end{tabular}
\end{center}
}
\end{table}

%\vspace{-0.1 in}
\subsection{Adversary Models and Competitive Ratio}

For online mechanism design,  there are some common models about the adversary (the power of the adversary from weakest to weak, to medium, to strong)
that could affect the performance of the mechanism: 
\begin{enumerate}
\item
\emph{Oblivious adversary} (also called weakest adversary):
The oblivious adversary knows the online mechanism, but not the coin toss of the mechanism if there is any.
The adversary has to generate the \emph{entire} request sequence in advance before any requests are processed by the online mechanism. 
The adversary is charged the cost of the optimum offline algorithm for that sequence.
Notice that the adversary could be \emph{valuation oblivious} or  \emph{discount-function oblivious}, or both.

\item
\emph{Adaptive-online adversary} (also called medium adversary): 
This adversary may observe the online mechanism and generate the next request based on the algorithm’s (randomized) answers to  \emph{all previous requests}. 
The adversary must serve each request online, \ie, without knowing the random choices made by the online algorithm on the present or any future request.
This adversary must make its own decision before it is allowed to know the decision of the online mechanism.

\item
\emph{Adaptive-offline adversary} (also called strong adversary):
This adversary knows  everything of the mechanism design, even the random number generator used by the randomized mechanism. 
This adversary is so strong that randomization does not help against it.
This adversary is allowed to make its own decision after it knows the decision of the mechanism.
The adversary then generates a request's values and its arrival sequence adaptively. However, it may serve the sequence to the mechanism offline.

\item
\emph{Arrival-sequence adaptive-online adversary} (also called weak adversary): 
The values of the bidding-price $\textbf{v}$ are chosen by the adversary in advance and the set of values are given to the mechanism.
Then the adversary can only decide the arrival sequence  of $\textbf{v}$ (and maybe the exact arrival time\footnote{Notice that in this work, we often assume that  the arrival times of all $n$ users are determined by the Poisson process. Thus it is not determined by the adversary.}) when the user $i$ will arrive, adaptively based on the historical decisions of the mechanism.

\item
\emph{Valuation adaptive-online adversary} (also called medium adversary):
 The adversary may observe the online mechanism historical decisions and can choose the  valuation sequence $\vec{\textbf{v}} \ge 0$ 
 adaptively based on the historical decisions of the mechanism.
 Here the value $v_i$ of the to-arrive user $i$ is dynamically chosen by the adversary.
 
\item
\emph{Discount-function adaptive-online adversary} (also called medium adversary): 
The adversary may observe the online mechanism's historical decisions and  can choose the values of the discounting function $d(t) \ge 0$   arbitrarily and adaptively based on the historical decisions of the mechanism.
\end{enumerate}

Similarly, we can define \emph{arrival-sequence  adaptive-offline adversary}, 
\emph{valuation  adaptive-offline adversary}, and \emph{discount-function adaptive-offline adversary}. 
An adversary is \textbf{fully adaptive-online} if it is both valuation  and discount-function adaptive-online.
An adversary is \textbf{fully adaptive-offline} if it is both valuation  and discount-function adaptive-offline.
In this work we will design various mechanisms under different adversary models, mostly under the adaptive-online adversary models.
Notice that the adaptive-offline adversary is more powerful than the adaptive-online adversary.

We use the well known \textbf{offline} truthful Vickrey auction~\cite{vickrey1961counterspeculation} mechanism $\mathcal{M}_V$ as the baseline for comparison. 
Given all the reported prices (or valuations) in advance, Vickrey mechanism $\mathcal{M}_V$ always selects the highest bidder but charges the second-highest bid, and this simple mechanism
 assures the truthfulness of the buyers. 
Let $\mathcal{M}( \vec {\textbf{v} })$ denote the revenue  achieved by a mechanism $\mathcal{M}$ given the valuation sequence $\vec{\textbf v}$.
Recall that  $\mathcal{M}( \vec {\textbf{v} })$ is a determined value here if the mechanism $\mathcal{M}$ is not random.
The competitive ratio of a mechanism $\mathcal{M}$ with respect to the performance of offline Vickrey auction mechanism $\cal M_V$ against an adversary
can be defined as follows based on the power of the adversary.

\begin{definition}
The (worst case) competitive ratio of a mechanism $\mathcal{M}$ against a \emph{fully adaptive-online} adversary, with respect to the Vickrey mechanism, is defined as
\begin{eqnarray}
\label{varrho}
\varrho({\cal M})= \max_{\vec {\textbf{v}} \ge 0, d(t) \ge 0} \frac{\mathcal{M}_V(\vec {\textbf{v}}, d(t)  ) }  {\mathcal{M} (\vec {\textbf{v}},  d(t) ) }
\end{eqnarray}
\end{definition}

Here ${\mathcal{M} (\vec {\textbf{v}},  d(t) ) }$ is the revenue produced by the mechanism $\mathcal{M}$ 
 given the sequence of arrived users with valuations  $\vec {\textbf{v}}$, and the discount function $d(t)$.
Notice that when the mechanism $\cal M$ is a randomized mechanism, 
 we use $E[\mathcal{M}]$ to denote the expected revenue gained by $\mathcal{M}$.
In this case,  we will use $E({\mathcal{M} (\vec{\textbf{v}},  d(t) ) })$ instead  in Eq.\eqref{varrho}.

Similarly, we can define
\begin{definition}
Given a discount function $d(t)$, known to the mechanism $\cal M$,  the competitive ratio $\varrho_{d}({\cal M})$
of the mechanism $\mathcal{M}$ against a \emph{valuation adaptive-online} adversary is defined as
\begin{eqnarray}
\label{varrho-d}
\varrho_d({\cal M})= \max_{\vec{\textbf{v}} \ge 0} \frac{\mathcal{M}_V(\vec{\textbf{v}},d(t) ) }  {\mathcal{M} (\vec{\textbf{v}},d(t) ) }.
\end{eqnarray}
\end{definition}

\begin{definition}
Given the valuation sequence $\vec{\textbf{v}}$, known to the mechanism $\cal M$, 
 the competitive ratio $\varrho_{\vec{\textbf v}}({\cal M})$ of the mechanism $\mathcal{M}$ against a \emph{discount-function adaptive-online} adversary is defined as
\begin{eqnarray}
\label{varrho-a}
\varrho_{\vec{\textbf{v}}}({\cal M})= \max_{d} \frac{\mathcal{M}_V(\vec{\textbf{v}} ,d(t) ) }  {\mathcal{M} (\vec{\textbf{v}} ,d(t) ) }.
\end{eqnarray}
\end{definition}

One goal of this work is to design some mechanisms $\mathcal{M}$ with smaller competitive ratios $\varrho$, $\varrho_d$, or $\varrho_{\vec{\textbf{v}}}$.
We will mainly focus on the design of mechanisms with good competitive ratios $\varrho$, $\varrho_d$.

When the adversary is arrival-sequence adaptive-online, we study the expected performance of the mechanism.
We consider the expected value over both the mechanism's random choices (if any) and the random order of buyers' arrival. 
Let $\mathbf{T}=\{t_1, t_2,\dots, t_n\}$ and $\mathbf{V}=\{v_1, v_2, \cdots, v_n\}$ be the sets of time instants and buyers' initial prices 
 (not the reported prices at time $t_i$), respectively. 
 And we denote by $\Pi$ the set of all the $n!$ bijective ordering functions from $\mathbf{T}$ to $\mathbf{V}$ (note $|\mathbf{T}|=|\mathbf{V}|$).
For each possible permutation $\pi\in\Pi$, we use $\pi(t_j)=i$ to denote that the buyer $i$ will appear at time $t_j$, 
 or simply $\pi(j)=i$.

We first define the \emph{expected} competitive ratio $ \rho_d({\cal M})$
of the mechanism $\mathcal{M}$ against  an \emph{arrival-sequence adaptive-online} adversary, with
 some  discount function $d(t)$ given to the mechanism  $M$ and the adversary $\cal A$.

\begin{definition}
Given a known discount function $d(t)$, and known set of initial values ${\textbf v}=\{ v_1, v_2, \cdots, v_n\}$,
 the \emph{expected} competitive ratio of a mechanism $\mathcal{M}$ 
 against  an \emph{arrival-sequence adaptive-offline} adversary is defined  as
\begin{eqnarray}
\label{rho-d-a}
\rho_{\textbf{v},d}({\cal M})=\frac{E[\mathcal{M}_V]}{E[\mathcal{M}]}=
 \frac{\frac{1}{n!}\sum_{\pi \in \Pi}\mathcal{M}_V(\pi, d(t))}{\frac{1}{n!}\sum_{\pi \in \Pi}\mathcal{M}(\pi, d(t))}.
\label{for-ratio-d-a}
\end{eqnarray}
\end{definition}

Recall that $\textbf v$ is the set of valuations, and  $\vec{\textbf v}$ is a permutation of $\textbf v$.
In this work we assume that the arrival of users follows a given distribution, such as Poisson arrival. 
We say the adversary chooses the arrival sequence $\vec{\textbf v}$  of all buyers,  it means that when the arrival times of all $n$ users are determined by the Poisson process,
 the adversary then map, using a mapping $\pi$, these users to a sequence  $\vec{\textbf v}$ of the original set  ${\textbf v}$ of valuations.

Then we consider the worst possible competitive ratio when $d(t)$ could be arbitrary, while the
set of initial values ${\textbf v}$ is known in advance.
\begin{definition}
Given the set of known initial values ${\textbf v}=\{ v_1, v_2, \cdots, v_n\}$,
the \emph{expected} competitive ratio of the mechanism $\mathcal{M}$ against  an \emph{arrival-sequence adaptive-offline} adversary,
 in the worst case of all possible discount functions $d$ is defined  as
\begin{eqnarray}
\label{rho}
\rho_{\textbf v}({\cal M})=\frac{E[\mathcal{M}_V]}{E[\mathcal{M}]}=
\max_{d(t) \ge 0} \frac{\frac{1}{n!}\sum_{\pi \in \Pi}\mathcal{M}_V(\pi, d(t))}{\frac{1}{n!}\sum_{\pi \in \Pi}\mathcal{M}(\pi, d(t))}.
\label{for-ratio}
\end{eqnarray}
\end{definition}

In other words, $\rho_{\textbf v}({\cal M})=\max_{d} \rho_{\textbf{v},d}({\cal M})$.
Similarly, we can define $\rho_{d}({\cal M})$ when the discount function $d(t)$ is given to the mechanism.
Our goal is to design a mechanism $\mathcal{M}$ with smaller expected competitive ratio $\rho({\cal M})$, $\rho_d({\cal M})$, 
and $\rho_{\textbf v}({\cal M})$ (close to 1).

\subsection{Concepts of IR, IC, CS, and Truthfulness}

A mechanism is incentive-compatibility (IC) if  each participating buyer will maximize its (expected) utility when the buyer truthfully arrives in time and truthfully reports
	 his discounted valuations. 
	 We consider two kinds of incentive compatible, \textit{value-bidding} and \textit{time-arrival} truthfulness.
\begin{definition} 
	\label{def:ic}
	A mechanism is \textbf{incentive-compatible (IC)} if it is value-IC and semi-time-IC  for any buyer $j$:
	\begin{compactenum}
		\item\textbf{value-truthful (or value-IC):} the bidder $j$'s utility $u_j$ (or expected utility $E[u_j]$) cannot be improved by submitting a price $r_j' \not=v_j*d(t_j)$,
		no matter what others submit, \ie, $u_j (v_j*d(t_j) )\geq u_j(r_j')$.
		\item\textbf{semi-time-truthful (or semi-time-IC):} any bidder $j$ cannot improve his utility by delaying his arrival to any time $t_q $ with $t_q> t_j$.
		 Notice that here we assume that the buyer cannot arrive earlier than  when he can actually arrive.
		 That is, $\forall t_q, t_q > t_j$, $v_i\cdot d(t_q)-\xi_q \leq v_i \cdot d(t_j)-\xi_j$, here $\xi_j$ is the payment charged to the buyer at time $t_j$.
	\end{compactenum}
\end{definition}

In addition to require the mechanism $M$ to be incentive compatible, sometimes, we also require that the mechanism to satisfy some additional properties
\begin{compactenum}
	\item \textbf{Individual Rationality (IR):} Each participating buyer will have a non-negative utility, \ie, $u_i \geq 0$.
	\item \textbf{Consumer Sovereignty (CS):} Each user will have a chance to win the auction if only its bid is sufficiently high while others arrived before are fixed.
	\item \textbf{Competitiveness}: 
	We use competitive analysis to theoretically evaluate the performance guarantee of our mechanisms.
	Following~\cite{hajiaghayi2004adaptive}, we use the offline {Vickrey mechansism}~\cite{vickrey1961counterspeculation} as the baseline for comparison.
	\item \textbf{Computational Efficiency:} All decisions should be computed in polynomial time as each buyer arrives.
\end{compactenum}

A mechanism satisfies IC is called IC-mechanism and a mechanism satisfies IR is called IR-mechanism.
Then we can call the mechanism to be truthful-mechanism if it is IR-mechanism and IC-mechanism.
In our online scenario, we want the mechanism to be strategy-proof (or called truthful) if it satisfies IR and IC.

\begin{definition} 
	\label{def:truthful}
	A mechanism is \textbf{truthful} if it is value-truthful, semi-time-truthful, and individual-rational  for any buyer $j$:
\end{definition}

In offline, it is well known that a single-parameter mechanism is value truthful (in dominant strategies) if it is bid monotonic (a winner would keep winning if she increases her bid). Thus a monotone algorithms can be converted to truthful mechanisms~\cite{nisan2007algorithmic}.
Notice that our definition of \textit{semi-time-truthfulness} is different from the traditional ones, as here we do not consider the case that a losing buyer delays his arrival to attend the auction again.
In our scenario, even this happens, the expected revenue obtained from our mechanism will not decrease.
The design of a truthful multi-parameter mechanism \cite{chawla2010multi} turns out to be more challenging when the agent (here the potential buyers) can 
 manipulate multiple inputs (in the setting studied in this work, a buyer can change the reported price and the arrival time), instead of a single parameter input.

\section{Lower Bound of Competitive Ratios}
\label{sec:bound}
\label{without_any_prior_knowledge}

We  show some lower bounds $\Omega(n)$  or $\Omega(\frac{\log n}{\log \log n})$,
 depending on the assumptions on the known information and adversary models, 
 on the revenue competitive ratio of truthful mechanisms against some commonly used adversary models.
Here, we study the case that the discount function $d(t)$ is arbitrary and unknown.
In Table~\ref{tab:lower-bounds}, we summarize all the lower bounds of the competitive ratios proved in this work.
Recall that $\varrho(M)$ is the worst case competitive ratio of the mechanism $M$, and $\rho(M)$ is the expected competitive ratio of the mechanism $M$ against arrival-sequence
 adaptive-online adversary.
% and we have the following theorems about the competitive ratios of truthful mechanisms.

\begin{table*}[t]
{\centering
\caption{Some Lower Bounds for Competitive Ratio of IR-Mechanisms  or Strategy-proof Mechanisms}
\label{tab:lower-bounds}
\vspace{-0.25in}
\begin{center}
	\small
	\begin{tabular}{c|c|c|c|c}
	\toprule
	\textit{Mechanism's infomation}  & Truth & \textit{Adversary's infomation} & \textit{Lower-Bound} & \textit{Theorem} \\	
	\hline		
	know $d(t)$, not know $\textbf v$ & IR&  valuation adaptive-online, \ie, can choose $\textbf v$, and $\vec{\textbf{v}}$  &  $\varrho(M)=\infty$ &Theorem \ref{theo:unknowdt-fixed}\\
	\hline		
	know $d(t)$, not know $n$,  & IR& valuation adaptive-online,, \ie, can choose $\textbf v$, and $\vec{\textbf{v}}$  & $\varrho(M)=\infty$ &Theorem \ref{theo:unknowdt-fixed-random-1}\\
	\hline	
	know $n$, not know $d(t)$,  ${\textbf{v}}$ &  IR& adaptive-online, \ie, can choose $d$, $\textbf v$ and $\pi$, thus $\vec{\textbf{v}}$  &  $\varrho(M)=n/4$ &
	Theorem 	\ref{theo:unknowdt-fixed-random}\\
	\hline	
	know $n$ and $d(t)$, not know  ${\textbf{v}}$  &IR & valuation adaptive-online, \ie, can choose $\textbf v$, $\pi$ & $\varrho(M)=n/4$ &
	Lemma \ref{theo:known-d-adaptive-a} \\
	\hline	
	know $n$ and ${\textbf{v}}$, not know  $d(t)$,&  IR& discount adaptive-online, \ie, can choose $d(t)$  &  $\varrho(M)=n/4$ &
	Theorem \ref{theo:known-a-adaptive-d} \\
	\hline		
	know $n$, $d(t)$ &  IR, IC&  adaptive-online, \ie, can choose $d(t)$, $\textbf v$  &  $\rho(M)=\infty$ &
	Theorem  \ref{theo:known-dt-truthful} \\
	\hline		
	know $n$, $\textbf v$ &  IR, IC&  adaptive-online, \ie, can choose $d(t)$, $\textbf v$  &  $\rho(M)=\infty$ &
	Theorem  \ref{theo:known-dt-truthful} \\
	\hline		
	know $n$, $d(t)$ &  IR, IC&  adaptive-online, \ie, can choose $d(t)$, $\textbf v$, and $n_i\ge 2$  &  $\rho(M)=\Omega(n)$ &
	Theorem  \ref{theo:known-dt-a-truthful-2} \\
	\hline		
	know $n$, $d(t)$ &  IR & adaptive-online, \ie, can choose $d(t)$, $\textbf v$  &  $\rho(M)=\Omega(\frac{\log n}{\log\log n})$ &
	Theorem~\ref{theo:lowerbound-non-inc} \\
	\hline		
	\end{tabular}
	\end{center}
}
\end{table*}

%%%%%%%%%%%%%%%%
\subsection{Bounds for General Discount Functions}
\label{subsec:general-lower-bound}

\begin{theorem}
\label{theo:unknowdt-fixed}
	For any deterministic online auction IR-mechanism $M$, 
	the mechanism $M$ has an arbitrarily large competitive ratio $\varrho(M)$ in the worst case against \emph{valuation adaptive-online} adversary,
	\ie, there is some initial valuation sequence $\vec{\textbf{v}}$ chosen by the adversary such that $\varrho_d(M)$ is unbounded for an arbitrary  $d(t)>0$.
\end{theorem}
\begin{proof}
Given an online deterministic mechanism $M$, assume that it has a bounded competitive ratio  $\varrho_d(M)\le \beta$.
Assume that for any bidding-price sequence $\vec{\textbf{v}}$, and any arbitrary discount function $d(t)$, 
 it makes a decision to sell the product to some user $i$ (depending on the historical values of $\vec{\textbf{v}}$ and $d(t)$) at time $t_i$. 
 Then we trigger the arrival of a new user $i+1$ at time $t_{i+1}$, with an sufficiently large bidding-price $v_{i+1}$  %(or sufficiently small discount $d(t_{i+1})>0$)
 such that 
 $v_{i+1} d(t_{i+1}) > \beta v_{i} d(t_i)$. We also trigger the arrival of a new user $i+2$ at time $t_{i+2}$ such that  $v_{i+1} d(t_{i+1})=v_{i+2} d(t_{i+2})$.
 Then the optimum revenue by Vickrey auction is at least $v_{i+1} d(t_{i+1})$, while the revenue by the mechanism $M$ is at most $v_i d(t_i)$ due to the property of individual rationality.
 Thus, the competitive ratio   $\varrho_d(M)$ of $M$ for this new instance is at least $\frac{v_{i+1} d(t_{i+1}) }{v_i d(t_i)} >\beta$. It is a contradiction.
\end{proof}

Notice that Theorem \ref{theo:unknowdt-fixed} only requires the mechanism to satisfy IR.
It still holds even if the mechanism $M$ knows  the discount function $d(t)$, but not the initial bidding price $\vec{\textbf{v}}$.
Similarly, if the mechanism knows the exact initial bidding-price sequence $\vec{\textbf{v}}$, we have

\begin{lemma}
\label{theo:unknown-a-fixed3}
For any deterministic online auction IR-mechanism $M$, 
 the mechanism $M$ has an arbitrarily large competitive ratio $\varrho(M)$ in the worst case against \emph{valuation adaptive-online adversary},
	\ie, when given the discount-function $d(t)$, the adversary can dynamically choose the valuation $\vec{\textbf{v}}$
	 such that $\varrho_{d}(M)$ is unbounded.
\end{lemma}

\begin{lemma}
\label{theo:unknowdt-fixed2}
For any deterministic online auction IR-mechanism $M$, 
 the mechanism $M$ has an arbitrarily large competitive ratio $\varrho(M)$ in the worst case against \emph{discount-function adaptive-online adversary},
	\ie, given  an arbitrary $\textbf{v}$, the adversary can choose some discount-function values $d(t)$ such that $\varrho_{\vec{\textbf{v}}}(M)$ is unbounded.
\end{lemma}

We then prove some lower bounds on the competitive ratios for randomized truthful mechanisms. 
Similarly, we can prove the following theorem.

\begin{theorem}
\label{theo:unknowdt-fixed-random-1}
If a IR-mechanism $M$, even randomized, does not know the number of users to arrive and the end time of the game (\ie, the time when the last user to arrive),
 any  randomized IR-mechanism $M$ cannot have a bounded competitive ratio $\varrho(M)$.
\end{theorem}
\begin{proof}
The basic idea to prove this theorem is that as the mechanism does not know how many users to arrive, it has to accept the early arrived users.
When a buyer $i$ arrives, if the mechanism accepts this user, then the adversary can trigger the arrival of the other users with a much larger $v_j d(t_j) \gg v_i d(t_i)$
 for $j > i$,
 and $v_{n-1} d(t_{n-1}) = v_n d(t_n) \gg v_j d(t_j)$ for all $j \le n-2$.
 Then the revenue by Vickrey auction is $v_n d(t_n)$,  while the revenue by $M$ is at most $v_i d(t_i)$.
 Then $\varrho(M)$ is unbounded.
 
 Similarly, for a randomized mechanism, assume that the mechanism $M$ accepts a user $i$  with some probability $p_i$ and $\sum_{i=1}^{m} p_i= 1$ for some $m$, 
 then the adversary can trigger the arrival of new users $m+1$ and $m+2$ with much larger $v_j d(t_j)$ for $j=m+1, m+2$.
 Then the revenue by Vickrey auction is $v_{m+1} d(t_{m+1})$, while the revenue by $M$ is at most $\sum_{i=1}^{m} p_i v_i d(t_i) \le \max_{i=1}^m v_i d(t_i) \ll v_{m+1}d(t_{m+1})$.
 Then  $\varrho(M)$ is unbounded.
\end{proof}

Then we consider the case for every newly arrived user $i$, the adversary knows the probability $p_i$ that the mechanism $M$ will accept
 the user $i$.
Notice that $M$ could compute $p_i$ using all historical values of $\vec{\textbf{v}}$ and $d(t_j)$ for $j \le i$.
Clearly, adaptive offline adversary is a sufficient condition  for this capability, \ie, the adversary has a strong power.
Then we have the following theorem.

%\begin{theorem}
%\label{theo:unknowdt-fixed-random}
%Assume that both the mechanism and the adversary know that there are $n$ bidders to arrive.
%	For any  online randomized auction IR-mechanism $M$, 
%	the mechanism $M$ has a competitive ratio $\varrho(M) \ge n/4 - \epsilon$ for any small value $\epsilon >0$ in the worst case against an adaptive-offline adversary
%	 or the adversary who can always know the value of the probability $p_i$.
%\end{theorem}

\begin{theorem}
\label{theo:unknowdt-fixed-random}
Assume that both the mechanism and the adversary know that there are $n$ bidders to arrive.
	For any  online randomized auction IR-mechanism $M$, 
	the mechanism $M$ has a competitive ratio $\varrho(M) \ge n/4 - \epsilon$ for any small value $\epsilon >0$ in the worst case against an adaptive-online adversary.
%	 adaptive-offline adversary or the adversary who can always know the value of the probability $p_i$.
\end{theorem}
\begin{proof}
Notice that the mechanism $M$ has to make an online and irrevocable decision on whether to sell the product to each of the arrived users $i$.
Let $p_i \in [0,1]$ be the probability that the mechanism will sell the product to user $i$ arrived at time $t_i$, where $\sum_{i=1}^{n} p_i =1$.  
Note that the probability $p_i$ could depend on all historical reported prices $b_j$ (where $b_j =v_j d(t_j)$) for $j \in [1,i]$, including the user $i$.
%Let $b_i = v_i d(t_i)$ for simplicity of the proof.

We will see how the adversary  $\cal A$ and the mechanism $M$ play the game as follows.
The mechanism $M$ first receives the bid $b_1=b$ and decides a probability $p_1$.
If $p_1 > 2/n$, then the adversary will trigger the arrival of a new user with $b_2 \gg b_1$.
The basic idea by the adversary $\cal A$  is that when the mechanism $M$ chooses a user with a ``large" probability $p_i$
 (large means that $p_i>2/n$), the adversary will keep triggering users with a much larger bid $b_{i+1} \gg b_i$. 
Thus, the mechanism $M$ will have a regret to allocate ``large" probability to previous users.
This process is repeated till at some time, the mechanism $M$ accepts a user $i$ with a probability $p_i \le 2/n$.
Whenever this happens (including $p_1 \le 2/n$), the adversary will trigger a new user $i+1$ with the same bid $b_{i+1}=b_i$ (this is to make sure that in the Vickrey auction, the first price and the second price are same).
To see what happens when $p_i \le 2/n$, let us start with $p_1 \le 2/n$ as an example: then the adversary will trigger the arrival of a new user $2$ with $b_2 =b_1$.
Then the mechanism $M$ will compute the probability $p_2$:\begin{itemize}
\item
If $p_2 \le 2/n$, then the adversary will trigger the arrival of remaining users with $b_i =0 $ for $i \ge 3$.
In this case, the maximum revenue achieved by the mechanism $M$ is at most $\sum_{i=1}^{n} p_i b_i =\frac{4}{n} b$,
 as when it sells the product to user $i$, it receives at most $b_i$ in return due to  the constraint of individual rationality.
 Notice that the revenue by the Vickrey auction in this case is $b_1$.
 Then $\varrho(M)= n/4$.
\item
Otherwise, we know that  $p_2 > 2/n$, then the adversary will trigger the arrival of a new user with $b_3 \gg b_2$.
 Then if $p_3 \le 2/n$, then the adversary will trigger the arrival of a new user with $b_4=b_3$.
 If $p_4 \le 2/n$, then the  adversary will trigger the arrival of remaining users with $b_i =0 $.
 In this case, the maximum revenue achieved by the mechanism $M$ is at most $\sum_{i=1}^{n} p_i b_i \sim \frac{4}{n} b_3$, as $b_3=b_4 \gg b_2$.
 Notice that the revenue by the Vickrey auction in this case is $b_3$ as $b_3=b_4 \gg b_i$ for $i \not = 3,4$.
 Then $\varrho(M)= n/4 -\epsilon$ for an arbitrary small $\epsilon >0$.
 \end{itemize}
 
 Formally, we can extend this design to a general game played between  $\cal A$ and  $M$ as follows.\\
 \textbf{Game between  the adversary $\cal A$ and the mechanism  $M$}:
\begin{enumerate}
\item
Whenever the mechanism $M$ decides to accept the user $i$ with a probability $p_i > 2/n$, the adversary will trigger the arrival of new users with arbitrarily large $b_{i+1} \gg b_j$ for all $j \le i$ such that  the mechanism $M$ will have sufficiently large regret to accept $i$ with not so-small probability $p_i >2/n$.
%If the mechanism makes too much $p_i$s that $p_i>2/n$, when a user with extremely large bid arrives, the mechanism has close to $0$ probability to accept this user at the end.
%It will result a very bad performance for the mechanism compared with the Vickrey auction.
\item
Whenever the mechanism decides to accept a user $i$ with $p_i \le 2/n$, the adversary will trigger the arrival of a new user $i+1$ with $b_{i+1} =b_i$.
The adversary hopes that the mechanism $M$ will again accept the user $i+1$ with probability $p_{i+1} \le 2/n$.
\item
Whenever the mechanism $M$ accepts  two consecutive users $i$ and $i+1$ with  $p_i \le 2/n$ and $p_{i+1} \le 2/n$ where $b_{i+1} = b_i \gg b_j$
 for $j \le i-1$, then the adversary will trigger the arrival of remaining users with $b_j =0$ for $j >i+1$.
 The game then stops.
 \end{enumerate}
  
The adversary $\cal A$ and  $M$ will play the aforementioned game till all users are triggered to arrive and are processed by $M$.
 In general, when the mechanism $M$ makes an online  decision and plays the aforementioned game with the adversary $\cal A$, there are two cases:
 
\textbf{Case 1}: there are two consecutively arrived users when the mechanism $M$ accepts both with a probability at most $2/n$.
Let  $i$ and $i+1$ be the first of these two accepted users during the process of the game played by $\cal A$ and $M$.
In our game design, both users  have the same bidding $b_{i+1} =b_i \gg b_j$ for $j < i$.
 Then the adversary will trigger the arrival of remaining users with $b_j =0$ for $j >i+1$.
 In this case, the   maximum revenue achieved by the mechanism $M$ is at most $\sum_{j=1}^{n} p_j b_j \le \frac{4}{n} b_i +\epsilon$, as $b_i=b_{i+1} \gg b_j$ for $j\not =i, i+1$.
 Notice that the revenue by the Vickrey auction in this case is $b_i$ as $b_i=b_{i+1}$.
 Then  the competitive ratio is $\varrho(M)= n/4 -\epsilon$ for an arbitrary small $\epsilon >0$.

\textbf{Case 2}: there are no two consecutively arrived users when the mechanism $M$ accepts both with a probability at most $2/n$.
In this case, for each user with $p_i \le 2/n$, it is followed by a user with $p_{i+1} >2/n$.
Then we know that at least $n/2$ users will be accepted with a probability $>2/n$. Then $\sum_{i=1}^{n} p_i >1$, which is a contradiction.
\end{proof}

When the discount function $d(t)$ is given and known to the mechanism $M$, then the adversary $\cal A$ can only choose the initial valuations $\textbf v$ to make the mechanism perform as worse as possible.
It is straightforward to prove that $\varrho_d(M) \ge n/4 - \epsilon$ by carefully and adaptively choosing ${\vec{\textbf{v}}}$, 

\begin{lemma}
\label{theo:known-d-adaptive-a}
Even the mechanism designer knows the discount valuation $d(t)$, 
every  randomized IR-mechanism $M$ has $\varrho_{d} (M) \geq \frac{n}{4} - \epsilon$ in the worst case against the \emph{valuation online-adaptive}
 adversary. 
 Here the adversary can  dynamically select the valuations $\vec{\textbf{v}}$ based on the observed decision strategy of the mechanism $M$ and the historical  known discount function values
  $d(t)$. 
 \end{lemma}

Similarly, we can show that  $\varrho_{\vec{\textbf{v}}}(M) \ge n/4 - \epsilon$ by carefully and adaptively choosing $d(t)$ when the valuations vector are given, 
\ie, the exact values of each arrived users are known to the mechanism.
Here  the only condition on $d(t)$ is that it is in the range $(0,1]$.

\begin{theorem}
\label{theo:known-a-adaptive-d}
Even the mechanism designer knows the  valuation sequence $\vec{\textbf{v}}$,
every  randomized IR-mechanism $M$ has $\varrho_{\vec{\textbf{v}}}(M) \geq \frac{n}{4} - \epsilon$ in the worst case against the discount-function online-adaptive 
 adversary. 
 Here the adversary can  dynamically select $d(t)$ based on the observed decision strategy of the mechanism $M$ and the historical arrival sequence of $\vec{\textbf{v}}$.
\end{theorem}
\begin{proof}
The proof is similar to the proof of Theorem~\ref{theo:unknowdt-fixed-random}.
The difference lies in that previously the adversary can adaptively choose $v_i$ and $d(t_i)$ to meet some conditions required by the game.
Here in this case, the adversary can only adaptively choose $d(t_i)$ to meet the conditions used in the game.
Recall that in the case 1) of the game between the adversary $\cal A$ and the mechanism $M$, the adversary needs to choose $b_{i+1} \gg b_j$ for all $j \le i$.
As during the game between $M$ and the adversary $\cal A$, there could have $2/n$ such cases, where the adversary needs to get a sufficiently large $b_{i+1}$.
However, unlike $v_i$ which could be unbounded, we know that $0< d(t) \le 1$ is bounded.
Let $A= \max_{i,j} \frac{v_i}{v_j}$ for all values in $\textbf v$.
To leave room for larger $d(t_i)$ values, we can choose sufficiently small $d(t_1) = d_1$ such that $d_1 \cdot (A\cdot K)^{n/2} \le 1$ for some  large number $K >0$.
Then in the step 3 of the game played between $M$ and $\cal A$, the adversary $\cal A$ needs to find $b_{i+1}$ such that $b_{i+1} \ge b_j \cdot K$.
This can be done if $d(t_{i+1}) \ge K \cdot A \cdot d_j $ for any $j \le i$.
We can choose $d(t_{i+1}) = A \cdot K \cdot \max_{j=1}^{i} d_j$ where $K >0$ is a sufficiently large value.
It is easy to show that $d(t_i) \le 1$ for all $i$ as $d_1 \cdot (A\cdot K)^{n/2} \le 1$ and there are at most $n/2$ such steps in the game.
Then the revenue achieved by the mechanism $M$ is at most $(\frac{1}{K} +2\frac{2}{n}) b_i$ when the mechanism $M$ computes $p_i \le 2/n$ and $p_{i+1} \le 2/n$;
 while the revenue by the Vickrey auction is $b_i$.
 Thus,  $\rho(M) \ge \frac{n/4}{1+\frac{n}{4K}} \ge \frac{n}{4} - \epsilon$ for $\epsilon = \frac{n}{4K}$.
 As $K$ could be arbitrarily large, $\epsilon> 0 $ could be arbitrarily small.
\end{proof}

Notice that in these aforementioned theorems, we assume that the adversary is valuation adaptive-online, or discount-function adaptive-online or both, \ie,
  the adversary selects the values $v_i$  or $d(t)$ dynamically and carefully  depending on the actions of the mechanism $M$ on the first $i-1$ users.
Then we consider the case when the adversary is arrival-sequence adaptive-online, \ie,
 the values of $\textbf{v} =\{v_1, v_2, v_3, \cdots, v_{n-1}, v_n \}$ are chosen by (then given to the mechanism \footnote{The buyers' initial valuations \textbf{v} could be obtained in advance if the mechanism could conduct a sufficient prior market survey among the potential buyers.}) the adversary in advance,
  but the arrival sequence ${\vec{\textbf{v}}}$ could be adaptively chosen by the adversary online.
The adversary can  dynamically select $d(t)$ based on the actual decisions of the mechanism $M$ and the historical arrival sequence of $\vec{\textbf{v}}$.
We also limit our attention to all \emph{online truthful} mechanisms where the payment to a selected user must be decided when a winner is selected.
In other words, the mechanism cannot delay the payment computation when the winner is selected.
We then have Theorem~\ref{theo:known-dt-truthful} for the competitive ratio bound.

Assume the mechanism designer knows the discount function $d(t)$, there is a worst ($d(t), \textbf{v}$) pair such that every randomized truthful \emph{online} mechanism $M$ has an unbounded competitive ratio $\rho(M) \sim \infty$. That is, we have the following theorem.
\begin{theorem}
\label{theo:known-dt-truthful}
Assume that the adversary chooses a discount function $d(t)$ (even it must be non-increasing) and gives it to the mechanism.
Then no randomized truthful \emph{online} mechanism $M$ can achieve a bounded competitive ratio, \ie, $\rho(M) \sim \infty $ 
  in the worst case against the arrival-sequence online-adaptive  adversary, for some special $d(t)$.
  Here the adversary can choose the valuations set $\textbf{v}$, but the mechanism $M$ does not know $\textbf{v}$.
 %$\rho(M) \geq n- \epsilon$
\end{theorem}
\begin{proof}
We prove this by constructing the following instance.
\begin{equation}
\small
\label{worstcase-n1}
\begin{aligned}
(d,V)\defeq &
\begin{cases}
d(t)=
\begin{cases}
1, \quad t\in [0,t_{1}]\\
\frac{1}{n^k}, \quad t\in (t_{1},t_{n}]
\end{cases}
\\
V= \quad
\begin{cases}
v_{1}=K \cdot n^k - \delta\\
v_{2}=K\\
v_{i}=\delta, i \in [3,n]
\end{cases}
\end{cases}
\end{aligned}
\end{equation}
where $K$ is a large integer, say $K=n^2$; $\delta$ is a sufficiently small positive number compared with $n$, say $\delta=n^{-k}$, where $k \ge 5$.
For this instance, the Vickery revenue $r_{(2)}$ is composed of the following four complementary cases:
\begin{enumerate}
\item both $v_1$ and $v_2$ appear at $[0,t_1]$:
 the probability that this event happens is  $0$.
 
\item both $v_1$ and $v_2$ appear at $(t_1, t_n]$:
 the probability that this event happens is  $\frac{n-1}{n}\cdot \frac{n-2}{n-1}$.
In this case, $r_{(2)}=v_2\cdot \frac{1}{n^k}$.

\item $v_2$ appears at $[0,t_1]$ and $v_1$ appears at $(t_1, t_n]$:
 the probability that this event happens is  $\frac{1}{n}\cdot \frac{n-1}{n-1}$.
 In this case, $r_{(2)}=v_2-\delta$.
 
\item  $v_1$ appears at $[0,t_1]$ and $v_2$ appears at $(t_1, t_n]$:
 the probability that this event happens is $\frac{1}{n}\cdot\frac{n-1}{n-1}$.
 In this case, $r_{(2)}=v_2\cdot\frac{1}{n^k}$.
\end{enumerate}

Then we have:
$r_{(2)}= \frac{n-1}{n}\cdot\frac{n-2}{n-1}\cdot\frac{v_2}{n^k}+ \frac{1}{n}\cdot(v_2-\delta) + \frac{1}{n}\cdot \frac{v_2}{n^k}$.
Due to the small discount value $\frac{1}{n^k}$ in time interval $(t_1, t_n]$, we can omit the case 2)'s and 4)'s contribution to $r_{(2)}$.
We then study the case 3), the largest reported price $r_{(1)}=v_2\cdot1$ (produced by the user with a value $v_2$) 
 always arrives before the second largest reported price $r_{(2)}=v_2-\delta$ (produced by the user with a value $v_1$), 
and all other reported prices $r_i$s are $\delta \cdot \frac{1}{n^k}$, which can be ignored compared with $r_{2}$.
Thus, there is no strategy-proof mechanism that can select $r_{(1)}$ when the corresponding user arrives at time $t_1$ and make the payment as $r_{(2)}$ at the time $t_1$.
Then there is no randomized truthful \emph{online} mechanism $M$ has a bounded competitive ratio.
\end{proof}

Given a constant $B$, we partition the discount function into discount classes where the discount value in each class is within a factor $B$ of each other.
In other words, in the $i$-th discount  class, the discount values are $(B^{-i}, B^{-i+1}]$.
Let $n_i$ be the number of arrived users in the $i$-th discount class.
\begin{theorem}
\label{theo:known-dt-a-truthful-2}
Assume that the adversary chooses a discount function $d(t)$ (even it must be non-increasing) and gives it to the mechanism.
If there is a constant $B$, such that $n_1\ge2$.
% $n_i \ge 2$ for all non-empty discount classes.
Then for any randomized truthful \emph{online} mechanism $M$, its  competitive ratio is at least $\rho(M)  = \Omega(n) $ 
  in the worst case against the arrival-sequence online-adaptive  adversary, for some special $d(t)$.
 Here the adversary can choose the valuations set $\textbf{v}$, but the mechanism $M$ does not know $\textbf{v}$.
\end{theorem}
\begin{proof}
We prove this by constructing the following instance.
\begin{itemize}
\item [\textbf{Instance Size}:]
There are $n-1$ instances, each instance ${\cal I}_x$ has $n$ buyers, where $x \in [2,n]$.
\item [\textbf{$n$ Different Discount Functions $d(t)$}:]
Each ${\cal I}_x$ is composed of  the following discount function: $d(t_j) = 1$ for $1 \le j \le x$, and $d(t_j) =\frac{1}{n^k}$ for  $x < j \le n$. 
Here $t_j$ denotes the $j$-th arrived buyer's arrival time.
Clearly, $d(t)$ is non-increasing.
\item [\textbf{Valuation set $\textbf{v}$}:]
Each ${\cal I}_x$'s valuation set is composed of  the following three parts: 1) $v_2=K$; 2) $v_1=n^k*(v_2-\delta)$; and 3) remaining $(n - 2)$ users each with a value $\delta$; 
where $K$ is a large integer, say $K=n^2$; $\delta$ is a sufficiently small positive number compared with $n$, say $\delta=n^{-k}$, where $k \ge 5$.
\end{itemize}
Let $t_x$ be the arrival time of the $x$-th user.
For each $I_x$, the Vickery revenue $r_{(2)}$ is composed of the following four complementary cases:
\begin{enumerate}
\item  both $v_1$ and $v_2$ appear at $[0,t_x]$:
	 the probability when this case happens is $P_1=\frac{x}{n}*\frac{x-1}{n-1}$.
	 In this case, $r_{(2)}=v_2*1$.

\item both $v_1$ and $v_2$ appear at $(t_x, t_n]$:
 the probability is  $P_2=\frac{n-x}{n}*\frac{n-x-1}{n-1}$.
In this case, $r_{(2)}=v_2*\frac{1}{n^k}$.

\item 	 $v_2$ appears at $[0,t_x]$ and $v_1$ appears at $(t_x, t_n]$:
 the probability is  $P_3=\frac{x}{n}*\frac{n-x}{n-1}$.
 In this case $r_{(2)}=v_2-\delta$.
 
\item  $v_1$ appears at $[0,t_x]$ and $v_2$ appears at $(t_x, t_n]$:
 the probability is $P_4=\frac{x}{n}\cdot\frac{n-x}{n-1}$.
 In this case $r_{(2)}=v_2\cdot\frac{1}{n^k}$.
\end{enumerate}

Thus, the expected revenue by the Vickrey mechanism $M_V$, \ie, the expected value of the second reported price $r_{(2)}$ is
\begin{equation*}
\begin{aligned}
E(r_{(2)} )& =v_2 \cdot \frac{x}{n}\cdot \frac{x-1}{n-1}+v_2\cdot\frac{1}{n^k}\cdot\frac{n-x}{n}\cdot\frac{n-x-1}{n-1}\\
& +(v_2-\delta)\cdot \frac{x}{n}\cdot\frac{n-x}{n-1}+v_2\cdot\frac{1}{n^k}\cdot\frac{x}{n}\cdot\frac{n-x}{n-1}\\
& \approx v_2\cdot \frac{x}{n}
\end{aligned}
\end{equation*}
Due to the small discount value in time interval $(t_x, t_n]$, we can omit case 2)'s and 4)'s contribution to $r_{(2)}$.

In case 3), $r_{(1)}=v_2\cdot1$ (produced by $v_2$) always arrives before the $r_{(2)}=v_2-\delta$ (produced by $v_1$), thus, there is no strategyproof mechanism that can achieve the payment as $\Theta(1)\cdot r_{(2)}$ when the user with value $v_1$ arrives.

In case 1), both $v_1$ and $v_2$ appears in the first discount class, and the probability when this happen and $v_1$ arrives after $v_2$ is $\Theta(1/n^2)$.
Thus, the maximum revenue received by any mechanism is at most $O(v_2 /n^2)$, while the revenue achieved by the offline Vickrey auction is $\Theta(v_2 /n)$.
Thus for all truthful mechanisms $M$,  its competitive ratio is at least  $\Theta(n)$.
\end{proof}

%%%%%%%%%%%%%%%
\subsection{Bounds for Non-increasing Discount Functions}
\label{subsec:non-increasing-lower-bound}

Recall that we do not have much limitation on the discount function $d(t)$ when we analyze some lower bounds on the competitive ratios of mechanisms satisfying IR and IC.
It has been proved that when all users are truthful in reporting their values, and do not delay their arrivals,  any algorithm has a smaller lower bound on the  competitive ratio.
We first review the theorem proved in ~\cite{babaioff2009secretary} for the discounted secretary problem when the users are assumed to be truthful, and the goal is to select
 a user with the maximum discounted values.
 Theorem~\ref{theo:lowerbound-non-inc-IR} holds even when $d(t)$ is non-increasing.

\begin{theorem} (Discounted Secretary) ~\cite{babaioff2009secretary}. 
\label{theo:lowerbound-non-inc-IR}
When all users are truthful in reporting their values and arrive honestly,
any algorithm for the discounted secretary problem (with the objective to maximize the discounted values of the accepted user)
 has a (worst-case) competitive ratio of  $\Omega( \frac{\log n} {\log \log n})$ when there are $n$ secretary candidates, 
 even when $d(t)$ is known to be non-increasing.
\end{theorem}

%\xue{Can we prove that this is true when $d(t)$ is non-increasing?}
We then show that a smaller lower bound for the competitive ratio of any truthful mechanism when the discount function $d(t)$ is non-increasing.
%Our proof is inspired by the proofs from ~\cite{babaioff2009secretary}  when the users are assumed to be always honest in reporting their true values and arrive honestly.

\begin{theorem} (\textbf{Non-Increasing Discount Function}) 
\label{theo:lowerbound-non-inc}
When the discount function $d(t)$ is non-increasing and known to the mechanism (also adversary),
 any IR-mechanism for the discounted auction problem has a (worst-case) competitive ratio at least
$\Omega( \frac{\log n} {\log \log n})$, compared with the offline optimum Vickrey auction $M_V$, against an arrival-sequence adaptive-online adversary,
 when there are $n$ buyers. 
\end{theorem}
\begin{proof}
The proof essentially is inspired by and extends the proofs in ~\cite{babaioff2009secretary}.
The main difference is that here we need to compare the auction mechanism with the Vickrey auction, while
 the proof in ~\cite{babaioff2009secretary} only needs to focus on finding the maximum discounted values, by assuming all users are always truthful
   in reporting their true values and arrive honestly.

We prove this by constructing some examples used by the arrival-sequence adaptive-online adversary.
We will construct a discounting function $d(t)$, and a family of input instances (\ie, the set of initial valuations ${\cal I} ={\cal I}_1, {\cal I}_2, \cdots, {\cal I}_t, \cdots$) such that no \emph{randomized} IR-mechanism $M$ can have a competitive ratio less than $\Theta(  \frac{\log n} {\log \log n} )$ for all possible instances.
We will prove this by contradiction.

More formally, we construct the following instances, each of $n$ users, as follows.
\begin{itemize}
\item [\textbf{Instance Size}:]
The number of users in each instance ${\cal I}_t$ is $n$, where $n=L^{4c}$ and $L=c$ are some positive integers.
Then it is easy to prove that $L=c= \Theta(\frac{\log n}{\log \log n})$.

\item[\textbf{Discount Function $d(t)$}:]
For any time $t \le 2c$, let $n_t = L^{2t}$. Clearly $n_t \le n$ for $t \le 2c$.
The discount function $d(t)$ used by the adversary is defined as follows.
$d(j) = L^{-1}$ for $1 \le j \le n_1$, and $d(j) =L^{-t}$ for  $n_{t-1} \le j \le n_t$.
Clearly, $d(t)$ is a step-function and is non-increasing.

\item [\textbf{$2c$ Instances}:]
Let $K$ be a sufficiently large integer, compared with $n$, say $n^3$.
The instance ${\cal I}_1$ is the following set, composed of $\frac{n}{n_1}$ numbers of value $K$, and $n -\frac{n}{n_1}$ zeros.
In general, the instance ${\cal I}_t$, for $1<t \le 2c$, is composed of  the following three parts
1) $\frac{n}{n_j} - \frac{n}{n_{j+1}}$ numbers, each of value $K^j$, for $j=1$ to $j=t-1$, 
2)  $\frac{n}{n_t}$ numbers, each of value $K^t$, and
3) remaining $n -\frac{n}{n_1}$ zeros.
\end{itemize}

Consider a randomized mechanism $M$ that will decide which user to accept, and how much to charge the user, so that users are strategyproof.
Our goal is to analyze the \emph{expected} competitive ratio of $M$ against an  arrival-sequence adaptive-online adversary.
Thus, given each of these instances ${\cal I}_t$, we need to consider the "average" performance of $M$ for all possible $n!$ permutations of users' arrival sequence.
We will show that no mechanism can perform well on average for each of these $2c$ instances.

Recall that, the analysis of~\cite{babaioff2009secretary} assume that the users are always honest.
In our proof here, we need to analyze the truthful mechanisms, and compared them with the Vickrey auction.
We first give a lower bound on the expected revenue gained by the Vickrey auction $M_V$ on instance ${\cal I}_t$.

\begin{lemma}[Lower bound on $E(M_V({\cal I}_t)$]
\label{lemm:bound-vickrey}
For $1\le t \le 2c$, $E(M_V({\cal I}_t) \ge \delta \cdot K^t L^{-t} $. 
Here $\delta= 1-\frac{2}{e}$ is a constant.
\end{lemma} 
\begin{proof}
For all permutations of ${\cal I}_t$, let $X_j$ be the event that there are exactly $j$ elements of ${\cal I}_t$ with the maximum value $K^t$ appeared
 in the first $n_t$ time-slots. 
 Let $Z_2$ be the event that there is at least $2$ elements of value $K^t$ appeared in the first $n_t$ slots.
 If $Z_2$ is true, then the Vickrey auction will get a revenue at least $K^t L^{-t}$ because for all elements in first $n_t$ slots,
  the second largest value is $K^t$, and the discount value is at least $L^{-t}$.
  Notice that $\prob{X_0}=(\frac{n-n_t}{n})^{\frac{n}{n_t}}  \le 1/e$ as there are $\frac{n}{n_t}$ elements of value $K^t$.
For the event $X_1$, we have $\prob{X_1}= \frac{n}{n_t} \cdot \frac{n_t}{n} (1- \frac{n_t}{n})^{\frac{n}{n_t} -1} = (1- \frac{n_t}{n})^{\frac{n}{n_t} -1}$.
  Thus,  $\prob{Z_2}=1- \prob{X_0} - \prob{X_1} =1- (1- \frac{n_t}{n})^{\frac{n}{n_t} }- (1- \frac{n_t}{n})^{\frac{n}{n_t} -1} \ge 1-\frac{2}{e}$.
%   if $n_t \ll n$, which is true in our setting. % as $t \le 2c$, $n_t \le L^{4c}$ and $n=L^{6c}$.
This finishes the proof of lemma~\ref{lemm:bound-vickrey}.
\end{proof}

\textbf{Proof by Contradiction:}
We then prove Theorem~\ref{theo:lowerbound-non-inc} by contradiction as follows.
Assume that there is a mechanism $M$ that is $\alpha c$-competitive, \ie, 
\[E(M_V({\textbf v}, d)) / E(M({\textbf v}, d)) \le \alpha c\]
 for any input sequence $\textbf v$
 and all non-increasing discount functions $d(t)$. 
Here $\alpha >0$ is a constant to be fixed by our proof later, and $c=\Theta(\frac{\log n}{\log \log n})$ is the number defined in the aforementioned instances.
We will prove that this is impossible by contradiction. 
First, we show that if there is such a mechanism $M$, $M$ needs to have some features.

\begin{lemma}[\textbf{No small probability}]
\label{lemma-no-small-prob}
Let $M$ be a $\alpha c$-competitive mechanism. 
On any instance ${\cal I}_t$, let $W_t$ be the event that $M$ will pick one element from the first $n_t$ items of the  input sequence permuted from ${\cal I}_t$.
Then $\prob{W_t} \ge \frac{t}{c}$ for some carefully chosen constant $\alpha$. 
Recall that here $c= \Theta(\log n / \log \log n)$.
\end{lemma}
\begin{proof}
We prove this by induction on the index $t$.
First consider the base case ${\cal I}_1$.  % \{ \frac{n}{n_1}\ elements\ of\ values\ K,  n-\frac{n}{n_1} \  zeros  \}$.
We define two variables $Y_1, Y_2$.
\begin{compactitem}
\item
Let $Y_1$ be the expected revenue that $M$ gets by picking some element from the first $n_1$ arrived elements.
\item
Let $Y_2$ be the expected revenue that $M$ gets from picking some element from later arrived elements (starting from $n_1+1$ and onwards).
\end{compactitem}
Notice that the revenue $Y_2$ is at most $K/L^2$ as the value of the element is at most $K$ and the discount value after time $n_1$ is at most $L^{-2}$.
Recall that for Vickrey mechanism $M_V$, $E(M_V) \ge \delta \frac{K}{L}$ from Lemma~\ref{lemm:bound-vickrey}.
To make sure that $M$ is $\alpha c$-competitive, we need $Y_1 +Y_2 \ge \frac{1}{\alpha c} E(M_V)$.
Thus, it is required that 
\[Y_1 \ge \frac{1}{\alpha c} E(M_V) -Y_2 \ge \frac{1}{\alpha c} E(M_V) - \frac{1}{c} K/L
 \ge ( \frac{\delta}{\alpha c} -\frac{1}{c} ) \frac{K}{L}.\]
Notice that $Y_1 \le \prob{W_1} \cdot K/L$ as the value of the element is at most $K$ and the discount function for time before $n_1$ is $1/L$.
Then $M$ is $\alpha c$-competitive implies that 
\begin{equation*}
\prob{W_1} \ge  \frac{\delta}{\alpha c} -\frac{1}{c} 
\end{equation*}
Thus if we have
\begin{equation}
\label{condition-alpha-1}
 \frac{\delta}{\alpha c} -\frac{1}{c}  \ge \frac{1}{c}, \quad \ie,  \quad 0< \alpha \le \delta/2
\end{equation}
We could conclude that  $\prob{W_1} \ge \frac{t}{c} $ for case ${\cal I}_t$, $t=1$.

We then prove the induction step.
Assume that  $\prob{W_j} \ge \frac{j}{c} $  is true for all input instances ${\cal I}_j$, $j=1, 2, \cdots, t$.
We then show that it is true for the input instance ${\cal I}_{t+1}$.

First, we will show that, with high probability, the online mechanism $M$ cannot distinguish whether the input sequence is ${\cal I}_t$ or ${\cal I}_{t+1}$
 by simply observing the first $n_t$ elements triggered by the adversary.
For the instance ${\cal I}_{t+1}$, recall that   $X_1$ is the event that any element of a value $K^{t+1}$  (there are $\frac{n}{n_{t+1}}$ such elements)
 appeared in some  of the first $n_t$ time-slots.
Then we have $\prob{X_1} \le  \frac{n}{n_{t+1}} \frac{n_t} {n} = \frac{1}{L^2} $.
Thus, with probability $\ge 1-\frac{1}{L^2} $, among all permutations of ${\cal I}_{t+1}$, the first $n_t$ elements in the sequence is same as some sequence
 permuted from the instance ${\cal I}_{t}$.
 Then with probability $\ge 1-\frac{1}{L^2} $, any mechanism $M$ cannot distinguish whether the input instance is ${\cal I}_{t}$ or ${\cal I}_{t+1}$ by only looking the first $n_t$ elements.
For the mechanism $M$ to be $\alpha c$-competitive, for the first $n_t$ elements,
  with probability  $\ge 1-\frac{1}{L^2} $, the behavior of the mechanism $M$ on ${\cal I}_{t}$ and ${\cal I}_{t+1}$ will be same.
To prove our lemma, we define three events when $M$ runs on some permuted sequence of ${\cal I}_{t+1}$.
\begin{compactitem}
\item
Let $Q$ be the event that $M$ will pick some element from the first $n_t$ slots,
for instance  ${\cal I}_{t+1}$. 
The total expected revenue that $M$ can get from event $Q$ is denoted as $E_Q(M)$.
\item Let $R$ be the event that $M$ will pick some element after the $n_{t+1}$ timeslot.
Let  $E_R(M)$ be the expected revenue received in these choices.
\item
 Let $P$ be the event that $M$ will pick some elements from the time interval $(n_t, n_{t+1}]$. 
Let  $E_P(M)$ be  the expected revenue that $M$ will get from this event $P$.
\end{compactitem}

Then $E(M)=E_Q(M) + E_R(M)+E_P(M)$.
It is easy to show that  $\prob{Q} \ge \frac{t}{c} \cdot  ( 1-\frac{1}{L^2})$.
Let us see the expected revenue that the mechanism $M$ can get from event $Q$, the decisions on  first $n_t$ users from the instance ${\cal I}_{t+1}$.
The selected element could be one of the elements with value $K^{t+1}$, or some other elements with values at most $K^t$.
When the selected element is $K^{t+1}$, it could be selected when it is from some interval $[n_{j-1}, n_{j})$, for $j=1,2,\cdots, t$.
The expected revenue is $K^{t+1} \cdot L^{-j} \cdot \frac{n}{n_{t+1}} \frac{n_j - n_{j-1}} {n} < \frac{K^{t+1} }{L^j} \frac{n_j}{n_{t+1}}$.
When select some other elements, its expected revenue is at most $K^t \cdot L^{-1}$ as the value is at most $K^t$ and discount is at most $L^{-1}$.
 Thus, the  expected revenue $E_Q(M)$ is at most
 \[ E_Q(M) \le K^t \cdot L^{-1} +\sum_{j=1}^{t}  \frac{K^{t+1} }{L^j} \frac{n_j}{n_{t+1}} \le \frac{3} {L} \frac{K^{t+1}}{L^{t+1}} \le \frac{3} {L} \frac{1}{\delta} E(M_V). \]
 Recall that for the optimum offline Vickrey auction $M_V$, its expected revenue $E(M_V) \ge \delta  \frac{K^{t+1}}{L^{t+1}}$
  for the instance ${\cal I}_{t+1}$.
 We carefully select a number $\beta >1$ such that 
 \begin{eqnarray}
 \label{required-cond}
 \frac{3} {L \delta} \le  \frac{1}{ \beta \alpha c}, \quad \ie, \quad 1 < \beta \le  \frac{\delta}{ 3 \alpha } 
 \end{eqnarray}
 Recall $L=c$ in our setting.
 Then, $E_Q(M) \le  \frac{1}{ \beta \alpha c} E(M_V)$.
 As the mechanism $M$ is $\alpha c$-competitive compared with $M_V$, \ie,  $E(M) \ge  \frac{1}{  \alpha c} E(M_V)$.
  So $M$ must get at least $(1-\frac{1}{\beta}) \frac{1}{ \alpha c} E(M_V)$ from the rest of the items, starting from $n_t+1$ elements.
  
 For event $R$, we know that the expected revenue received in this event 
 \[E_R(M) \le K^{t+1} /L^{t+2} \le \frac{1}{\delta c} E(M_V)\]
  as the discount values for these later time-slots is at most $L^{-(t+1)}$, the value of the element is at most $K^{t+1}$, and $L=c$.
  
Thus, to assure that $M$ is $\alpha c $-competitive, as $E(M)= E_Q(M) + E_R(M)+E_P(M) $, we need $E_P(M)$ to satisfy
\[ \frac{1}{\alpha c} E(M_V) \le E(M) \le E_P(M)  +(   \frac{1}{ \beta \alpha c} + \frac{1}{\delta c}) E(M_V) \]
Thus, $E_P(M) \ge (\frac{1}{\alpha} -\frac{1} {\alpha \beta} - \frac{1}{\delta}) E(M_V) /c$.
Notice that $E_P(M) < \prob{P} \cdot \frac{K^{t+1}}{L^{t+1}} \le \prob{P}  E(M_V) / \delta$ 
 as the element picked by the mechanism has a value at most $K^{t+1}$ and the discount value is always $1/L^{t+1}$.
 Then we conclude that
 \begin{eqnarray}
 \prob{P} \ge \frac{\delta}{c}(\frac{1}{\alpha} -\frac{1} {\alpha \beta} - \frac{1}{\delta})
 \end{eqnarray}
 
 Notice that event $W_{t+1}$ is composed of $Q$ and $P$.
 By carefully selecting $\alpha$ and $\beta$ such that 
 \begin{eqnarray}
 \label{cond-ind}
 (1-\frac{1}{L^2}) t+   (\frac{\delta}{\alpha} -\frac{\delta} {\alpha \beta} - 1) \ge t+1 
  \end{eqnarray}
  Then $\prob{W_{t+1}} = \prob{Q} +\prob{P} \ge \frac{t+1}{c}$. 
  As $t \le 2c$, the above inequality~(\ref{cond-ind}) will always be true from the following sufficient condition
  \begin{eqnarray}
  \label{cond-ind-beta}
  3 \le \frac{\delta}{\alpha} -\frac{\delta} {\alpha \beta}, \quad \ie, \quad \beta \ge \frac{\delta}{\delta - 3 \alpha}
 \end{eqnarray}
 Thus, Lemma~\ref{lemma-no-small-prob} is true for all $t \le 2c$ if all the sufficient conditions are met.
 To assure that all the sufficient conditions are met, in summary, the following conditions are sufficient for the lemma to hold.
  \begin{eqnarray}
 \left \{ 
 \begin{aligned}[rcl]
  \alpha &  \le  &  \delta /2 \\
  \beta  &\le &  \frac{\delta}{ 3 \alpha } \\
  \beta & \ge & \frac{\delta}{\delta - 3 \alpha}
\end{aligned}
\right .
  \end{eqnarray}
We can choose $\alpha = \delta /6$ and $\beta =2$ to meet all these conditions.
This finishes the proof of Lemma~\ref{lemma-no-small-prob}.
\end{proof}

From Lemma ~\ref{lemma-no-small-prob}, for the instance ${\cal I}_t$, where $t >c$, we have 
 $\prob{W_t} \ge \frac{t}{c} >1$, which is impossible.
 Thus, the assumption that there is a IR-mechanism $M$ with competitive ratio  at most $\alpha c= \delta c /6 =\Theta(\frac{\log n}{\log \log n})$ cannot hold.
This finishes the proof of Theorem~\ref{theo:lowerbound-non-inc}.
\end{proof}

Notice that in the proof of Theorem ~\ref{theo:lowerbound-non-inc}, we do not require that the mechanism to always be truthful. 
The claim holds for all mechanisms that satisfy the individual rationality, \ie, the payment charged from the user is at most his reported bid.

\section{Online Truthful Mechanism Design}
\label{sec:alg}

In this section, we present some revenue-competitive online truthful mechanisms with time-sensitive valuations. 
By assuming that $d(t)$ is an \emph{arbitrary} non-increasing  function:
1) we present an online \textit{ truthful} and $O(\log^2 n)$-competitive (under some additional assumptions) mechanism $\mathcal{M}_R$ in \S\ref{alg_unknown_random} and provide detailed theoretical analysis in \S\ref{analysis_M_R};
2) We further propose an heuristic mechanism $M_W$ in \S\ref{alg_unknown_weighted}.

Recall that the algorithm proposed in~\cite{babaioff2009secretary} assume that all users were truthful in advance.
Their method selects the user with the largest reported price and charge this user its reported price.
Then Babaioff et al~\cite{babaioff2009secretary} proved that 
for arbitrary \emph{known} discount function $d(t)$, and a set of arbitrary initial valuations $\textbf V$,
when all buyers are truthful in reporting their bidding price and arrive truthfully, 
 any online algorithm will have a competitive ratio  at least $\Omega(\frac{\log n}{\log \log n})$;
and there is an online auction mechanism with a competitive ratio $\Theta(\log n)$~\cite{babaioff2009secretary}. 

%%%%%%%%%%%%%%%
\subsection{Random-Selection Mechanism for Arbitrarily Known Non-increasing Discount Function}
\label{alg_unknown_random}

In the previous subsection~\ref{subsec:general-lower-bound},
 we show that in most cases, the lower bound on the competitive ratio is at least $\frac{n}{4}$ for general discount functions
 when the adversary is discount-function adaptive-online,
 valuation adaptive-online, or arrival-sequence adaptive-online.
 In many practical applications, the discount functions may exhibit some additional nice properties such as
 non-increasing, or concave, or convex, or with bounded derivatives.
%As we already showed that the competitive ratio on the expected revenue could be $\Theta(n)$ for general discount functions,
 In the rest of the work, we assume that the discount functions have some   additional properties, listed as follows.

 \begin{assumption}[Non-increasing discount function]
We assume that $d(t)$ is an arbitrary known and non-increasing.
\end{assumption}

 \begin{assumption}[Non-increasing buyer density function]
 Let $t(d)$ be the inverse of the function $d(t)$.
 The \emph{buyer density} at a value $d$ is defined as ${\tilde t}(d)= t(d/B) - t(d)$ for some constant $B >0$.
A discount function $d(t)$ is called a non-increasing buyer density function if for any $d_1 <d_2$, we have
 ${\tilde t}(d_2)  \ge {\tilde t}(d_1) $.
\end{assumption}

For simplicity of analysis, we sometimes assume that the minimum value in the valuation set ${\textbf V} =\{v_1, v_2, \cdots, v_n \}$ is 1.
We assume that the set is non-decreasing, \ie, $v_i \ge v_{i+1}$ for $i\in [1,n-1]$.
As shown in \S\ref{pro:formulation}, the problem studied here is formulated as a single-round online auction with discounting valuations.
This is quite similar to the \textit{discounted secretary problem}~\cite{babaioff2009secretary},
 in which there is a discount function $d(t)$ for the value of each candidate. 
 Given a sequence of randomly arrived applicants, we need to hire the one having the largest ``value''.
The benefit derived from selecting an element with the value $v$ at time $t$ is $d(t) \cdot v$. 
 Note that~\cite{babaioff2009secretary} proposed an $O(\log n)$-competitive online algorithm for this discounted-secretary problem.
While without any prior knowledge, they exhibit a lower bound of $\Omega(\frac{\log n}{\log \log n})$.
Notice that when there is no discount,  the classical secretary problem is to select an element (a ``secretary") with maximum value online.
There is a well-known $e\verb|-|$competitive algorithm~\cite{dynkin1963optimal} which contains two phases, \ie, observation and decision.
%(which is also the state-of-the-art) contains two phases: (1) directly drop the first $1/e$ fraction of the elements but record the maximum value of them; (2) for the remaining elements, select the first one whose value is larger than those of all the past elements (if such element does not exist, then select the final element). Based on some simple calculations, the probability of selecting the best applicant converges toward $1/e$.

Comparing with the discounted secretary problem,  the problem studied in this work need to address the challenges caused by the non-truthfulness of all users. \ie, 
 we need to design a strategy-proof mechanism satisfying IR, IC, computational efficiency and bounded competitive ratios. 
The revenue (or benefit) gained in the mechanism  for our setting is defined differently. 
Namely, the benefit  is not the \textit{value} of the buyer, but the price that the buyer pays. 
Consequently, we need to develop some new ideas for designing the selection and payment rules.
Another closely related work is by Hajiaghayi et al.~\cite{hajiaghayi2004adaptive}, which presented a $4$-competitive strategyproof online auction, 
with respect to offline \textit{Vickrey} auction for revenue, 
but without considering discount factors. 
Their mechanism is as follows: observe the first $1/2$ fraction of bids and record the largest price among them as $v_{max}$; 
 then select the first one from the remaining bids who reports the price larger than $v_{max}$ and ask him to pay at the price $v_{max}$. 
The probability of second-largest one appearing at the first $1/2$ fraction is $1/2$, and the probability that the largest one appearing 
 at the last $1/2$ fraction is $1/2$.
 Hence, with probability of $1/4$ the mechanism chooses the largest one and pay at the second-largest price. 
 Thus achieve $4$-competitive ratio.
However, when considering the discount function $d(t)$, each bidder's true valuation is changing and thus the existing mechanism may result in an arbitrarily large competitive ratio. 

\noindent\textbf{Key idea:} 
To tackle these challenges, we propose the following Randomized-Select Mechanism $(\mathcal{M}_R)$.
Intuitively, the smaller the discount value, the lower the probability to get a large reported price.
As a consequence, the probability that we choose from the bidders at the time with larger $d(t)$ should be higher.
Inspired by the work~\cite{babaioff2009secretary}, we divide the values of $d(t)$ into multiple classes, 
 where each class has similar discount values, say bounded by $B>1$ of each other. 
Then we treat each class as a sub-instance 
  having the same discount value by ignoring the discounts within this class. 
Then we drop those classes with significantly small values of $d(t)$, due to their minor contribution to the expected second-largest reported price. %\ie, $E[r_{(2)}]$.
We uniform and randomly select a class or carefully select a class (denote as $c^*$) from the reserved classes, and conduct a two-stage observing-selecting process.

\begin{algorithm}[t!]
	\LinesNumbered
	\small
	\caption{Randomized-Select Mechanism $\mathcal{M}_R$}
	\label{alg:M_R}   		
	\KwIn{report price vector $\mathbf{r}$; discount function $d(t)$; time horizon $T$;  arrival rate $\lambda$.} 
	\KwOut{winner determination $\mathbf{y}$; payment $\mathbf{p}$}
%	${n}=\lambda*T, c^*=0, {n}_{c^*}=0$\;
%	$winner=\emptyset$, $payment=0$, $\mathbb{R}=\emptyset$\;
	%\While {${{n}_{c^*}}<2$}{
	%Let time $t_{c1}$ and $t_{c2}$ satisfy $d(t_{c1})=2^{-c^*}d_{max}$ and $d(t_{c2})=2^{-(c^*-1)}d_{max}$ % \Comment{discount classes} \;
	%$I_{c^*}=(d(t_{c1}), d(t_{c2})], T_{c^*}=(t_{c2},t_{c1}]$,	${n}_{c^*} = \lambda*(t_{c1}-t_{c2})$\;
	%\State Calculate $I_{c^*}$ and $T_{c^*}$ respectively. %\gets (2^{-c}d_{max}, 2^{-(c-1)}d_{max}]$
	%\State $T_{c^*} \gets \{t:d(t)\in I_{c^*}\}$.
	%}   
%	For $c\geq1$, calculate $T_{c}$ according to Equation (\ref{discount_class}) and (\ref{time_interval}) \;
	$c^* \gets$ Choose a class from the reserved class set $ \mathbb{C}$ (as in Eq.(\ref{reserved_classes})) uniformly at random,
	 where $c^*$ called \textbf{Star-Class} hereafter \;
         ${n}_{c^*}=\lambda*|T_{c^*}|$; $t \gets \text{current time}$; ${j}^*=0$ \;
	Calculate $T_{c^*}$ according to Equation (\ref{discount_class}) and (\ref{time_interval}) \;
	\While {$t \in [0,T]$}
	{
		%Record the $j$-th reported price  $r_j$ by $\mathbb{R}.add(r_j)$\;
		Update $\mathbf{r}$ when receiving the $j$-th reported price $r_j$\;
		\While {$t \in T_{c^*}$}
		{ 
			$(winner, payment, j^*) \gets \mathcal{M}_O( \mathbf{r}, {n}_{c^*}, j^*)$\; 
		}
		Let $r_{(1)}^{c}=\max\{r_j | 1\leq j \leq j^*\}$ \; %r_j \in \mathbf{r}
		\While { $t \in T_{c>\hat{c}}$ and no winner selected}
			{	
%				Obtain the $j$-th arrived buyer's report price  $r_j$\;
				\If{the $j$-th buyer arrives and $r_j\geq r_{(1)}^c$} 
				{
					{
					Select this buyer as $winner (=j)$, and charge this buyer at $payment= r_{(1)}^{c}$\;	   			
					}
				}
			}
		$j^*\gets j $, record the total number of observed price at $t$\;
	}
	$y_{winner}=1$; $p_{winner}=payment$\; 
%	Update $\mathbf{y}$ and $\mathbf{p}$ \;
	Update and Return $\mathbf{y},\mathbf{p}$\;
\end{algorithm}

\noindent\textbf{Randomized-Select Mechanism $(\mathcal{M}_R)$:}
let $d_{\max}$ be the maximum discount value and we set ${d}_{max} = 1$ without loss of generality.
The details of $\mathcal{M}_R$ is shown in Algorithm~\ref{alg:M_R}. 

We first divide the values of $d(t)$ into multiple classes, numbered as $1,2,3,\cdots$.
For any class with ID $c\geq1$, let $d_{c}= B^{-(c-1)}$ for some constant $B>1$, and the interval indicating the $c$-th discount class, for $c \ge 1$, is denoted as:
\begin{eqnarray}
\label{discount_class}
I_c = (B^{-c}, B^{-(c-1)}] =(d_{c+1}, d_c]
\end{eqnarray}	
And the corresponding time interval whose discount value belongs to class $I_c$ is calculated by:
\begin{eqnarray}
\label{time_interval}
T_c=\{t:d(t)\in I_c\} 
\end{eqnarray}	
When $d(t)$ is monotone,  $T_c$ will be a continuous time interval.
Notice that in our mechanism design, we assume that the discounting function is non-increasing and choose $B=2$.
Let $t(d)$ be the inverse function of the discount function $d(t)$, and ${\textbf t}_c=t(d_c)$.
Then the time-intervals $T_c$ partitioned are continuous and in sequel, 
 \ie, $T_c=[t(d_c), t(d_{c+1}))=[{\textbf t}_c, {\textbf t}_{c+1})$ for $c \ge 1$.
In this work, we assume that the users arrive by a uniform random process or the Poisson process with an arrival rate  $\lambda >0$.
We assume that $\lambda$ is not too small, \ie, the number of arrived users in each reserved class is at least some constant.
Let ${n}_c$ be the expected total number of buyers appearing at $T_c$ time interval.
 
 \begin{assumption}[Dense  Arrival Users]
Assume that all buyers arrive by a Poisson process (or uniform random) and the number of  arrived users, denoted as $n_c$, in each discount class $c$  satisfies
$n_c \ge 2$.
\end{assumption}

Observe that here when $B$ is small, the aforementioned assumption may not hold.
In our mechanism design, we only focus on the case when we can carefully choose a large enough constant $B$ such that the
 aforementioned assumption of dense  arrival of users holds.
 In other words, we assume that the function $d(t)$ has bounded derivatives, \ie,  $-\delta_2 \le d'(t) \le -\delta_1$ for some positive constant $\delta_1>0$
  and $\delta_2>0$.

% \xuenote{it is better to draw a figure here}

As the initial valuations are random, it is impossible to find the interval $T_c$ to which the potential maximum bid $a_i d(t_i)$ belongs.
One approach is to choose a class from the \textbf{reserved class set} $\mathbb{C}$  (as in Equation (\ref{reserved_classes})) 
by some carefully designed strategy (or uniformly at random), where the selected class $c^* $ is called \textbf{Star-Class} hereafter.
\begin{eqnarray}
\label{reserved_classes}
\mathbb{C}=\{1,2,3, \cdots,\hat{c} \}
\end{eqnarray}	
Here $\hat{c}$ is of order $\Theta(\log n + \log \frac{v_{1}} {v_{2}})$ and the more detailed bound will be computed as in Eq.~\eqref{hatc}.
Here $v_{1}$ is the maximum initial valuation observed in observation phase, and $v_{2}$ is the second-largest initial valuation.
In most cases, we can assume that $\log \frac{v_{1}} {v_{2}}=O(\log n)$.
For example, this assumption holds when $\frac{v_{1}} {v_{2}} =  O(n^k)$ for some constant $k>0$.
The reason we only focus on the reserved class set is that for other classes, the discount function values is too small (of order $O(1/n)$), thus with high probability, 
 the bidding-price will be extremely  small compared with the maximum bidding-price.

Next, we run the ``Observe-then-Decide" mechanism $\mathcal{M}_O$ (as shown in Algorithm~\ref{alg:M_o})
  in the \textbf{Star-Class} to decide the winner and payment.
Notice that the $j$-th arrived buyer's initial valuation is $\frac{r_j}{d(t_j)}$.
In the mechanism $\mathcal{M}_O$, we will ignore the discount values for the moment as the discount values of all users in 
 this discount-class is within a constant $B$ factor of each other.
 The mechanism  $\mathcal{M}_O$ mainly is composed of two phases:
\begin{compactitem}
\item \textbf{Observation phase} (line 4-7): in this class, we observe the first $1/2$ fraction of buyers, and record the largest initial valuation among them as $v_{(1)}({c^*})$.
For simplicity of notations, we often abbreviate $v_{(1)}({c^*})$ as $v_{(1)}$ if no confusion\footnote{Here $v_{(1)}({c^*})$ denotes the maximum value of users in the class $c^*$, while $v_{(1)}$ is the maximum value among all users.} is caused.
To prevent users from delaying arrival, we select each arrived user as \emph{winner} with a probability $\frac{1}{\lfloor n_{c^*}/2 \rfloor +1}$ and charge the user at $0$. 
If no one is selected in this phase, we continue to the next phase; or we return the \emph{winner} and the \emph{payment} for this winner.
\item \textbf{Decision phase} (line 8-11): select the first one (denote his arrival time as $t_j$) from the remaining buyers whose initial valuation is larger than $v_{(1)}$ and charge him at $v_{(1)}*d(t_j)$.
\end{compactitem}
Although we assume that no two buyers arrive at the same time, we need to make clear that how we do when there are multiple buyers delayed their arrivals which result in receiving multiple report prices at the same time.
\begin{compactitem}
\item In the Observation phase, when receiving multiple bids at the same time, we select the highest one as winner with probability $\frac{1}{\lfloor n_{c^*}/2 \rfloor +1}$ and charge him at $0$.
\item In the Selection phase, 1) { if $\mathcal{M}_O$ has just observed $(x+1)-m$ bids when receiving $m$ bids simultaneously, let the highest bid be the $(x+1)$-th one, and the others are considered as appearing at the Observation phase; 2) if $\mathcal{M}_O$ has just observed $j-1$ bids when receiving $m$ bids at $t_j$, then let the highest bid be the $j$-th one.}
% Then we run the Decision part in Algorithm~\ref{alg:M_o}). %to decide the $winner$ and $payment$ .
\end{compactitem}

Finally, if no winner is selected by $\mathcal{M}_O$, we select the winner from $T_{c\geq \hat{c}}$ whose reported price is larger than the previously observed ones (line 8-12 in Algorithm~\ref{alg:M_R}).
Notice that initially, $\mathbf{y}=\mathbf{p}=\{0,0,\dots,0\}$.
%$\mathcal{M}(\mathbf{r}, d(t))=(\mathbf{y},\mathbf{p})$,
%\\ \xue{ $\mathbf{y}=\mathbf{p}=\{0,0,\dots,0\}$;  initialized with zero vector. Update line number in four places (line.}

%%%%%%%%%%%%%%%%%%%%%%%
\begin{algorithm}[t!]
	\small
	\LinesNumbered
	\caption{ {Observe-then-Select $\mathcal{M}_O( \mathbf{r}, {n}_{c^*}, j^*)$}}
	\label{alg:M_o}	   		
	\KwOut{ $winner$ and $payment$.}
	$winner=0$, $payment=0$, $x=\lfloor n_{c^*}/2 \rfloor$\;
		\If{the $j$-th buyer arrives}
		{
			$v_j \gets \frac{r_j}{d(t_j)}$\;
			\If(\tcp*[h]{Observation phase}){$1 \leq j-j^* \leq x$}
			{
				Select this buyer as $winner$$(=j)$ with probability $\frac{1}{x+1}$ and charge him at $payment=0$\; 
				\If{$winner\not=0$}
				{
					Return $winner$, $payment$ \;
				}
			}		
			\If(\tcp*[h]{Decision phase}) {$x < j-j^* \leq {n}_{c^*}$}
			{
				{Let $r_{(1)}=\max\{r_j | 1\leq j-j^* \leq x\}$ \;}
				\If {$r_j \geq r_{(1)}$ }
				{
					{Select this buyer as $winner (=j)$ and charge this buyer at $payment=r_{(1)}$\;}
					{Return $winner$, $payment$\;}
				}
%				Let $v_{(1)}=\max\{v_j | 1\leq j-j^* \leq x\}$\;
%				\If{$v_j \geq v_{(1)}$}
%				{				
%					Select this buyer as $winner (=j)$ and charge this buyer at $payment=v_{(1)}*d(t_j)$\;
%					%\frac{r_j}{d(t_j)}   			
%					Return $winner$, $payment$\;	
%				}
			}		
		}

	Return $winner$, $payment$, $j$\;  	
\end{algorithm}
%%%%%%%%%%%%%%%%%%%%%%%

%%%%%%%%%%%%%%%
\subsection{Analysis of  Random-Selection Mechanism $\mathcal{M}_R$}
\label{analysis_M_R}

In this subsection, we analyze the truthfulness and competitive ratio of the mechanism $\mathcal{M}_R$ respectively in Theorem~\ref{MA:truthfulness} and \ref{M1:ratio}.

\subsubsection{Mechanism $M_R$ is Truthful}
\label{subsubsec:truthful}

Before we detail our proof, we introduce the following lemma~\cite{goldberg2001competitive}.

\begin{lemma}
\label{lemma:bid-independence-truth}
	An auction is value truthful iff it is equivalent to a bid-independent auction if it is a single parameter game.
\end{lemma}

However, the game studied in this work is a multiple-parameter game where an agent (each potential buyer) can manipulate his reported discounted valuation,
 and  his arrival time.

\begin{theorem}	
	\label{MA:truthfulness}
If the discount function $d(t)$ is non-increasing, and is a non-increasing buyer density function,
 then the mechanism $\mathcal{M}_R$ is \textbf{truthful}.
\end{theorem}
\begin{proof}
There are three possible strategies for the users to improve his utility:
1) just changing his reported valuation, 2) just delaying his arrival, 3) both delaying his arrival and changing his valuation.
We need to prove that it is both value-truthful and sem-time-truthful for all three cases.
Notice that the star-class is chosen randomly, thus, a user cannot affect the probability whether he is in the star-class or not.
Recall that there are two possible ways to select a user: in the observation phase, and in the decision phase.

\textbf{First Case}:  the user does not delay his arrival, but may change his reported valuation.
It is easy to see that $\mathcal{M}_R$ is a bid independent auction.
In the observation phase, each user observed will be selected with a  probability and charged a payment $0$, which are bid-independent.
In the decision phase, it is just a simple variation of the online reserved price auction. It is thus truthful in this phase.
Thus, it is value-truthful, \ie, the buyer's utility cannot be improved by bidding untruthfully.

\textbf{Second Case}: the user delays his arrival, but does not manipulate his reported valuation for better proft.
Then we show that a user cannot improve its utility by simply delaying his arrival.
To prevent the buyers from delaying the arrival, in our mechanism, we applied two \textbf{random procedures}: 
	1) the {Star-Class} $c^*$ is chosen randomly at uniform 
	and 2) for the buyers who appear at the observation phase of the Star-Class, we randomly choose one as the winner  and gets a positive utility, 
	 which we will show that is larger than the utility by delaying arrival time. 
	 
\textbf{Case 2.1}, 
the buyers have no incentives to delay their arrival to another class. 
	As the buyers arrive in a random order, each buyer has the equal  possibility to appear at the randomly selected Star-Class.
Recall that the discount function is non-increasing, the Star-class is then from the first $\hat{c} $ time-intervals.
A buyer delaying his arrival may end-up in the discarded classes and thus loses the chance to be selected eventually. 
Then delaying arrival will not increase the probability chosen in the star-class.

\textbf{Case 2.2}, the buyers who appear at the Star-Class $c^*$ have no incentives to delay their arrival to another time instance in $T_{c^*}$ (within the same Star-Class). 
	Recall that we have $x=\lfloor n_{c^*}/2 \rfloor$ in $\mathcal{M}_O$ as the divide line between the Observation and Decision phase. %For each $1\leq j < x$ and $j>x$ 
There are three subcases here: 1) from being at Observation-phase to being at Observation-phase; 
2) from being at Observation-phase to being at Decision-phase; 3) from being at Decision-phase to being at Decision-phase.
We will prove that the buyers  cannot improve his utility by delaying his arrival to another time instance that belongs to the same class by the following three subcases:

\begin{enumerate}
\item[Subcase \textbf{1)}:] For each buyer {$1\leq j \le x$}  who appears at the observation phase,
 $\mathcal{M}_O$ selects the $j$-th buyer as winner with the same probability $p_x = \frac{1}{x+1}$. 	
Clearly a buyer cannot improve his winning probability by delaying his arrival to the next time instance in the observation phase.
Note that when receiving multiple bids at the same time, the Mechanism selects the highest one as winner with probability $\frac{1}{x +1}$.
Actually it will have a \emph{smaller chance} to be the winner by delaying his arrival.
The reason is that  the mechanism will only select one winner, and process all  arrived users in the order of arrival.
Later arrived users need to compete with earlier arrived buyers.

\item[Subcase \textbf{2)}:] In the decision phase,  when the  buyer delays his true arrival time, it may lose the chance to be selected as the winner while cannot reduce its payment. 
Thus, users in the decision phase has no incentives to delay.
Again, delaying the arrival time actually results a smaller probability of being selected as the final winner of the game.

\item[Subcase \textbf{3)}:]
 The buyers who appear at the Observation phase delayed into the Decision phase, we need to show that the expected utility gained in the decision phase is no larger than that from the observation phase. Notice that when a user $j$ were at the observation phase, its expected utility is $\frac{1}{x+1} r_j$ as its payment at this stage is $0$.
 When it delays its arrival from observation-phase to the decision phase, the probability that it will be accepted as the winner is no more than $\frac{1}{x+1}$ (the probability is maximum when it is the first element of the decision phase and it is the largest element among all elements in observation phase and all elements in the decision phase observed so far).
 Thus, the expected utility at the decision phase is at most $\frac{1}{x+1} (r_j - r_o^*)$, where $r_o^* >0$ is the maximum observed reported price at the observation phase of the Star-Class.
This is clearly less than the expected utility from the observation phase. 
\end{enumerate}	

\textbf{Case 2.3}: We then analyze the case when a user $j$ delays his arrival from the observation phase in one class $c_{i_1}$
 to the observation phase of another later class $c_{i_2}$ with ${i_2} >{i_1}$.
 Notice that the expected utility at the observation-phase of $c_{i_1}$ is $\frac{1}{1+x_1} r_j = a_j \frac{1}{1+x_1} d(t_{j_1}) $, 
  and the expected utility at the observation-phase of $c_{i_2}$ is $\frac{1}{1+x_2} r_j =  a_j \frac{1}{1+x_2} d(t_{j_2}) $.
  Here $x_1$ and $x_2$ are the number of users at these observation phases respectively, 
   and $t_{j_k}$ ($k=1,2$) are some time-instances of the observation phase of the discount class $c_{i_k}$.
  Recall that we assume the buyers arrive follow a Poisson process, the value $x_k $ is proportional to the time-duration of the discount class $i_k$.  
  Let $d_{i_k}$ be the maximum discount value of the discount class $c_{i_k}$.
  Then as the discount function is non-increasing, we know that the length of the time interval $T_{i_k}$ (corresponding to the discount class $c_{i_k}$) 
   is $t(d_{i_k}/2) - t(d_{i_k})$.
   Here $t(d)$ is the inverse function of $d(t)$.
Assume that the number of users in a time-interval is proportional to its duration.
A sufficient condition for preventing a user from delaying his arrival from the observation phase in one class $c_{i_1}$
 to the observation phase of another later class $c_{i_2}$ with ${i_2} >{i_1}$ is then
  $a_j \frac{1}{1+x_1} d(t_{j_1}) \ge a_j \frac{1}{1+x_2} d(t_{j_2}) $. 
  Then a sufficient condition is 
 \[
 \frac{d(t_{j_1})}{1+\frac{t(d_{i_1}/2) - t(d_{i_1})}{2}}   \ge  \frac{d(t_{j_2})}{1+\frac{t(d_{i_2}/2) - t(d_{i_2})}{2}}   
 \]
 Here the time $t_{j_k}$ is some time-instance during the time-interval $T_{i_k} =[ t(d_{i_k}), t(d_{i_k}/2)]$ when the user arrived online.
 This sufficient condition is easily satisfied when the discount function is non-increasing, and non-decreasing density function.
  
 By combing the aforementioned analysis and the delaying within the star-class, we can show that
 a user has no incentives to delay his arrival from the observation phase in one class $c_{i_1}$
 to the decision phase of another later class $c_{i_2}$ with ${i_2} >{i_1}$.
 
\textbf{Third Case:}
We then show that a buyer cannot improve his utility by delaying its arrival and changing his reported valuation.
The proof is a simple combination of the proofs for the first two cases.
We omit the details here.

Moreover,  every buyer has a chance to be the winner and the payment is always no more than his valuation.
	In summary, the mechanism $\mathcal{M}_R$ is {truthful} and satisfies IR and CS.
\end{proof}

To further prevent a user from manipulating his arrival time or reported bidding-price, in our mechanism we could add the following design without affecting the
 competitive ratio and truthfulness of the mechanism.
1) the duration of the observation phase is randomly chosen with carefully selected probability, 
2) the boundaries of the intervals are randomly selected within a small perturbation of the fixed interval boundary. 
Then the agent cannot know the exact boundaries of the intervals and whether it is in the observation phase or selection phase.
It would be difficult for the user to decide whether to delay its arrival to increase its expected revenue.

\subsubsection{Competitive Ratio of  Mechanism  $M_R$}
\label{subsubsec:ratio}

Recall that in Section~\ref{sec:bound}, we have proved that no mechanism satisfying IR and IC properties is able to obtain a bounded competitive ratio
  without any restrictions on $d(t)$, see Theorem~\ref{theo:known-dt-truthful} for details.
Especially, there is a worst case (see Eq.~\eqref{worstcase-n1}) such that:
1) there is no  constant $ B>1 $ such that $n_1\ge 1$ according to our \textbf{Discounts Division Algorithm};
2)  the rest of the $n-1$ buyers have extremely small discount value. In such case, no mechanism will have a bounded competitive ratio.
Here, we relax some restrictions on $d(t)$ such that we can explore possible competitive mechanisms. 
 \begin{assumption}
 \label{assumption-n-1}
There exists a constant $ B>1 $ such that $n_1\ge2$.
\end{assumption}

We then analyze the competitive ratio of the mechanism $M_R$, and 
 use the well known offline truthful Vickrey auction~\cite{vickrey1961counterspeculation} mechanism $\mathcal{M}_V$ as the baseline for comparison. 
Let $\eta=\max\{ \frac{n_c}{n_1} , \frac{n_1}{n_c} \mid c \in[1, \hat{c}] \}$.
Assume that there is a constant $ B>1 $ such that all discount classes have  comparative number of buyers, that is 
 $\frac{1}{\eta} \le \frac{n_c}{n_1}\leq \eta$, where $c\in [1, \hat{c}]$ and $\eta >1 $ is a constant.

In the rest of the section, we will prove that the mechanism $\mathcal{M}_R$ is {$O(\eta \log^2 n)$}-competitive,
 \ie, $\frac{E[\mathcal{M}_V]}{E[\mathcal{M}_R]} \leq O(\eta \log^2 n)$.
 Here $\eta = \max \frac{n_i}{n_j}$, and $n_i$ is the number of users in the $i$-th discount class.
The proof of this result is composed of following lemmas and theorems.
Given the reported prices $\bm{r}=(r_1,r_2,\dots, r_n)$ in sequence,
  the Vickrey mechanism $\mathcal{M}_V$ always selects the highest bidder but charges the second-highest bid. 
  Thus we have $E[\mathcal{M}_V]=E[r_{(2)}]$, where $r_{(2)}$ is the second-largest reported price among all potential buyers.
That is, the expected revenue achieved by mechanism $\mathcal{M}_V$ is the expected value the second-largest reported price.

To prove Theorem~\ref{M1:ratio}, we first show that the globally second-largest expected report price $r_{(2)}$
  gets most of its value from top $O(\log n)$ classes  (see proof in Eq.~\eqref{equ:E_c(2)}). 
Then we prove that at each reserved class $c\in \mathbb{C}$,
$\mathcal{M}_R$ has $O(\log^2 n)$-competitive ratio with respect to the expected value of the second-largest reported price achieved in class $c$
 in Lemma~\ref{lemma:MU-lossInClass}.
Recall that the number of reserved classes 
\begin{eqnarray}
\label{hatc}
\hat{c} =4\log_{B} n +3 +\log_{B} \frac{v_{1}}{v_{2} } 
\end{eqnarray}
 if we partition the discount function to a sequence of intervals such that the discount values in each interval is within a factor $B$ of each other.
% When $B=2$, we know that the number of reserved classes $\hat{c}$ is
% $\hat{c}= 4\log_{2} n +3 +\log_{2} \frac{v_{1}}{v_{2}}$.

First we show that the expected revenue received from classes other than the reserved discount-classes is at most a constant fraction of the optimum.
Thus, focusing on the first $\hat{c}$   discount classes, \ie, reserved classes, will not impact the order of the competitive ratio.

\begin{lemma}
	\label{bound-r(2)-first-hat-c}	
	\small
	%$\sum_{c \geq 3\log n +2} E_c[r_{(2)}] \leq E[r_{(2)}]/2$.	
	$\sum_{c \geq \hat{c}+1} E_c[r_{(2)}]  < \frac{E[r_{(2)}]}{B(B-1)}$, where $B>1$ is some constant  used to partition the discount classes. 
	%See $\hat{c}$ in Equation~(\ref{hatc}). %$4\log_{B} n +3 +\log_{B} \frac{v_{(1)}}{v_{(2)}} E_c[r_{(2)}]$
\end{lemma}
\begin{proof}
As a general extension,	we divide the discount values into different classes as
$[B^{-(c-1)}d_{max}, B^{-c}d_{max})$, where integer $c\ge 1$,  and $B>1$ is any constant.
Notice that in our designed mechanism $B =2$.
Let ${\textbf V}= \{ v_1, v_2, v_3, \cdots, v_i, v_{i+1}, \cdots, v_n \}$ be the set of initial valuations of $n$ buyers with $v_i \ge v_{i+1}$.
Then $v_{j}$ is the $j$-th largest values in the set $\textbf V$.
As the adversary is arrival-sequence adaptive online, given $n$ arrival time sequence ${\textbf T}=\{ t_1, t_2, \cdots, t_n \}$ with $t_i < t_{i+1}$ chosen by the Poisson process,
 the adversary will  map  ${\textbf T}$ to one of the possible $n!$ permutations  $\Pi$ of $\textbf V$.
 Let $\pi \in \Pi$ be a mapped permutation used by the adversary.
 Let $n_c$ be the number of time-instance $t_k$ from the $c$-th time-interval $T_c$ corresponding to the $c$-th discount class.
 Recall that in our design, we assumed that $n_c \ge 2$ for any $1 \le c \le \hat{c}$.
 We define two events as follows
 \begin{itemize}
 \item[\textbf{Event $X_{j,c}$:}]
We let $X_{j,c}$ be the event that  $v_{j}$ appears at some time $t_k$ from the $c$-th time interval $T_c$  (\ie, $\pi(k)=j$), and
  $v_{j}d(t_k)$ is the second-largest reported price $r_{(2)}$ among all reported prices globally.
  Denote the probability that this event $X_{j,c}$ happens as $P_{jc}$.
\item[\textbf{Event $Y_{j,c}$:}]
We let $Y_{j,c}$ be the event that  $v_{j}$ appears at some time $t_k$ from the $c$-th time interval $T_c$  (\ie, $\pi(k)=j$), and
  $v_{j}d(t_k)$ is the largest reported price $r_{(1)}$ among all reported prices globally.
  \end{itemize}
  
For the event $X_{j,c}$, as  $d(t_k)v_{j} $ is the second largest reported price, assume that $ d(t_q)v_{\pi(q)}$ is the largest reported price for some time-instance $t_q$.
Then by considering all possible $q$, $\prob{X_{j,c}}$ can be computed as 
$ \prob{\pi(k)=j} \cdot \prob{ \lor_{q=1,q\not= k}^{n} { \left( d(t_k)v_{j} \le  d(t_q)v_{\pi(q)}  \land_{l=1,l\not= q,j}^n d(t_k)v_{j} \geq  d(t_l)v_{\pi(l)} \right)} } $
%=\prob{\pi(k)=j} \cdot \prod_{l=1,l\not= j}^{n}\prob{ d(t_k)v_{j} \geq  d(t_l)v_{\pi(l)} } $
 where $v_{\pi(l)}$ is the initial valuation of the buyer who appears at time $t_l$ from a $c$-th discount class.
 
 For the event $Y_{j,c}$, its probability $\prob{Y_{j,c}}$ can be computed as 
\begin{eqnarray*}
&& \prob{\pi(k)=j} \cdot \prob{  d(t_k)v_{j} \geq  d(t_l)v_{\pi(l)}, \forall l \in [1,n] } \\
&=&\prob{\pi(k)=j} \cdot \prod_{l=1,l\not= j}^{n}\prob{ d(t_k)v_{j} \geq  d(t_l)v_{\pi(l)} } 
\end{eqnarray*}
% where $v_{\pi(l)}$ is the initial valuation of the buyer who appears at time $t_l$ from a $c$-th discount class.

Let $E_c[r_{(2)}]$ denote the expected value of second-largest price that is obtained from the $c$-th class (\ie, $r_{(2)}$ appears at class $c$).
Here the expectation is taken over all possible permutations of $\Pi$, while the arrival times ${\textbf T}$, the initial valuations ${\textbf V}$, and the
 discount functions $d(t)$  (thus all discount classes) are given.
Note that $E[r_{(2)}] =\sum_{c=1}^{n} E_c[r_{(2)}] $ as there are at most $n$ discount classes that contain users. 
Then 
\[
E_c[r_{(2)}] =\sum_{j=1}^{n} \prob{X_{j,c}} v_j d(t_k) \ge \sum_{j=1}^{n} \prob{X_{j,c}} v_j \cdot \frac{d_{\max}}{B^c}
\] 
as $ \frac{d_{\max}}{B^c} \le d(t_k) \le \frac{d_{\max}}{B^{c-1}}$ for a time-slot  $t_k$ from the $c$-th discount class.
Recall that $d_c=  \frac{d_{\max}}{B^{c-1}}$   is the maximum discount values at the reserved-class $c$.

Consider the second largest initial value $v_2$ from $\textbf V$, we give a lower bound on the probability $\prob{X_{2,c}} $.
When $c=1$,  one case for  the event $X_{2,c}$ to happen is when  $v_2$ is mapped to the time $t_2$ in the time interval $T_1$,
  and $v_1$ is mapped to the time instance $t_1$. Then $\prob{X_{2,1}} \ge \frac{(n-2)!}{n!} \ge 1/n^2 $.
Then the expected revenue (\ie, the expected value of the globally second largest reported price) by the Vickrey auction is
\begin{eqnarray}
\label{lemma-lower-expected-r2}
E[r_{(2)}] > E_1[r_{(2)}]  > \frac{d_{\max}}{B n^2} v_2
\end{eqnarray}

%\xue{
%\begin{equation}
%\label{lemma-lower-expected-r2-2}
%\begin{aligned}	
%E[r_{(2)}] =\sum_c E_c[r_{(2)}]   & > \sum_c\left( \frac{B^{-c}*d_{max}*v_{(2)}}{n^2}  \right) \\ & > \frac{d_{\max}v_{(2)}}{B(1-1/B) n^2} 
%\end{aligned}
%\end{equation}
%}
We then prove some upper bound on the expected value of $E_c[r_{(2)}] $.
		\begin{equation}
		\small
		\label{equ:E_c[r_{(2)}]}
		\begin{aligned}	
		E_c[r_{(2)}] 
		& =	\sum_{t_k \in T_c} \left( d(t_k) \cdot \sum_{j=1}^{n}  \left(v_{j} \cdot \prob{X_{j,c}} \right) \right)	\\
		& \leq  \sum_{t_k\in T_c} B^{-(c-1)}d_{max} \sum_{j=1}^n v_{j}  \\
		& \leq B^{-(c-1)} d_{max} \cdot n_c \cdot n \cdot v_1
		\end{aligned}
		\end{equation}
		
%		\xuenote{We lose too much here.}\\
Note that ${n}_c < n$, and we further set 
	\begin{eqnarray}
	\label{hatc_B}
	\tilde{c} = 4\log_{B} n +3 +\log_{B} \frac{v_{1}}{v_{2}}
	\end{eqnarray}
	
%	\xue{	\begin{eqnarray}
%	\label{hatc_B-2}
%	\tilde{c} = 4\log_{B} n +2 +\log_{B} \frac{v_{1}}{v_{2}}
%	\end{eqnarray}
%	}
From (\ref{equ:E_c[r_{(2)}]}), we obtain the following inequality:
	\begin{equation}
	\small
	\label{equ:sum:E_c[r_{(2)}]}
	\begin{aligned}	
	\sum_{c \geq \tilde{c}+1} E_c[r_{(2)}] 
	< \sum_{c \geq \tilde{c}+1 } B^{-c} Bn^2 d_{max}v_{1}
    < \frac{d_{max} v_{2}}{n^2 B^3 (1-1/B)} 
	\end{aligned}
	\end{equation}
	
By combining Eq.~(\ref{lemma-lower-expected-r2}), (\ref{hatc_B}) and  (\ref{equ:sum:E_c[r_{(2)}]}),
%\xue{$\sum_{c \geq \tilde{c}} E_c[r_{(2)}] < \frac{E[r_{(2)}] }{B}$, we get that $\sum_{c \geq \hat{c} +1} E_c[r_{(2)}]  < \frac{E[r_{(2)}]}{B(B-1)} $. }	
we finish the proof of the lemma.	
\end{proof}

This directly implies that the expected revenue takes more than half of its value from the top $\hat{c}$  discount classes,
if $B \ge 2$.
Furthermore, we have the following theorem
\begin{theorem}
By  assuming $\frac{v_{1}}{v_{2}}\leq  n^k$ for some number $k>0$,  
 then more than half of $E[r_{(2)}]$ is from the top $ O(\log n)$  discount classes, \ie,
\begin{eqnarray}
\label{equ:E_c(2)}
\sum_{c=1}^{\hat{c}} E_c[r_{(2)}]=E[r_{(2)}]-\sum_{c \geq (\hat{c}+1)} E_c[r_{(2)}] > E[r_{(2)}]/2
\end{eqnarray} 
\end{theorem}

Recall that we consider the arrival-sequence adaptive online adversary.
Let $\Pi$ be the set of $n!$ permutations and 
 $\Pi'$ denote the arrival sequence set (permutations) that the second-largest reported price $r_{(2)}$ appears at the top $\hat{c}$ classes.
Let $\mathcal{M}(\pi)$ (or $\mathcal{M}(\Pi')$) denote $\mathcal{M}$'s total expected revenue achieved from the  arrival sequence defined by mapping $\pi$
 (or the sequence set $\Pi'$).
Then we have: $$\mathcal{M}_V (\Pi')=\sum_{c=1}^{\hat{c}} E_c[r_{(2)}]=\frac{1}{|\Pi'|} \sum_{\pi \in \Pi'} r_{(2)}(\pi)> E[r_{(2)}]/2 $$
This directly implies the following lemma: %deduces
\begin{lemma}
	\label{lemma:MVPi}		
	$\mathcal{M}_V (\Pi') > {\mathcal{M}_V (\Pi)}/{2}$.
\end{lemma}

Since $\mathcal{M}_R$ discards the classes whose ids are larger than $\hat{c}$, we have $\mathcal{M}_R(\Pi) \ge \mathcal{M}_R(\Pi')$.
In the following, we focus on
% the value of $\mathcal{M}_R(\Pi')$
the performance of $\mathcal{M}_R$ on the sequence set $\Pi'$.
\begin{lemma}
	\label{lemma:MU-lossInClass}		
	$ {\mathcal{M}_R(\Pi')} \geq \frac{ \mathcal{M}_V (\Pi') }{O(\log^2 n)} $.
	%in each of the top $\hat{c}$ classes.
	%We achieve more than $\min\{1/2,\frac{v_{(n)}}{v_{(1)}}\}$ competitive ratio in each discount class.
\end{lemma}
\begin{proof}
Given the set $\textbf V$ of valuations for all $n$ buyers, the discount function $d(t)$, and the constant $B$ to partition the discount values into discount classes, 
we discuss the performance of a mechanism in the following five complementary sets of permutations:
	\begin{compactenum}
		\item $\Pi_0 $: the sequence set that both $r_{(2)}$ and $r_{(1)}$ appear at the top $\hat{c}$ but different classes;
		\item $\Pi_1$: the sequence set that $r_{(2)}$ appears at the top $\hat{c}$ classes and $r_{(1)}$ appear at the same class with $r_{(2)}$; 
		\item $\Pi_2$: the sequence set that $r_{(2)}$ appears at the top $\hat{c}$ classes while $r_{(1)}$ at some class whose id is larger than  $\hat{c}$. 
		\item $\Pi_3$: the sequence set that  both $r_{(2)}$ and $r_{(1)}$ are at some classes whose id are larger than  $\hat{c}$. 
		\item $\Pi_4$: the sequence set that   $r_{(1)}$ appears in the first $\hat c$ classes, and  $r_{(2)}$ is at some class whose id is larger than  $\hat{c}$. 
	\end{compactenum}
	Clearly, we have $\Pi'= \Pi_0	\cup \Pi_1 \cup \Pi_2$. 
	
	To prove this lemma, we first introduce the following claims.
	Let $n_{(c)}$ denote the $c$-th largest class in terms of the number of buyers.
	Assume there are at least $2$ buyers for each reserved class $c$, \ie, $n_c\geq 2, \forall c\in \mathbb{C}$.
	Moreover, we assume that each class has a comparative number of buyers, \ie, we have
	 $\frac{n_{(1)}} {n_{c}} \le \eta$, for any class $c \le \hat{c}$, where $\eta >1$ is a constant.
%	\xue{Assume $\xi=\frac{  \sum_{c=1}^{\hat{c}} n_{c} } {n_{ \hat{c}+1}} $, where $\xi$ is a constant number.}
	%To prove Lemma~\ref{lemma:MU-lossInClass}, we first introduce the following claim:
For any subset of permutations $\Pi$, let $r_{(2)}(\Pi)=\sum_{\pi \in \Pi} r_{(2)}(\pi)$. 
Here $r_{(2)}(\pi)$ denotes the second largest reported price in a reported-price sequence defined by the mapping $\pi$.
	
Then we have the following claim:
	\begin{claim}
		\label{claim1}
If $\eta$ is a bounded value, then $ r_{(2)}(\Pi_1) \geq \frac{1}{\eta \cdot \hat{c}}  {r_{(2)}(\Pi_0)} $. \\
	\end{claim}
	\noindent\textbf{Proof Sketch}: 
	For any sequence $\pi$ in $\Pi_0$, there is always a sequence $\pi'$ from $\Pi_1$ with the following properties: 
	1) $r_{(1)}(\pi') \geq r_{(1)}(\pi)$ and  2) $r_{(2)}(\pi') \geq r_{(2)}(\pi)$. 
	Moreover, for each $\pi'\in \Pi_1$, there is at most $\eta \cdot \hat{c}$ sequences $\pi$ in $\Pi_0$ that can be mapped from. 
	Thus, we have ${r_{(2)}(\Pi_1)} \geq \frac{1}{\eta \cdot \hat{c}} \cdot {r_{(2)}(\Pi_0)}$.	

	\begin{proof}
		Let $\Pi_0 ^{2,1}$ denote the subset of $\Pi_0$ where $r_{(2)}$'s class id is smaller than  $r_{(1)}$'s;
		and $\Pi_0 ^{1,2}=\Pi_0 \setminus \Pi_0 ^{2,1}$ denote the sequence set where $r_{(2)}$'s class id is larger than  $r_{(1)}$'s.
		For simplicity, we use $c^{(k)} (\pi)$ to denote the class id of $r_{(k)}$ in sequence $\pi$, % \ie, $c^{(2)} (\pi) < c^{(1)} (\pi)$; \ie, $c^{(1)} (\pi) < c^{(2)} (\pi)$.
		and $r_{(j)}(\pi)$ to denote the $j$-th largest price in $\pi$. 
		
		For each $\pi \in \Pi_0$, let $t_q$ denote the largest reported price $r_{(1)}$'s corresponding buyer's arrival time, and his initial valuation
		 $v_{\pi(q)}$ is the $y$-th largest one in ${\textbf V}=\{v_1, v_2, \dots, v_n\}$.
		Similarly, $t_p$ denotes the $r_{(2)}$'s corresponding buyer's arrival time, and his initial valuation $v_{\pi(p)}$ is the $z$-th largest one among all initial valuations.
		Formally, we have:
		\begin{equation*}
		\begin{aligned}
		\begin{cases}
		r_{(1)}=v_{\pi(q)} \cdot d(t_q)=v_y \cdot d(t_q)\\
		r_{(2)}=v_{\pi(p)}\cdot d(t_p)=v_z \cdot d(t_p)
		\end{cases}
		\end{aligned}
		\end{equation*}
		
		For any permutation sequence $\pi$ in $\Pi_0^{2,1}$ or $\Pi_0^{1,2}$, we show that there is always a sequence $\pi'$ from $\Pi_1$ with following properties: 1) $r_{(1)}(\pi') \geq r_{(1)}(\pi)$ 
		and  2) $r_{(2)}(\pi') \geq r_{(2)}(\pi)$. 
		%In other words, the revenue that $\mathcal{M}_V$ obtained from $\pi'$ is at least as the value obtained from $\pi$.
		We prove this separately for $\Pi_0^{2,1}$ and $\Pi_0^{1,2}$.
		
\textbf{Case 1)}
		For each $\pi \in \Pi_0 ^{2,1}$, 
		there is a $\pi' \in \Pi_1$ that satisfies that $r_{(1)}(\pi')$ and $r_{(2)}(\pi')$ are from the same discount class.
		We consider the discount class $c^{(2)}(\pi)$ from which the second-largest reported price $r_{(2)}(\pi')$ produces.
		Recall that there are at least two time-instances in this class with users arrive, \ie, $n_{c^{(2)}(\pi)} \ge 2$, where one time-instance produces the second-largest
		 reported price $r_{(2)}(\pi)$.
		 Let $t_j$ be any other time-instance from this discount class. 
		 We can define a new mapping $\pi'$ by switching the values at time $t_q$ and time $t_j$. 
		 In other words we swap the values $v_{\pi(q)}$ with $v_{\pi(j)}$. 
		 For all other time-instances, $\pi$ and $\pi'$ are same.
		\begin{equation}
		\begin{aligned}
		\begin{cases}
		\pi' (t_j)= \pi (t_q)=v_{y}\\
		\pi' (t_q)=\pi (t_j)\\
		\pi' (t_l)=\pi (t_l), \text{ for all } t_l \not = t_j, t_q
		\end{cases}
		\end{aligned}
		\end{equation}
		where $t_j \in T_{c^{(2)}(\pi)}$.			
		Since $t_j \in T_{c^{(2)}(\pi)}$ and $t_q \in T_{c^{(1)}(\pi)}$, we have $d(t_q) \leq d(t_j)$.
		Thus, $v_{y}  d(t_j) \geq v_{y} d(t_q)$.
		The newly reported prices at the time-instance $t_q$ in sequence $\pi'$ will be 
		$v_{\pi'(t_q)} d(t_q)= v_{\pi(t_j)} d(t_q) \le v_{\pi(t_j)} d(t_j)$.
		Recall that for all other reported prices, the values in $\pi$ and $\pi'$ are the same.
		Thus, we have $r_{(1)}(\pi') = v_{y}  d(t_j) \geq r_{(1)}(\pi)$ and $r_{(2)}(\pi')=r_{(2)}(\pi)$.
		Values of $r_{(1)}(\pi')$ and $r_{(2)}(\pi')$ now come from the same discount class in $\pi'$.
		
\textbf{Case 2)}
	For each $\pi \in \Pi_0 ^{1,2}$, in this case, we find a time-instance $t_j$ from the class $c^{(1)}(\pi)$, where $t_j \not = t_q$.
	Then we exchange the mapping between $t_j$ and $t_p$ from the mapping $\pi$ to produce another mapping $\pi'$.
	Thus, the new mapping $\pi' \in \Pi_1$ is defined as 
		\begin{equation}
		\begin{aligned}
		\begin{cases}
		\pi'(t_j)= \pi (t_p)=v_{z}\\
		\pi'(t_p)=\pi(t_j)\\
		\pi'(t_l)=\pi(t_l), \text{ for all } t_l \not = t_j, t_p
		\end{cases}
		\end{aligned}
		\end{equation}
		where $t_j \in T_{c^{(1)}(\pi)}$.	
		%and $\pi'(t_l)=\pi(t_l)$ for each $t_{l\not = j,p}$.			
		Since $t_j \in T_{c^{(1)}(\pi)}$ and $t_p \in T_{c^{(2)}(\pi)}$, we have $d(t_p) \leq d(t_j)$.
		Thus, $v_{x}*d(t_j) \geq v_{x}*d(t_p)$.
		Finally we have $r_{(1)}(\pi') \geq r_{(1)}(\pi)$ and $r_{(2)}(\pi')\geq r_{(2)}(\pi)$.
Notice that it is possible that in the new mapping $\pi'$, the value $ r_{(1)}(\pi)$	is less than the value 
 $v_z d(t_j)$. Thus, we have a new $r_{(1)}$ in the mapping $\pi'$.
		
Observe that, for each mapping $\pi$ in $\Pi_0$ and the corresponding mapping $\pi'$ from $\Pi_1$ produced by our proof,
 we have $r_{(1)}(\pi') \ge r_{(1)}(\pi) $ and  $r_{(2)}(\pi') \ge r_{(2) }(\pi) $.
Notice that it is possible that a number of permutation sequence $\pi$ from $\Pi_0$ could be mapped to the same sequence $\pi'$ from the set $\Pi_1$.
For each $\pi'$, we need to reduce the number of to-be-mapped permutations $\pi$. 
To do so, for each $\pi \in \Pi_0^{1,2}$, assume that $r_{(2)}(\pi)$ is from the $g$-th time-slot in the discount class. Then we swap this arrival with the $\lfloor{g/ \eta}\rfloor$-th timeslot
 in the discount class where $r_{(1)}(\pi)$ lies.
As $n_i /n_j \le \eta$ for any discount class $i$ and $j$,  $\lfloor{g/ \eta}\rfloor$ is a valid timeslot.
Similar argument holds for each $\pi \in \Pi_0^{2,1}$.
Then for each $\pi'$ from $\Pi_1$, there are at most $\eta\cdot \hat{c}$ permutations $\pi$ from 
$\Pi_0$ that can be mapped to $\pi'$ by our design.
%For each $\pi'\in \Pi_1$, there is at most $\eta*\hat{c}$ sequences in $\Pi_0$ that can be mapped from. 
		Thus, ${r_{(2)}(\Pi_1)}\geq \frac{1}{\eta \cdot \hat{c}} \cdot {r_{(2)}(\Pi_0)}$.
		This finishes the proof of this claim.
	\end{proof}

To prove a  competitive ratio bound on $M_R$, we then compare the value of $ r_{(2)} (\Pi_2)$  with $ r_{(2)} (\Pi_0)$ and 
	 $ r_{(2)} (\Pi_1)$.
When given the discount function $d()$,
 let $\Pi^{i, l}$ be the set of all permutations where the user with the $i$-th largest initial valuation $v_i$ appearing at the time $t_l$, \ie, 
  the user is the $l$-th arrived user; and the value $r_l= v_i \cdot d(t_l)$ is the \emph{second} largest reported price among $n$ users.
   We use $\Pi^{i}$ to denote all permutations $\pi$ where $v_i$ arrived at some time-slot and  results the second largest reported price.
  Then clearly, $\Pi = \bigcup_{i=1}^{n} \bigcup_{l=1}^{n} \Pi^{i, l} = \bigcup_{i=1}^{n} \Pi^i$, which are all $n!$ possible permutations.
 Let $\Pi_0^{i, l} = \Pi_0 \bigcap \Pi^{i, l}$, $\Pi_1^{i, l} = \Pi_1 \bigcap \Pi^{i, l}$, and $\Pi_2^{i, l} = \Pi_2 \bigcap \Pi^{i, l}$.
 Similarly, we define $\Pi_0^{i}$, $\Pi_1^{i}$, and $\Pi_2^{i}$.
 Let   $ \| \Pi_2^{i, l} \|$ be the cardinality of the set $ \| \Pi_2^{i, l} \|$.
 Then we have
 \begin{eqnarray}
 \label{total-revenue-pi-2}
  r_{(2)} (\Pi_2) = \sum_{i=1}^{n} \sum_{l \in \text{  top $\hat{c}$ classes} } v_i \cdot d(t_l) \| \Pi_2^{i, l} \|.
 \end{eqnarray}
	
	\begin{claim}
	\label{claimPi2}
	If $\frac{v_1} {v_2} \le n^k$ for some number $k>0$, then we have
	\[ r_{(2)} (\Pi_2) \leq  (k+4) \log n \cdot r_{(2)}(\Pi_1).\]
	\end{claim}
	\begin{proof}
 Recall that we assumed that $v_1 \ge v_2 \ge v_3 \cdots \ge v_n$ and $\frac{v_2}{v_1} \ge 1/n^k$ for some number $k>0$.
  Then we partition all the values $v_i$ into two subsets: 1) $V_L= \{v_i \mid v_i \ge v_2/n^3 \}$, and 2) $V_S= \{v_i \mid v_i < v_2/n^3 \}$.
 
 We restrict our attention to all permutations from $\Pi_2$.
 For all users in the set $V_L$ (called users with "large" valuations), clearly, when user $i$ arrives at the time $t_l$ (belonging to some class $q$)
  and produces the second largest report price,
  then there is a user $j$ with the largest reported price and user $j$ appears at some time $t_z$ from a class $p$ after the class $\hat{c}$.
  Then $v_j \le v_1 \le n^k v_2 \le n^{k+3} v_i$.
  On the other hand, 
  \[v_j B^{-p+1} \ge v_j d(t_z) \ge  v_i d(t_l) \ge v_i B^{-q}.\]
  Then $B^{p-q-1} \le v_j / v_i \le n^{k+3}$ implies that $p-q \le (k+3) \log_B n +1$.
In other words, the class $p$ producing the largest reported price, and the class $q$ that produces the second largest price is at most $(k+3) \log n$ separated.
For any permutation $\pi$ from $\Pi_2^{i,l}$,
  we can swap the element $v_j$ with some element from the $q$-th class.
 If  then element $v_j$ is the $y$-th arrived user from the class $p$, then the new position of user $j$ is the $y/\eta$-th user from the $q$-th class. 
 Obviously this mapping is a valid mapping as we assume that the number of users at the class $q$ is at least $1/\eta$ fraction of the number of users from the class $p$.
 We call this new permutation as $\pi'$.
 For this new permutation $\pi'$, the user $i$ still produces the second largest reported price as only the largest price increases.
 Then $\pi'$ now is in $\Pi_1^{i,l}$.
 For the aforementioned mapping from a permutation $\pi \in \Pi_2^{i,l}$ to a permutation $\pi' \in \Pi_1^{i,l}$, there are at most $(k+3) \log n$ permutations $\pi$ produces the same permutation $\pi'$.
Then $r_{(2)} (\Pi_2^{i,l}) \le (k+3)\log n \cdot r_{(2)} (\Pi_1^{i,l}) $.
Then it is easy to show that  the revenue produced by all users $v_i$ from $V_L$ is at most
  \begin{eqnarray}
  \label{large-vi}
  r_{(2)} (\Pi_2 \bigcap V_L) \le (k+3) \log n \cdot r_{(2)} (\Pi_1)  .
  \end{eqnarray}
  Here we abuse the notation $\Pi_2 \bigcap V_L$ to denote the set of permutations from $ \Pi_2$, 
  each of these generates a second largest reported price by some element from $V_L$.

For users $v_i$ from the set $V_S$  (called users with "small" valuations), clearly, $ i\ge 3$.
Then for each permutation $\pi$ from $\Pi_2^{i,l}$, we can define a unique permutation $\pi'$ as follows:
\begin{enumerate}
\item Swap the arrival time-slots of $v_i$ and $v_2$, \ie, $v_2$ from some time-instance $t_p$ to the time-slot $t_l$, and move $v_i$ to $t_p$;
then $v_2 d(t_l) >  v_i d(t_l)$ and $v_i. d(t_p) <  v_2 d(t_p)$. Values of  $v_2 d(t_l)$ could be the second largest report price, or the largest report price.
\item
If  $v_2 d(t_l)$ is the second largest report price, we continue to swap elements that produces the largest reported price. 
Let $v_j$ be the element arrived at sometime $t_q$ from a discount-class $c > \hat{c}$, and it still produces the largest reported price.
Clearly, this element must be $v_1$ as $v_j \le v_2$ for all $j \ge 3$, and $d(t_q) \le d(t_l)$.
Then we swap the element $v_1$ at time $t_q$ with an element $v_j$ appeared at time $t_{l+1}$ from the same discount-class with $v_2$.
Then obviously now both the first and the second largest reported price are from the same discount class (although there is a situation when $v_2$ is the last element of this discount class, we can use time-slot $t_{l-1}$ instead.).
\item
Otherwise, $v_2 d(t_l)$  now is the largest reported price.
Assume that the element $v_1$ arrives at time $t_q$ in the permutation $\pi$.
Then we swap $v_1$ with some element, say $v_j$ appeared at the time $t_{l-1}$.
When $l=1$, we can put $v_1$ at $t_1$, $v_2$ at $t_2$, and $v_j$ at time $t_q$.
After such swap,  $v_2 d(t_l)$  now is the second-largest reported price.
\end{enumerate}
In both cases, we can show that $v_2$ is the element that will generate the second-largest reported price, and $v_1$ is the element that
 results the largest reported price.
Both the first and the second largest reported price are from the same discount class.
Let the resulted new permutation be $\pi'$.
Clearly, for permutation $\pi'$, it is either in the permutation set $\Pi_2^{2,l}$ (when we only swap element  $v_2$ with $v_i$), or in the permutation set
  $\Pi_1^{2,l}$.
  Furthermore, each permutation $\pi$ is mapped to one permutation $\pi'$ in our definition.
  For one $\pi'$, there are at most $n^2$ possible permutations $\pi$ that can be mapped to this $\pi'$ as there there are at most $n^2$ pairs of locations for $v_1$ and $v_2$.
  Then $\| \Pi_2^{i,l} \|  \le n^2 \| \Pi_1^{2,l}\| $.
  Thus,  for all elements $v_i \in V_S$, the revenue produced by these elements is at most 
\[r_{(2)} (\Pi_2^{i,l}) =  v_i d(t_l) \| \Pi_2^{i,l} \| \le  \frac{v_2 }{n^3} d(t_l) \cdot n^2 \| \Pi_1^{2,l} \| = \frac{r_{(2)} (\Pi_1^{2,l}) }{ n}.\]
Then it is easy to show that  the revenue produced by all users $v_i$ from $V_S$ is at most
 \begin{eqnarray}
  \label{small-vi}
 r_{(2)} (\Pi_2 \bigcap V_S) \le r_{(2)} (\Pi_1^{2}) \le r_{(2)} (\Pi_1) .
 \end{eqnarray}
  Here we abuse the notation $\Pi_2 \bigcap V_S$ to denote the set of permutations from $ \Pi_2$, 
  each of these generates a second largest reported price by some element from $V_S$.
 We use $\Pi_1^{i}$ to denote all permutations $\pi$ where $v_i$ arrived at some time-slot and  results the second largest reported price.
 The claim then follows from the inequalities~\eqref{large-vi} and ~\eqref{small-vi} as
 $ r_{(2)} (\Pi_2 ) = r_{(2)} (\Pi_2 \bigcap V_S) +  r_{(2)} (\Pi_2 \bigcap V_L)$.
 This finishes the proof of the claim.
	\end{proof}

Recall that Lemma~\ref{bound-r(2)-first-hat-c} implies that, if $B \ge 2$, 
$r_{(2)}(\Pi)= r_{(2)}(\Pi_0)+ r_{(2)}(\Pi_1)+r_{(2)}(\Pi_2)+r_{(2)}(\Pi_3) +r_{(2)}(\Pi_4) \le 2( r_{(2)}(\Pi_0)+ r_{(2)}(\Pi_1)+r_{(2)}(\Pi_2)) $.
Then $r_{(2)}(\Pi) \le 2( (k+3) \log n+1+ \eta \hat{c}) r_{(2)}(\Pi_1)$ if $v_1 /v_2 \le n^k$.
Then for the Vickrey mechanism, if $v_1 /v_2 \le n^k$, we have
	\begin{equation}
	\label{equ:M_V}
	\begin{aligned}			
	\mathcal{M}_V(\Pi') & =\mathcal{M}_V(\Pi_0)+\mathcal{M}_V(\Pi_1)+\mathcal{M}_V(\Pi_2)\\
	&=  r_{(2)}(\Pi_0)+ r_{(2)}(\Pi_1)+r_{(2)}(\Pi_2)\\
	&\le  2( (k+3) \log n+1+ \eta \hat{c}) r_{(2)}(\Pi_1)
	\end{aligned}
	\end{equation}

It seems that it is straightforward that $M_R$ will have a good competitive ratio as our mechanism will work on a discount-class, and 
 previous work showed that \emph{observe-then-decide} mechanism will get a constant fraction of the second largest price.
 However, the subtle difference here is that $r_{(2)}$ and $r_{(1)}$ are different from the values $v_2$ and $v_1$.
 Previous mechanisms only care about the values of $v_1$ or $v_2$.
 
	Next, we study the performance of $\mathcal{M}_R$ on each reserved class $c\in \mathbb{C}$.	
	Let $d_{max}^c$ and  $d_{min}^c$ denote the maximum and minimum discount value in class $c$ respectively.
	Let $\Pi_1^{(c)} \in \Pi_1$ denote the sequence set that 
	 both the global largest reported price $r_{(1)}$ and  the global second largest report price $r_{(2)}$ appear at class~$c$.
	 For each $\pi \in \Pi_1^{(c)} $, let $V^c(\pi)=\{v_{(j)}^c | 1\leq j \leq n_c\}$ denote the initial valuation set of class $c$ in the permutation $\pi$,
 where $v_{(j)}^c$  is the user appearing at the class $c$ with the $j$-th largest initial valuation. 
Here $n_c$ is the number of arrived users in class $c$.
We further use $\Pi(V^c)$ to denote all the permutations in which the initial valuations appear at the class $c$ is $V^c \subseteq {\textbf V}$. 
%and $V^c(\pi)$ denote the valuations set of class $c$ in the permutation $\pi$. 
For a permutation $\pi$, $V^c(\pi)$ is the set of initial valuations of users appearing at the class $c$ under arrival sequence $\pi$.
Then $\Pi(V^c(\pi))$ is the permutations set in which each $\pi$ has the \emph{same} valuation set in class $c$.

	Then we have the following two claims for each $c\in [1,\hat{c}]$.

\begin{claim}
\label{Fact1}
For each discounted class $c$ and each arrival sequence $\pi \in \Pi_1^{(c)}$, $\mathcal{M}_R(\Pi(V^c(\pi))) \geq  \frac{1}{4e\hat{c}} \cdot v_{(2)}^c \cdot d_{min}^c$.
\end{claim}
\begin{proof}
Recall that in $\mathcal{M}_R$, 
the class $c$ is selected with probability $\frac{1}{\hat{c}}$, 
then the mechanism $\mathcal{M}_O$ runs the winner decision in $T_{c^*}$. 
The mechanism $\mathcal{M}_O$ consists of two phases: \textit{observation} phase 
	and \textit{decision} phase.
The probability that $\mathcal{M}_O$ enters into the decision process is 
		$$\left(1-\frac{1}{\lfloor \frac{n_{c}}{2} \rfloor +1}\right)^{\lfloor \frac{{n_{c}}}{2} \rfloor}\geq 1/e$$
In addition, 
 the proportion of arrival sequences (permutation of users) $\pi$ in $\Pi(V^c(\pi) )$ that 
  $v_{(2)}^c$ appears at the first half of buyers and $v_{(1)}^c$ appears at the second half is $1/4 $. 
  Therefore, at class $c$, with more than $\frac{1}{e} \cdot \frac{1}{4}$ probability, 
   the mechanism $\mathcal{M}_O$ selects the buyer with initial valuation $v_{(1)}^c$ as winner and charges  the user at  at least $v_{(2)}^c$
(when considering the discount values, it will charges  at more than $v_{(2)}^c \cdot d_{min}^c$).
This proves the Claim~\ref{Fact1}. 
\end{proof}

\begin{claim}
\label{Fact2}
For each arrival sequence $\pi \in \Pi_1^{(c)}$, the Vickrey mechanism $\mathcal{M}_V( \pi ) \leq v_{(2)}^c \cdot d_{max}^c$.
\end{claim}
\begin{proof}
Recall that the Vickrey mechanism $\mathcal{M}_V$ always selects the highest bidder and charges at the second-highest bid value.
Then for any arrival sequence $\pi\in \Pi_1^{(c)}$,  $r_{(2)}\leq v_{(2)}^c \cdot d_{max}^c$.
Then we have $\mathcal{M}_V(\pi) \leq v_{(2)}^c \cdot d_{max}^c$.
\end{proof}

\begin{lemma}
\label{lemma-bound-ratio-MR}
$\frac{E[\mathcal{M}_R]}{\mathcal{M}_V(\Pi_1)} \ge \frac{1}{4B \mathrm{e}\cdot \hat{c}}$.
\end{lemma}
\begin{proof}	
Notice that
\begin{equation}
\begin{aligned}
	\begin{cases}	
	E[\mathcal{M}_R]
	& = \sum_{\pi \in \Pi'}  M_R(\pi) / \| \Pi' \|\\
	&=\sum_{c=1}^{\hat{c}} E_c[\mathcal{M}_R]  \\
	& =\sum_{c=1}^{\hat{c}}\sum_{\pi \in \Pi^{(c)}}\mathcal{M}_R(\Pi(V^c(\pi))) \\
	& \ge \sum_{c=1}^{\hat{c}}\sum_{V^c\in V^c(\Pi_1^{(c)})}\mathcal{M}_R(\Pi(V^c))\\
	\mathcal{M}_V(\Pi_1) %If $\exists \pi', \pi \in \Pi_1^{(c)}$ such that $V^c(\pi)=V^c(\pi')$ 
	& = \sum_{c=1}^{\hat{c}}  \mathcal{M}_V(\Pi_1^{(c)})\\
	& = \sum_{c=1}^{\hat{c}}  \sum_{\pi \in \Pi_1^{(c)}} \mathcal{M}_V(\pi)\\
	\end{cases}
\end{aligned}
\end{equation}

By combining Claim~\ref{Fact1} and Claim~\ref{Fact2},   for each $\pi \in \Pi_1^{(c)}$, we have:
$$\mathcal{M}_R(\Pi(V^c(\pi))) \geq  \frac{1}{4 B \cdot e \cdot \hat{c}} \cdot \mathcal{M}_V(\pi)$$
	Therefore, $$\sum_{V^c\in V^c(\Pi_1^{(c)})}\mathcal{M}_R(\Pi(V^c)) \geq  \frac{1}{4B \cdot e \hat{c}} \sum_{\pi \in \Pi_1^{(c)}} \mathcal{M}_V(\pi)$$
This proves the Lemma~\ref{lemma-bound-ratio-MR}.
\end{proof}

Then combining the inequality~\eqref{equ:M_V}, we have
\begin{eqnarray}
\begin{aligned}	
 \frac{E[\mathcal{M}_R]}{\mathcal{M}_V (\Pi')} 
 &\geq \frac{ r_{(2)}(\Pi_1)}{4 B\cdot e\cdot \hat{c} (\eta \hat{c}+1+(k+3)\log n) \cdot {r_{(2)}(\Pi_1)}} \\
& \geq \frac{1}{\Theta(\eta  \log^2 n)}.
 \end{aligned}	
 \end{eqnarray}
	This finishes our proof of Lemma~\ref{lemma:MU-lossInClass}.
\end{proof}

By combining Lemma~\ref{lemma:MVPi} and Lemma~\ref{lemma:MU-lossInClass},
we have ${\mathcal{M}_R(\Pi)} \ge  {\mathcal{M}_R(\Pi')} \geq \frac{ \mathcal{M}_V (\Pi) }{\Theta(\eta \log^2 n)} $, \ie,
 mechanism $\mathcal{M}_R$ achieves the expected value at least $\frac{E[\mathcal{M}_V]}{\Theta(\eta \log^2 n)}$.
This  proves the following Theorem~\ref{M1:ratio}.

\begin{theorem}
	\label{M1:ratio}
Let $\eta = \max_{i,j} \frac{n_i}{n_j}$ for all discount classes in the reserved discount class set.
If $n_i \ge 2$ for all discount classes, $\frac{v_1}{v_2} \le n^k$ for some constant $k>0$,
 then the mechanism $\mathcal{M}_R$ is {$O(\eta \log^2 n)$}-competitive, \ie, $\frac{E[\mathcal{M}_V]}{E[\mathcal{M}_R]} \leq O(\eta \cdot \log^2 n)$.
\end{theorem}

\begin{theorem}
	\label{M1:ratio-at-least}
With a bounded $\eta$, $n_i \ge 2$ and $v_1 /v_2 \le n^k$ for some bounded value $k$,
there is some input of the non-increasing discount function and initial valuation set, such that the mechanism $\mathcal{M}_R$ will have a competitive ratio at least
 {$\Theta(\log^2 n)$}, \ie, $\frac{E[\mathcal{M}_V]}{E[\mathcal{M}_R]} \ge \Theta(\log^2 n)$.
\end{theorem}
\begin{proof}
%\textbf{An input example of $(d,V)$ for worse  competitive ratio of $M_R$}: 
We'll show that the above bound is tight by giving the following worst case. 
\begin{equation}
\small
\label{worst_case}
\begin{aligned}
(d,V)\defeq &
\begin{cases}
d(t):
\begin{cases}
d(t_0)=d_{max}=1\\
d(t)=2^{-(c-1)}, \quad t\in (t_{2*c-2},t_{2*c}] \\
d(t)=\epsilon, \quad t > 2 \log n
\end{cases}
\\
V=\{v_{1}=\mathcal{K}^4, v_{2}=\mathcal{K}, v_{3}=\cdots=v_{n}=0\}
\end{cases}\\
&\text{where $ 1\le c \le  \log n, j\geq 0, t_j=j/\lambda$ and $\mathcal{K}=n$.}
\end{aligned}
\end{equation}
Here $\epsilon \in(0, \log n)$ is a sufficiently small value.
As shown in this special input $(d,V)$,
 there are only two positive initial valuations and $T_c = (t_{2*c-2},t_{2*c}]$ which indicates there are expected two arrivals for each discount class $c$. 
Then the first and second-largest reported prices are always the corresponding report price of buyers whose initial valuations are $v_{1}$ or $v_{2}$.
Thus, $\mathcal{M}_R$ obtains revenue of $r_{(2)}>0$ when both $v_{1}$ and $v_{2}$ appear at the same class and $v_{2}$ arrives before $v_{1}$. 
Notice that for the mechanism $M_R$, there are $\hat{c}=\Theta(\log n)$ reserved discount classes.
The probability that both $v_1$ and $v_2$ appear in the selected star-class is thus $\Theta(1/\log^2 n)$.
For the mechanism $M_V$, the probability that $v_2$ appears at some class $c$ with $c \le \log n$
 is $\Theta(1/n)$ and $v_1$ appears at some class with discount value larger than $\epsilon$ is $\Theta(\log n /n)$. 
 The expected revenue produced by $M_V$ is $E(M_V)= \Theta(\frac{\log n}{n^2} v_2)$.
 The expected revenue produced by $M_R$ is $E(M_R)= \Theta(\frac{1}{n^2\log n} v_2)$.
Thus, we have $E(M_R)= E({M}_V)/{\Theta(\log^2 n)}$ for this special input.
\end{proof}

Thus, for the mechanism $M_R$, its competitive ratio is $\Theta(\log^2 n)$ under the assumptions
 required by Theorem~\ref{M1:ratio-at-least}.
 
  \begin{assumption}
 \label{assumption-n-2}
Assume that there is a constant $ B>1 $ such that $n_c\ge 2$ and $\frac{n_c}{n_1}\leq \eta$ for each $c\in [1,\hat{c}]$, where $\eta >1 $ is a constant.
\end{assumption}

Then we have the following Theorem:
\begin{theorem}
	\label{M_R:ratio1}
If Assumption~\ref{assumption-n-2} is true, the mechanism $\mathcal{M}_R$ is {$\Theta(\log n)$}-competitive if we always pick the class $c=1$ to do the Observe-then-Select algorithm $\mathcal{M}_O$. Such mechanism is also called $M_1$ for simplicity.
\end{theorem}
\begin{proof}
Previous proofs showed that the expected revenue generated by the Vickrey auction on all permutations $\Pi_1$ satisfying
$r_2(\Pi_1) \ge \Theta(\log n)(r_2(\Pi_0)+ r_2(\Pi_2)) $.
When assumption~\ref{assumption-n-2} is true, for each $v_y$ and $v_z$ appearing at some class $c>1$ and resulting in the largest-reported price and second-largest reported price respectively,
 we can swap $v_y$ and $v_z$ with some elements from the first discount-class.
 This will result new largest reported price and second-largest reported price respectively.
 As the number of users in each class $c$ is assumed to be within a factor $\eta$ of the value $n_1$ (the size of the first discount class),
 then the expected revenue by the Vickrey auction on $\Pi_1$ is at most
  $r_2(\Pi_1) \le \sum_{c=1}^{\hat{c}} v_z  B^{-c+1} \frac{n_c^2}{n^2}=\Theta(\eta^2 B v_z \frac{n_1^2}{n^2})
   \le \Theta(\eta^2 B v_2\frac{n_1^2}{n^2})$.
  On the other hand, the mechanism $M_O$ running on the first discount class will results in an 
   expected revenue at least  $\Theta( v_2\frac{n_1^2}{n^2})$.
  Thus, the mechanism that selects the first discount-class and runs the $M_O$ scheme on this first discount-class will produce
   an expected revenue at least  $\frac{1}{\Theta(\eta^2 B \log n)}$ of the expected revenue by the Vickrey auction.
  In other words, the competitive ratio of the mechanism $\mathcal{M}_1$ (running $M_O$ on the first first discount class) is 
   at most $O(\eta^2 B \log n)$.

Using the example from the Theorem~\ref{M1:ratio-at-least}, we can show that there is some input of the non-increasing discount function and initial valuation set, such that the mechanism $\mathcal{M}_1$ will have a competitive ratio at least $\Omega(\log n)$.

This finishes the proof of the theorem.
\end{proof}

%%%%%%%%%%%%%%%
\subsection{Discussion on Number of Buyers Per Class}
\label{subsec:num-buyers}

Recall that we assume: 1) there are at least $2$ buyers for each reserved class $c$, \ie, $n_c\geq 2$ for all $ c \le \hat{c}$;
2) each class has a comparative number of buyers, \ie, $\frac{n_{(1)}} {n_{c}} \le \eta$, for any class $c \le \hat{c}$, where $\eta >1$ is a constant.
Here $n_{(1)}$ is the class with the largest number of buyers.
%Here $n_{(1)}$ is the size of the class where the largest reported price $r_{(1)}$ appears.
We then analyze the performance of $M_R$ if we relax these conditions.

If we remove all the restrictions about the number of buyers per class, then we have the following lemmas:
\begin{lemma}
\label{claim-n}
If $n_1 \ge 2$, the mechanism $M_R$ could induce an $\Omega(n)$ competitive ratio in the worst case, if there is no restrictions on the number of users per discount class.
\end{lemma}
\begin{proof}
To prove this claim, we construct the following input instance:
 1) there are only two classes with non-zero number of buyers: $n_1=2$, $n_{l}=n-2$ for $l=k\log_B n$; 
 2) the discount value in the first class  ($c=1$) is $1/B$, and $d(t)=\frac{1}{n^k}$ in the class $l=k\log_B n$; 
  3) there are only two large valuations $v_{1}=n^k \cdot v_{2}+\delta$, $v_{2}=K$, and $v_{i}= \delta$ for all $3 \le i \le n$, where $\delta>0$ is a sufficiently small number.
In other words, the input is composed of the following.
\begin{equation}
\small
\label{worst_case-2}
\begin{aligned}
(d,V)\defeq &
\begin{cases}
d(t)=
\begin{cases}
B^{-1}, \quad t\in [0,t_{2}]\\
\frac{1}{n^k}, \quad t\in (t_{2},t_{n}]
\end{cases}
\\
V= \quad
\begin{cases}
v_{1}=n^k\cdot v_2+\delta\\
v_{2}=K\\
v_{i}=\delta, i \in [3,n]
\end{cases}
\end{cases}
\end{aligned}
\end{equation}
Here $t_j=j/\lambda$ denotes the arrival time of the $j$-th arrived buyer.
Note that $k\log_B n=\log_B \frac{v_{1}}{v_{2}} < \hat{c}$, thus the class ($c=k\log_B n$) is a reserved discount class.
As shown in this special input $(d,V)$, we have:
\small{
$$r_{(2)}(\Pi_{1})=\frac{2}{n}*\frac{1}{n-1}*v_2*\frac{1}{B}+ \frac{n-2}{n}*\frac{n-3}{n-1}*v_2*\frac{1}{n^k}$$
$$r_{(2)}(\Pi_0)= \frac{2}{n}*\frac{n-2}{n-1}*v_2*\frac{1}{B} + \frac{n-2}{n}*\frac{2}{n-1}*v_2*\frac{1}{n^k} $$
}
thus, $r_{(2)}(\Pi_{1})\le \Theta(\frac{1}{n}) r_{(2)}(\Pi_0)$, then the Claim~\ref{claim-n} is proved.
\end{proof}

Even if we select the first discount class as the star-class and then run the mechanism $M_O$, its competitive ratio could still be $\Omega(n)$.

\begin{lemma}
\label{claim-n-M1}
If $n_1 \ge 2$, the mechanism $M_1$ could induce an $\Omega(n)$ competitive ratio in the worst case, if there is no restrictions on the number of users per discount class.
\end{lemma}

If we maintain the first assumption $n_c \ge 2$ and further assume that any two adjacent classes have a comparative number of buyers, \ie, we have the following assumption:
 \begin{assumption}
 \label{assumption_class}
 $\frac{n_{c+1}}{n_{c}} \le \eta$ and $\frac{n_{c+1}}{n_{c}} \ge \frac{1}{\eta}$, for any class $1\le c \le \hat{c}-1$, where $\eta >1$ is a constant.
\end{assumption}

Then similar to the previous lemma, we still can construct some input instances such that the mechanism $M_R$ still has a competitive ratio at least 
$\Omega(n)$.
We prove the following lemma:
\begin{lemma}
 There is a case such that $r_{(2)}(\Pi_{1})\le \Theta(\frac{1}{n}) r_{(2)}(\Pi_0)$.
\end{lemma}
\begin{proof}
We construct a special instance composed of the following valuation set and discount function.
The number of users in class $c$ is 2 times of the number of users in class $c-1$.
\begin{equation}
\small
\begin{aligned}
& n_c=2^c, c\in[1, \log n -1] \\
& d(t)= B^{-c+1}, \quad t\in T_c\\
& V= \quad
\begin{cases}
v_{1}=n^2 \cdot K\\
v_{2}=K\\
v_{i}=v_{2}, i \in [3,n]
\end{cases}
\end{aligned}
\end{equation}
Note that in this case, 
1) we have $r_{(2)}=v_2$ since there is always a ($v_i=v_2$) appears at the first class, in which $d(t)$ is $1$.
2) $r_{(1)}$ is always produced by ($v_1=n*v_2$) since $v_1*d(t)$ with $t\in T_{\log_B n -1}$ is always larger than $v_2*d_{max}$.
\ie: $$n^2 *v_2*B^{-(\log_B n -1)+1}>v_2$$
We can show that $r_{(2)}(\Pi_{1})=\Theta(\frac{2}{n}*v_2)$, as the second largest reported price is always from the discount-class $c=1$.
On the other hand, we can show that $r_{(2)}(\Pi_0)= \frac{n-2}{n}*v_2$ as $v_1$ could be in any other class and still resulting in the largest reported price.
This finishes the proof of the claim.
\end{proof}

\begin{lemma}
%There is a case such that $r_{(2)}(\Pi_{1})\le \Theta(\frac{1}{n}) r_{(2)}(\Pi_0)$.
$M_R$ could induce $\Theta(n)$ competitive ratio even when Assumption~\ref{assumption_class} holds.
\end{lemma}
\begin{proof}
Let $B=2; n_1=2, \frac{n_{c+1}}{n_{c}} =\eta$. Then $\sum_c n_c \le n$, we have $c \le \log_{\eta} \frac{n*(\eta-1)}{2}+1$. %n_c=2*\eta^{c-1}; 
The total number of class is at most $\hat{c}=\log_{\eta} (\frac{(n-2)*(\eta-1)}{2*\eta}+1)+1$. $n_c=2*\eta^{c-1};  n_{\hat{c}}\approx\frac{\eta-1}{\eta}n$.
\begin{equation}
\small
\begin{aligned}
%& n_c=2*\eta^{c-1}, c\in[1, \log n -1] \\
& d(t)= 2^{-c}, \quad t\in T_c\\
& V= \quad
\begin{cases}
v_{1}=n^5*K\\
v_{2}=K\\
v_{i}=\delta, i \in [3,n],\\
\end{cases}\\
& \text{where $\delta \lll K$ is any small enough constant.} 
\end{aligned}
\end{equation}
When $v_1$ appears at class $c=\hat{c}$, his report price becomes $v_1*2^{-\hat{c}}=v_2*n^2*2^{-\hat{c}}$.
We further set $\eta=2^{\frac{1}{4}}$, then $2^{\hat{c}}=(\frac{\eta-1}{2})^2*n^2$. Thus, $v_1*2^{-\hat{c}}>v_2$ which means even $v_1$ appears at the last class, his discounted valuation is still larger than $v_2$'s no matter where $v_2$ is.
We have $r_{(2)}(\Pi_1)=\Theta(\frac{1}{n^2})$ and $r_{(2)}(\Pi_0)=\Theta(\frac{1}{n})$.
$M_R$ have $\Theta(n)$ competitive ratio on this case.
\end{proof}

\textbf{Observe-then-Decision mechanism}:
For the case shown in Equation~(\ref{worst_case-2}), it seems that there is a very simple and intuitive method that could easily obtain a constant competitive ratio
 by adopting the \emph{observe-then-decision} type of mechanisms~\cite{hajiaghayi2004adaptive}. 
That is, we first observe $1/2$  fraction number of buyers then we select the one from the second half  of buyers whose reported price is larger than or equal to the largest one among the first half of buyers, and charge at this price. 
Since the probability that $v_2$ appears at the first half of buyers and $v_1$ appears at the second half is $1/4$, in this case, $v_2$ has the second largest reported price while $v_1$ has the largest reported price.
Therefore, we can obtain at least $1/4$ fraction of optimal revenue.
This kind of mechanism is firstly given by Hajiaghayi et al.~\cite{hajiaghayi2004adaptive} without considering the discount factors.
That is, if we know there is a sufficiently large initial valuation who will induce the largest report price no matter where he appears, then we just run the 
$1/2$-observation-$1/2$-selection mechanism.

\begin{lemma}
\label{claim-no-observe-desision}
In the worst case, the \emph{observe-then-decision} mechanism has a competitive ratio at least $\Theta(n)$.
\end{lemma}
\begin{proof}
We construct an input example such that the $\frac{1}{2}$-observation-$\frac{1}{2}$-selection mechanism could have a bad enough competitive ratio.
\begin{equation}
\small
\label{worst_case-3}
\begin{aligned}
(d,V)\defeq &
\begin{cases}
d(t)=
\begin{cases}
1, \quad t\in [0,t_{2}]\\
\frac{1}{n^k}, \quad t\in (t_{2},t_{n}]
\end{cases}
\\
V= \quad
\begin{cases}
v_{1}=K \cdot n^k - \delta\\
v_{2}=K\\
v_{i}=\delta, i \in [3,n]
\end{cases}
\end{cases}
\end{aligned}
\end{equation}
then we have:
\small{
\[r_{(2)}(\Pi_{1})=\frac{2}{n}*\frac{1}{n-1}*v_2*1+ \frac{n-2}{n}*\frac{n-3}{n-1}*v_2*\frac{1}{n^k}\]
\[r_{(2)}(\Pi_0)= \frac{2}{n}*\frac{n-2}{n-1}*(v_2-\delta) + \frac{n-2}{n}*\frac{2}{n-1}*v_2*\frac{1}{n^k}\]
}
%\begin{equation}
%\begin{aligned}
%\begin{cases}
%& r_{(2)}(\Pi_{1})=\frac{2}{n}*\frac{1}{n-1}*v_2*\frac{1}{B}+ \frac{n-2}{n}*\frac{n-3}{n-1}*v_2*\frac{1}{n^k}\\
%& r_{(2)}(\Pi_0)= \frac{2}{n}*\frac{n-2}{n-1}*(v_2-\delta) + \frac{n-2}{n}*\frac{2}{n-1}*v_2*\frac{1}{n^k}
%\end{cases}
%\end{aligned}
%\end{equation}

Here, again we have $r_{(2)}(\Pi_{1})\le \Theta(\frac{1}{n}) r_{(2)}(\Pi_0)$.
For this special instance, the optimal revenue from the case that both $v_1$ and $v_2$ appear at class ($c=1$) is $v_2*\frac{1}{B}$
 (or $v_2*\frac{1}{n^k}$ when both appeared at the class $c=k\log_B n$).
When $v_1$ appears at class ($c=1$) and $v_2$ appears at class ($c=k\log_B n$), we have $r_{(2)}=v_2*\frac{1}{n^k}$, 
which is small enough to omit when compared with the optimal revenue ($r_{(2)}=v_2-\delta$) of the case 
that  $v_2$ appears at class ($c=1$) and $v_1$ appears at class ($c=k\log_B n$).

\begin{enumerate}
\item $r_{(2)}=v_2*1$: 
	when both $v_1$ and $v_2$ appear at class ($c=1$) with probability $P_1=\frac{2}{n}*\frac{1}{n-1}$;
\item $r_{(2)}=v_2*\frac{1}{n^k}$: 
	when both $v_1$ and $v_2$ appear at class (c=$k\log_B n$) with probability $P_2=\frac{n-2}{n}*\frac{n-3}{n-1}$;
\item $r_{(2)}=v_2-\delta$: 
	when $v_2$ appears at class ($c=1$) and $v_1$ appears at class ($c=k\log_B n$) with probability $P_3=\frac{2}{n}*\frac{n-2}{n-1}$;
\item $r_{(2)}=v_2*\frac{1}{n^k}$:
	when $v_1$ appears at class ($c=1$) and $v_2$ appears at class ($c=k\log_B n$) with probability $P_4=\frac{2}{n}*\frac{n-2}{n-1}$.
\end{enumerate}

From the above analysis, we can see that the dominant expected revenue is from the case 3).
Observe that in case 3) $r_{(1)}$ always arrives earlier than $r_{(2)}$, 
Notice that in this case, the largest reported price $r_{(1)}$ is $v_2$, produced by the  user appearing at the first class,
 while the second largest reported price $r_{(2)}$ is $v_2 -\delta$, produced by the user appearing at the class $k\log_B n$.
 For this case, for any observe-then-decision mechanism, it cannot find any element with a value larger than the maximum observed one.
 Thus, it will not accept any user. Thus, it gets revenue $0$ in this case.
 Consequently, the total expected revenue produced by this mechanism is at most $\Theta(\frac{v_2}{n^2})$, while the optimum revenue by the Vickrey auction mechanism is 
 $\Theta(\frac{v_2}{n})$.
 Thus the competitive ratio of such mechanisms is at least $\Theta(n)$.
\end{proof}

\textbf{Modified-Observe-then-Decision Mechanism}:
Notice that  we  can modify the observe-then-selection mechanism to accept the last user when all users in the decision phase has a reported price less than the largest reported price
 from the observation phase.
 Such mechanism is called \emph{Modified-Observe-then-Decision} auction mechanisms.
 Then, for the example constructed in previous proof, the expected revenue by such modified mechanism is at most $\Theta(\frac{v_2}{n^2})$ as the probability that 1) the user with the valuation $v_1$ appearing
  as the last user, and 2) the  user with the valuation $v_2$ appearing  at the first class is $2/n^2$, and the revenue in this case is $v_2 - \delta$.
 Thus the competitive ratio of such modified mechanisms is still at least $\Theta(n)$.
Thus there is no observation-then-selection mechanism that can achieve an expected revenue at least $\Theta(\frac{1}{n})$  of the Vickrey auction mechanism for any input instance $(d,V)$.
Then we have the following lemma.

\begin{lemma}
\label{claim-no-modified-observe-desision}
In the worst case, the \emph{modified-observe-then-decision} mechanism has a competitive ratio at least $\Theta(n)$.
\end{lemma}
 
 \textbf{Instant-Decision-and-Pay-at-End Mechanism}:
 We then analyze the other set of mechanisms.
For each of the arrived users, we receive its reported price; then we must decide immediately on whether to accept this user as a winner or not.
However, the payment by this user can be decided at the end of game, \ie, after all users have arrived.
We call this kind of mechanisms as \emph{Instant-Decision-and-Pay-at-End}.
Then a simple application of the mechanism~\cite{babaioff2009secretary} is as follows:
1) uses the mechanism from~\cite{babaioff2009secretary} to find the user with the maximum reported price with some probability, and then accepts this user as the winner;
2) at the end, among all users, we compute the second-largest reported price $r_{(2)}$, and charge the winner a payment $r_{(2)}$.
Then it is easy to show that this mechanism is truthful and has a good competitive ratio.

\begin{lemma}
\label{lemma:decision-payment-at-end}
For arbitrary \emph{known} discount function $d(t)$, and a set of arbitrary initial valuations $\textbf V$,
and the aforementioned \emph{Instant-Decision-and-Pay-at-End} auction mechanism has a competitive ratio $\Theta(\log n)$~\cite{babaioff2009secretary}. 
\end{lemma}

\textbf{The first non-empty class has only one user}:
In all previous analysis, we assumed that $n_c \ge 2$ for all discount classes, especially $n_1 \ge 2$.
 We then show that if this condition does not hold, there is some special input such that 
 no truthful auction mechanism has a bounded competitive ratio on the expected revenue.
 We construct a set of special input cases as follows.
 The first discount class has only 1 users and the discount value is $1$.
 All the other $n-1$ users appear at the other class with a discount value $1/n^k$.
 The valuations are $v_1 = n^k v_2 -\delta$, and  $v_i=\delta$ for $i \ge 3$.
 
 \begin{lemma}
\label{lemma:decision-and-payment}
For arbitrary \emph{known} non-increasing discount function $d(t)$, and a set of arbitrary initial valuations $\textbf V$,
 there is a  set of special input instances such that all the  \emph{Observe-and-Decision} auction mechanisms have an unbounded competitive ratio. 
 \end{lemma}
 \begin{proof}
 We prove this as follows.
 Using the aforementioned example, we show that any  \emph{Observe-and-Decision} auction mechanism cannot make a good choice.
 Assume that a given  \emph{Observe-and-Decision} auction mechanism decides to accept a$i$-th arrived user with a probability $p_i$, based on all historically observed users' reported prices.
Recall that for any  \emph{Observe-and-Decision} auction mechanisms, it must see at least two users to make a decision.
Otherwise, by simply accepting the first user, the mechanism cannot assure the truthfulness of the first arrived user.

 As there is only one user in the class 1, the probability that $v_1$ arrives in this class is $1/n$.
 If $v_1$ arrives in this class 1,
 then each  \emph{Observe-and-Decision} mechanism cannot accept any users as all other users will have reported prices less than $v_1$.
Similarly if $v_2$ arrives in the class 1, such  mechanism cannot accept any users.
The only case an \emph{Observe-and-Decision} mechanism can accept a user is some user $v_i$ ($i \ge 3$) arrives at the class 1.
In this case, the revenue is $\delta$. Thus, the expected revenue by any \emph{Observe-and-Decision} auction mechanism is at most $\delta$.
Notice that the expected revenue by the Vickrey auction is $\Theta(v_2/n)$.
Thus the competitive ratio by any \emph{Observe-and-Decision} is  at least $\frac{v_2}{n \delta}$, which could be arbitrarily large as $\delta$ could be sufficiently close to $0$. 
 \end{proof}
 
 Similar to the proofs of previous lemmas, we can prove the following lemma.
 \begin{lemma}
\label{lemma:decision-and-payment-modified}
For arbitrary \emph{known} non-increasing discount function $d(t)$, and a set of arbitrary initial valuations $\textbf V$,
 there is a  set of special input instances such that all the  \emph{Modified-Observe-and-Decision} auction mechanisms have a competitive ratio at least $\Omega(n)$. 
 \end{lemma}

 We then show that there is a simple mechanism achieving an expected revenue at least $\Theta(1/n^2)$ of the expected revenue by the Vickrey auction.
 We only need  Assumption~\ref{assumption-n-1} to hold, \ie, there is a constant $ B>1 $ such that $n_1\ge2$.
The mechanism called \textbf{$M_1$} is similar to $M_R$:
 it selects the first class $c=1$ as the star-class  $c^*$, then runs the Observe-then-Select algorithm $M_O$ on this class.

\begin{algorithm}[t!]
	\LinesNumbered
	\small
	\caption{Fixed-Select Mechanism $\mathcal{M}_1$}
	\label{alg:M_1}   		
	\KwIn{report price vector $\mathbf{r}$; discount function $d(t)$; time horizon $T$;  arrival rate $\lambda$.} 
	\KwOut{winner determination $\mathbf{y}$; payment $\mathbf{p}$}
	$c^* \gets 1$\;
	Calculate $T_{c^*}$ according to Equation (\ref{discount_class}) and (\ref{time_interval}) \;
	${n}_{c^*}=\lambda*|T_{c^*}|$\; 
	$winner, payment \gets \mathcal{M}_O( \mathbf{r}, {n}_{c^*})$\;
	Return $winner$, $payment$\;  
\end{algorithm}

\begin{theorem}
	\label{M_1:ratio1}
If $n_1 \ge 2$, then the mechanism $\mathcal{M}_1$ is {$\Theta(n^2)$}-competitive.
\end{theorem}
\begin{proof}
Firstly, we have $E[M_V]=E[r_{(2)}] \le v_2 $.
Secondly, $M_1$ is able to obtain $v_2d(t_2)$ revenue when $v_2$ is the first arrival and $v_1$ is the second arrival.
That is $E[M_1] \ge \frac{1}{n(n-1)}\cdot v_2 \cdot \frac{1}{B}$. 
Therefore $\frac{E[M_1]}{E[M_V]} \ge 1/\Theta (n^2)$.

$M_1$ performs worst when there exists a class $c'$ whose number of arrivals is $\Theta (n)$ times of $n_1$ 
	and class $c'$'s minimum discount value is constant times of class $c=1$.
Here is an example, as show in Equation (\ref{equation:worst_case_M_1}).
There are only two classes, $n_1=2$ and $n_2=n-2$; 
	$d(t)=1$ when $t \in T_1$ and $d(t)=\frac{1}{B}$ while $t \in T_2$. 
	There are two positive valuations, $v_1=BK+1, v_2=K$ where $K$ is any positive integer.
\begin{equation}
	\small
	\label{equation:worst_case_M_1}
        \begin{aligned}
            (d,V)\defeq &
                    \begin{cases}
                        d(t)=
                            \begin{cases}
                                	1, 			\quad t\in [0,t_{2}]\\
                                	\frac{1}{B}, 	\quad t\in (t_{2},t_{n}]
                            \end{cases}\\
                         V= \quad
                             \begin{cases}
                                    v_{1}=BK+1	\\
                                    v_{2}=K		\\
                                    v_{i}=0, i \in [3,n]
                    		\end{cases}
        	  	     \end{cases}
	\end{aligned}
\end{equation}

In this case, 
	$E[M_1] \le \frac{1}{n^2} \cdot v_2$ and 
	$E[M_V] \ge \frac{v_2}{B}$, 
	thus $M_1$ is $\Theta(n^2)$ competitive. 
\end{proof}

 Even we apply the modified observe-then-select mechanism (see Alg.~\ref{alg:observe-select-1-class}) ,
  it still has a bad competitive ratio.
  
 \begin{algorithm}[thbp]
 \label{alg:observe-select-1-class}
	\LinesNumbered
	\small
	\caption{Modified Obs-Sel-in-Class-1 Mechanism $MOD_{1}$}
	\KwIn{report price vector $\mathbf{r}$; discount function $d(t)$; time horizon $T$;  arrival rate $\lambda$.} 
	\KwOut{winner determination $\mathbf{y}$; payment $\mathbf{p}$}
	
	For all users in the first class, applies the traditional \emph{Modified-Observe-and-Decision} auction mechanism to decide the winner and the payment of this winner.
\end{algorithm}

 \begin{lemma}
\label{lemma:decision-and-payment-not-too-bad}
For arbitrary \emph{known} non-increasing discount function $d(t)$, and a set of arbitrary initial valuations $\textbf V$,
\emph{Modified-Observe-and-Decision} auction mechanism $MOD_1$  (see Alg.~\ref{alg:observe-select-1-class})  has a competitive ratio $O(n^2)$.
 \end{lemma}
\begin{proof}
We prove this lemma as follows.
Let $l \ge 1$ be the first discount class that is not empty of users. For simplicity of proof, we assume that $l =1$.
Then with probability at least $\Theta(1/n^2)$, the two users with the largest two initial valuations $v_1$ and $v_2$ will appear in this class.
Then with a probability at least $1/4$, $v_2$ appears in the observation phase and  $v_1$ appears in the decision phase.
Then \emph{Modified-Observe-and-Decision} auction mechanism $MOD_1$ will select $v_1$ as the winner and charges a payment of order $\Theta(v_2)$ with probability of
 $\Omega(1/n^2)$.
 Notice that the maximum revenue by the Vickrey auction is at most $v_2$.
 Thus, mechanism $MOD_1$ achieves a competitive ratio at most $O(n^2)$.
\end{proof}

%%%%%%%%%%%%%%%
\subsection{Weighted-Selection Mechanism}
\label{alg_unknown_weighted}

The design of online approximate algorithm, due to its worst-case nature, can be quite pessimistic when the input instance at hand is far from worst-case. 
Therefore, we further propose an heuristic mechanism, named the ``Weighted-Select Mechanism" $\mathcal{M}_W$.
The mechanism $\mathcal{M}_W$ is almost identical to $\mathcal{M}_R$, however,
instead of selecting the {Star-Class} $c^*$ from the reserved classes randomly and uniformly (line $1$ in Algorithm~\ref{alg:M_R}),
$\mathcal{M}_W$ selects each class $c \in \mathbb{C}$ with different weights.

\textbf{Design of weights:}
intuitively, each class should be chosen by the proportion of its contribution to the expected second-largest reported value, \ie,  $E[r_{(2)}]$,
  the offline baseline Vickrey mechanism $\mathcal{M}_V$'s revenue. 
Formally, each class $c$'s contribution ratio is $\frac{E_c[r_{(2)}]}{\sum_c E_c[r_{(2)}]}$.
Recall that $$E_c[r_{(2)}]=\sum_{t_k \in T_c}\left(d(t_k) * \sum_j \left(v_{j} * P_{jk} \right) \right).$$
Here $v_j$ is the $j$-th largest initial valuation, and $P_{jk}$ is the probability that the user $v_j$ arrives at the $k$-th time-slot $t_k$.
However, without known the buyers' initial valuations set, we can not calculate  exactly the value of each $r_{(2)}^c$.
Our selection of weights are designed based on the following two insights:
\begin{enumerate}
	\item The class with larger number of arrived buyers should be chosen with larger probability.	
	\item The class with higher discount values should be chosen with larger probability.
\end{enumerate}

\textbf{Weighted Selection Mechanism $\mathcal{M}_W$}:
\begin{compactenum}
\item
For a given discount function $d(t)$ and the arrival rate $\lambda$, the selection weight for each class $c \in \mathbb{C}$ is calculated
  by $w_c=\sum_{t_j \in T_c} d(t_j)$. 
Notice that as the exact number of users to-arrive is unknown in advance, we will compute the value $w_c$ using the expected number of users to arrive $n_c$
 and the expected time-slot of the arrived time from the Poisson arrival process.
Here $w_c$ can be approximated as $\frac{n_c}{B^c}$ as the discount values $d(t_j)$ for all users in one discount class is in the range $[B^{-c}, B^{-c+1} ]$.
\item
$\mathcal{M}_W$  chooses each each discount class $c \in \mathbb{C}$ as the {Star-Class} with probability $\beta_c = \frac{w_c}{\sum_{c=1}^{\hat{c}}{w_c}}$.
 \item
 Run the mechanism $M_O$ on the selected discount class.
\end{compactenum}

\textbf{Analysis of $\mathcal{M}_W$:}
$\mathcal{M}_W$ is also a bid independent auction. 
Thus, it is value-truthful.
However, each class $c \in \mathbb{C}$ is chosen as the {Star-Class} $c^*$ with different probabilities which may result in a buyer (whose true arrival class is $c_p$) delaying his arrival to a class $c_p'$ that has a larger selection weight than $c_p$.
But even this event happens, the delayed buyer needs to compete with more number of buyers due to the fact that $n_{c_p}>n_{c_p'}$.
Thus, delaying a buyer's arrival to a class with a larger selection weight cannot indeed increase his winning probability.
Therefore, we have $\mathcal{M}_W$ is truthful.

Our experimental results demonstrate that the mechanism $\mathcal{M}_W$ performs much better than $\mathcal{M}_R$.
We later will show that a modified weighted selection mechanism can achieve a good competitive ratio when we relax the restrictions on the number of users $n_c$ for each discount-class.
Essentially, the mechanism to be presented in subsection~\ref{modified-weight} will select the discount class with the largest $n_c/B^c$ for all $c \le \hat{c}$.

%%%%%%%%%%%%%%%
\subsection{Modified Weighted Selection Mechanism for Non-increasing Discount Function}
\label{subsec:select-first}
\label{modified-weight}

Recall that in Section~\ref{sec:bound}, we have proved that no strategy-proof  mechanism (satisfying IR \& IC)
 can obtain a bounded competitive ratio without any restrictions on $d(t)$, see Theorem~\ref{theo:known-dt-truthful} for details.
Especially, there is a worst case (see Eq.~\eqref{worstcase-n1}) such that:
1) there is no a constant $B>1 $ such that $n_1\ge2$ according to our \textbf{Discounts Division Algorithm};
2) extremely small discount values for rest of the $n-1$ buyers.
In the previous subsection~\ref{alg_unknown_random}, we showed that a random selection mechanism $M_R$ can achieve a bounded competitive ratio
 $\Theta(\log ^2 n)$ with respect to the Vickrey mechanism.
 However, this is achieved under several additional assumptions 1) $n_i \ge 2$ for each discount class, 2) $\frac{n_i}{n_j} \le \eta$ for any discount class $i$ and $j$ in the 
  reserved discount classes, and  $v_1 / v_2 \le n^k$.
The first two conditions restrict the discount function $d(t)$.
Here, we try to relax this restrictions on $d(t)$ and explore possible competitive mechanisms. 
We also assume $d_{\max}=1$ for simplicity.

Observe that for the example constructed in the proof of Theorem~\ref{M_1:ratio1},
 it is better to choose the class $c=2$.
Based on this observation, we propose the following mechanism, and we can prove that it achieves a competitive ratio $\Theta(n \log n)$ under simple assumption.

\textbf{Most-Weighted Selection Mechanism $M'_W$:}
\begin{compactenum}
\item
Calculate $w_c= \frac{n_c} { B^{c}}$ for each class $c$; 
\item
Select the discount-class with the largest $w_c$ as the winner class and denote it as $c^*$, then run $M_O$ on this class.
\end{compactenum}

%If $\exists c^*$ such that $\frac{n_{c^*}^2}{B^{-c^*}}/ \frac{n\cdot n_c}{B^{-c}}$ equals $\Theta(1)$ for any $c$; then select class $c^*$ as the winner class. 
%Otherwise we select each class with equal probability and denote the selected class as $c^*$, then run $M_O$ on this class.

%\input{alg-xue-proof-0.tex}
\begin{theorem}
\label{M_W:new}
If $n_c \ge 2$ for each discount-class, then the mechanism $M‘_W$ is {$\Theta(n \log n)$}-competitive.
\end{theorem}
\begin{proof}
We use $\hat{c}$ to denote the number of all  possible discount-classes.
Here, any class has a positive chance to be selected.
Let 	$\Pi_{c,i}$ denote the reported price sequence set that $v_i$ appears at class $c$ and it produces the global second largest reported price; 
 $\Pi_{c}=\bigcup_i \Pi_{c,i} $ denote the reported price sequence set that the global second largest reported price appears at class $c$; 
	$r_{(2)}(\Pi_{c,i})=\sum_{\pi\in\Pi_{c,i}} r_{(2)}(\pi)$ denotes the total value of $r_{(2)}$ obtained from $\Pi_{c,i}$.
Here $r_{(2)}(\pi)$ denotes the second largest reported price in a reported-price sequence defined by the mapping $\pi$.
Let $c^{(k)} (\pi)$ represent the class id of $r_{(k)}$ in sequence $\pi$.

Let $E_c[r_{(2)}]= \frac{1}{n!}r_{(2)}(\Pi_{c}) =\frac{1}{n!}\sum_i r_{(2)}(\Pi_{c,i})$, 
	which denotes the expected value of $r_{(2)}$ obtained from class $c$. 
Note that $r_{(2)}(\Pi_{c,i})=0$ if $\Pi_{c,i}=\emptyset$. %We only consider $\Pi_{c,i}\not=\emptyset$ in the following part.
Furthermore, we use $E_c[M_O]=\frac{1}{n!} \sum_{\pi\in \Pi} M_O(\pi_c(\pi))$ to denote the Observe-then-Select mechanism $M_O$'s achieved revenue
 from class $c$; and $\pi_c(\pi)$ denotes the subset price sequence appeared in class $c$ when the whole price sequence is $\pi$. 
 $M_O( \pi_c(\pi) )$ is the revenue when run $M_O$ on subset price sequence $\pi_c(\pi)$.

For each $\pi \in \Pi$, let $t_q$ denote the largest reported price $r_{(1)}$'s corresponding buyer's arrival time, and his initial valuation
 $v_{\pi(q)}$ is the $y$-th largest one $v_y$ in ${\textbf V}=\{v_1, v_2, \dots, v_n\}$.
Similarly, $t_p$ denotes the $r_{(2)}$'s corresponding buyer's arrival time, and his initial valuation $v_{\pi(p)}$ is the $z$-th largest one $v_z$
  among all initial valuations.
Formally, we have:
		\begin{equation*}
		\begin{aligned}
		\begin{cases}
		r_{(1)}=v_{\pi(q)} \cdot d(t_q)=v_y \cdot d(t_q)\\
		r_{(2)}=v_{\pi(p)}\cdot d(t_p)=v_z \cdot d(t_q)
		\end{cases}
		\end{aligned}
		\end{equation*}
For each $\pi$, let $c_1=c^{(1)} (\pi)$ and $c_2=c^{(2)} (\pi)$ for simplicity.
In other words, $r_{(1)}$ appeared in class $c_1$, 
 and $r_{(2)}$ appeared in class $c_2$.
 We then define two sets of arrival sequences (permutations of valuations) as follows
 \begin{compactitem}
\item
Let  $\Pi_{c_2,c_1,y, z}$ denote the reported price sequence set that 
$v_y$ appears at class $c_1$ and it produces the largest reported price and
$v_z$ appears at class $c_2$ and it produces the second largest reported price.
\item
Let $\Pi_{c_2,z}= \bigcup_{c_1,y} \Pi_{c_2,c_1, y,z}$ denote the price sequence set that 
$v_z$ appears at class $c_2$ and it produces the global second largest reported price.
\end{compactitem}
Here the class id $c_1$ could be equal to, less than or larger than $c_2$. 
Thus, for each $v_z$, the number of permutations in $\Pi_{c_2,z}$ is at most $\|\Pi_{c_2,z}\| \le n \cdot n_{c_2} $.

We focus on studying the class $c$ that $E_c[r_{(2)}]>0$. \ie, the reported price sequence set $\Pi_{c}$ 
	%	(that the second largest reported price $r_{(2)}$ appears at class $c$) 
	is not empty. %, \ie, $\Pi_{c} \not=\emptyset$.
Before our detailed proof, we firstly introduce Claim~\ref{claim:MO_c*}, ~\ref{claim:c*} and ~\ref{claim:E_1}.

\begin{claim}
\label{claim:MO_c*}
For any class $c$, $E_{c^*}[M_O] \ge  \frac{1}{4Bn} E_c[r_{(2)}]$.
\end{claim}
\begin{proof}
For each arrival sequence $\pi \in \Pi_{c_2,c_1,y, z}$, %there is a $\pi'$ such that  $M_O( \pi_{c^*}(\pi') ) \ge B^{-c^*} \cdot r_{(2)}(\pi)$. 
we can always find a price sequence set $\Pi_{c_2,c_1,y, z}'(\pi)$ by switching $\max\{ v_z, v_y\}$ with some one who appears at the second half of class $c^*$; and switching  $\min\{ v_z, v_y\}$ with some one who appears at the first half of class $c^*$.
Let the resulted permutation be $\pi'$.
For each $\pi' \in \Pi_{c_2,c_1,y,z}'(\pi)$, $M_O$ will always select some one $>\min\{ v_z, v_y\}$ and get a revenue $\ge \min\{ v_z, v_y\} \cdot B^{-c^*}$ on $\pi_{c^*}(\pi')$.
There are two complementary cases here:
\begin{enumerate}
\item $v_z < v_y$: in this case, we have:
	$$r_{(2)}(\pi) 		\le v_z \cdot B^{-c_2+1};  \quad  \text{and }
	M_O( \pi_{c^*}(\pi') ) \ge v_z \cdot B^{-c^*}$$
%	\begin{equation*}
%		\begin{aligned}
%			\begin{cases}
%			 r_{(2)}(\pi) \le v_z \cdot B^{-c_2+1}\\
%			M_O( \pi_{c^*}(\pi') ) \ge v_z \cdot B^{-c^*}
%			\end{cases}		
%		\end{aligned}
%	\end{equation*}
%	 $c_1$ could be equal to, less than or larger than $c_2$. Thus, for each $v_z$, $\|\Pi_{c_2,z}\| \le n*n_{c_2} $.
	 \begin{equation*}
		 \begin{cases}
			M_O( \Pi_{c_2, z}^* ) \ge \frac{n_{c^*}^2}{4}\cdot v_z \cdot B^{-c^*}\\
			r_{(2)}(\Pi_{c_2,z}) \le n \cdot n_{c_2} \cdot v_z \cdot B^{-c_2+1} \\
		\end{cases}
	\end{equation*}

\item $v_z \ge v_y$: in this case, we have:
	$$r_{(2)}(\pi) 		\le r_{(1)} \le v_y \cdot B^{-c_1+1}; 
	M_O( \pi_{c^*}(\pi') ) \ge v_y \cdot B^{-c^*} $$
	Here it must be the case that $c_1 \le c_2$.
	 Then $\|\Pi_{c_2,c_1,y,z}\| \le n_{c_1} \cdot n_{c_2} $.
	Let $\Pi_{c_2,c_1, y, z}^*=\bigcup_{\pi \in \Pi_{c_2,c_1,y,z}} \Pi_{c_2,c_1,y,z}'(\pi)$ denote all the distinct mappings produced by swapping $v_y$ with some element in the first half of $c^*$-discount class, and swapping $v_z$ with some element in the second half of $c^*$-discount class.
	Then, we have $ \|\Pi_{c_2,c_1, y,z}^*\| \ge {\frac{n_{c^*}^2}{4}}$.
%Notice that actually, when swapping elements $v_y$ and $v_z$ with some elements in the discount class $c^*$,
%	 there are also $n_{c_1}$ and  $n_{c_2}$ positions to put the swapped-out elements from the discount class $c^*$.
%	 Thus, $ \|\Pi_{c_2,c_1, y,z}^*\| \ge {\frac{n_{c^*}^2}{4}} n_{c_1} n_{c_2}$.
	\begin{equation*}
		\begin{cases}
			M_O(\Pi_{c_2,c_1, y, z}^* ) \ge \frac{n_{c^*}^2}{4}\cdot v_y \cdot B^{-c^*}\\
			r_{(2)}(\Pi_{c_2,c_1,y, z}) \le  n_{c_1}\cdot n_{c_2} \cdot v_y \cdot B^{-c_1+1}\\
		\end{cases}	
	\end{equation*}
%	\begin{equation*}
%		\begin{aligned}
%			\begin{cases}
%			r_{(2)}(\pi) \le r_{(1)} \le v_y \cdot B^{-c_1+1} \\
%			 M_O( \pi_{c^*}(\pi') ) \ge v_y \cdot B^{-c^*}
%			\end{cases}		
%		\end{aligned}
%	\end{equation*}
\end{enumerate}

%Thus, we have: 
%	\begin{equation*}
%		\begin{aligned}
%			&
%			\begin{cases}
%			M_O( \Pi_{c_2, z}^* ) \ge \frac{(n_{c^*})^2}{4}\cdot v_z \cdot B^{-c^*}\\
%			r_{(2)}(\Pi_{c_2,z}) \le n \cdot n_{c_2} \cdot v_z \cdot B^{-c_2+1} \\
%			\end{cases}
%			\text{for case 1)}\\
%			&
%			\begin{cases}
%			M_O(\Pi_{c_2,c_1, z}^* ) \ge \frac{(n_{c^*})^2}{4}\cdot v_y \cdot B^{-c^*}\\
%			r_{(2)}(\Pi_{c_2,c_1,z}) \le  n_{c_1}\cdot n_{c_2} \cdot v_y \cdot B^{-c_1+1}\\
%			\end{cases}	
%			\text{for case 2)}		
%		\end{aligned}
%	\end{equation*}

We then analyze the competitive ratio for these two cases separately.

For case 1),  we have:
	\begin{equation}\small
	\label{equ_MO_r2-case1}
			 \frac{M_O( \Pi_{c_2, z}^* ) }{r_{(2)}(\Pi_{c_2,z})}  
			  \ge \frac{1}{4B \cdot n}\frac{n_{c^*}^2}{B^{c^*}}/ \frac{n_{c_2}}{B^{c_2}} \ge \frac{1}{4Bn} 
	\end{equation}
	The inequality comes from the fact that $\forall c, \frac{n_{c^*}}{B^{c^*}} \ge \frac{n_c}{B^c}$;
	$r_{(2)}(\Pi_{c_2,z})=\sum_{c_1, y}  r_{(2)}(\Pi_{c_2,c_1, y, z})$ and 
	$M_O( \Pi_{c_2, z}^* ) = M_O \left(\bigcup_{c_1, y} \Pi_{c_2,c_1, y, z}^* \right)$.

In the following, we focus on discussing the case 2). 
For case 2), let $\mathbb C_1$ denote the set of class ids that $r_{(1)}$ could appear. 
	That is, $\forall c_1 \in \mathbb C_1, \Pi_{c_2,c_1, y,z}\not=\emptyset$. Let $c_1^*=\max\{c_1|c_1\in \mathbb C_1\}$.
Then, we have:
	\begin{equation}\small
	\label{equ_MO_r2-case2}
			 \frac{M_O( \Pi_{c_2, z}^* )}{r_{(2)}(\Pi_{c_2,z})} 
%			 \ge\frac{M_O \left(\bigcup_{c_1} \Pi_{c_2,c_1, z}^* \right)}{\sum_{c_1}  r_{(2)}(\Pi_{c_2,c_1, z})}{}
			 \ge \frac{ n_{c^*} \cdot \frac{n_{c^*}}{B^{c^*}}}{4B\cdot n_{c_2}\cdot \sum_{c_1 \in \mathbb C_1} \frac{n_{c_1} }{B^{c_1}}}
%			  \frac{M_O(\Pi_{c_2,c_1, z}^*) }{r_{(2)}(\Pi_{c_2,c_1,z})}  
%			  \ge \frac{1}{4B  \cdot n_{c_2}}\frac{n_{c^*}^2}{B^{c^*}}/ \frac{n_{c_1}}{B^{c_1}} \ge \frac{n_{c^*}}{4Bn_{c_2}} 
	\end{equation}
	
	Since $\forall c_1 \in \mathbb C_1, \frac{n_{c1}}{B^{c1}}\le \frac{n_{c^*}}{B^{c^*}} $, 
	then $n_{c_1} \le B^{c_1}\cdot \frac{n_{c^*}}{B^{c^*}} \le B^{c_1^*}\cdot \frac{n_{c^*}}{B^{c^*}} $. 
	Therefore, $\frac{n_{c_1} }{B^{c_1}} \le \frac{B^{c_1^*}}{B^{c_1}}\cdot \frac{n_{c^*}}{B^{c^*}} $.
	Then for case 2), we have:
	$$\small{ \frac{r_{(2)}(\Pi_{c_2,z})}{M_O( \Pi_{c_2, z}^* )} 
	\le  \frac{4B\cdot n_{c_2}\cdot \sum_{c_1} \frac{n_{c_1} }{B^{c_1^*}}}{ n_{c^*} \cdot \frac{n_{c^*}}{B^{c^*}}}
	\le \frac{4B\cdot n_{c_2}\cdot \sum_{c_1} \frac{B^{c_1} }{B^{c_1^*}}}{ n_{c^*}}
	%{4B\cdot n_{c_2} \cdot \hat{c} \cdot \max\{ \frac{n_{c_1} }{B^{c_1}} \} } 
	}$$
	$\frac{n_{c2}}{n_{c^*}}\le n$ and $\sum_{c_1} \frac{B^{c_1} }{B^{c_1^*}} \le 1+\frac{1}{B-1} $.
	Finally, we have 
	$$\frac{M_O( \Pi_{c_2, z}^* )}{r_{(2)}(\Pi_{c_2,z})} \ge \frac{1}{4B\Theta(n)}$$
This finishes the proof of Claim~\ref{claim:MO_c*}.
	\end{proof}

%\xue{there is a bug in the previous proof. it seems that the fraction is opposite.}

%\begin{claim}
%	\label{claim:r_(2)}
%	$E[r_{(2)}]$ gains at least $1/2$ fraction of the total value from the top $\Theta(\log_B n)$ classes.
%\end{claim}
\begin{claim}
	\label{claim:c*}
The selected star-class $c^*$ satisfying $c^* \le \log_B \frac{n}{2}$.
\end{claim}
\begin{proof}
	Since $n_1 \ge 2$ and $\forall c, n_c \le n$, then $\forall c > \log_B \frac{n}{2}$,
		we have $\frac{n_c}{B^c} < \frac{n}{B^{\log_B \frac{n}{2}+1}}$.
		Therefore we have $\frac{n_c}{B^c} < \frac{n_1}{B^1}$ for any $c > \log_B \frac{n}{2}$. 
	Note that $\frac{n_{c^*}}{B^{c^*}} \ge\frac{n_c}{B^c} $ for any $c \not=c^*$, 
		thus $M_W$ will never select some class $c> \log_B \frac{n}{2}$ as $c^*$.
\end{proof}

\begin{claim}
	\label{claim:E_1}
	$E_1[r_{(2)}] \ge \frac{2}{B}\frac{v_2}{n^2}$.
\end{claim}
\begin{proof}
	Note that we assumed that $n_1\ge2$.
	When both $v_1$ and $v_2$ appear at class $c=1$, then the global second largest report price $r_{(2)}$ is at class $c=1$; and $r_{(2)} > v_2\cdot B^{-1}$ (we assume $d_{max}=1$ for simplicity).
	Therefore, the expected second largest report price $r_{(2)}$ obtained from class $c=1$ is
	$E_1[r_{(2)}] =r_{(2)}(\Pi_{1}) \ge \frac{n_1(n_1-1)}{n^2}\cdot v_2\cdot B^{-1} \ge \frac{2}{B}\frac{v_2}{n^2}$.
\end{proof}

Hereafter, we will prove that the mechanism $M_W$ is $\Theta(n\log n)$ competitive separately for the following two complementary cases:
\begin{compactenum}
\item for any class $c$, $\max\{ r_{(2)}(\pi) | \pi \in \Pi_{c} \} \le v_2 \cdot B^{-c}$.
		In other words, no matter where $v_1$ is, it always produces the largest reported price.
\item $\exists c$, $\max\{ r_{(2)}(\pi) | \pi \in \Pi_{c} \} > v_2 \cdot B^{-c}$.
		In other words, there is some class that $v_1$ appears at, it produces the second largest reported price.
\end{compactenum}

	\textbf{Case 1):}
	for any $c > 2\log_B n$ and $\Pi_{c} \not=\emptyset$, we have
		$$E_c[r_{(2)}]=\frac{1}{n!} \cdot r_{(2)}(\Pi_{c}) \le \frac{n_{c}}{n}\cdot \frac{v_2}{B^{c}} < \frac{n_{c}}{n}\frac{v_2}{n^2B}$$
	Then the total value of expected $r_{(2)}$ from classes whose id is larger than $ 2\log_B n$ is:
		$\sum_{c > 2\log_B n} E_c[r_{(2)}] < \frac{v_2}{n^2B}$. 
	Note that $E_1[r_{(2)}] > \frac{v_2}{n^2B}$.
	Thus, the total expected $r_{(2)}$ from classes $ c > 2\log_B n$ 
		is less than the expected $r_{(2)}$ obtained from the first class. 
		\ie, $ \sum_{c > 2\log_B n} E_c[r_{(2)}] < E_1[r_{(2)}] $.
	Thus $$E[r_{(2)}]< E_1[r_{(2)}] + \sum_{c=1}^{2\log_B n} E_c[r_{(2)}]$$
	Besides,
		%$c^* \le \log_B \frac{n}{2}$ and 
		for each $c$, $E_{c^*}[M_O] \ge  \frac{1}{4Bn} E_c[r_{(2)}]$, thus,
		 $$\small{(2\log_B n+1) \cdot E_{c^*}[M_O] \ge \frac{1}{4Bn} \sum_c E_c[r_{(2)}]}$$
		
	\textbf{Case 2):}
	let $c_s$ be the first class that the maximum value of 
		second largest reported prices appeared at class $c_s$ is produced by $v_1$.
		That is to say:
		1) $\max\{ r_{(2)}(\pi) | \pi \in \Pi_{c_s} \} = v_1 \cdot B^{-c_s} > v_2 \cdot B^{-c_s}$;
		2) $v_1 \cdot B^{-c_s} \le v_2$ (if $v_1 \cdot B^{-c_s} $ is larger than $v_2$, then it becomes the largest reported price);
		3) for any $c<c_s$, $\max\{ r_{(2)}(\pi) | \pi \in \Pi_{c} \} \le v_2 \cdot B^{-c}$, there is no chance to be the second largest reported price when $v_1$ appears at class $c$ since $v_1 \cdot B^{-c} > v_2 \cdot d(t_1)$.
		
	\begin{itemize}		
		\item $c_s < 2\log_B n$: 
		note that for any $\pi \in \Pi_{c_s}, r_{(2)}(\pi)\le v_1\cdot B^{-c_s} \le v_2$. 
		Then for each $c >c_s+ 2\log_B n$:
		$$E_c[r_{(2)}]=r_{(2)}(\Pi_{c}) \le \frac{v_2}{B^{c-c_s}} < \frac{v_2}{n^3B}$$
		Thus, the total expected $r_{(2)}$ from classes $ c > c_s+2\log_B n$ 
			is less than the expected $r_{(2)}$ obtained from the first class $c=1$. 
			\ie, $\sum_{c > c_s+2\log_B n} r_{(2)}(\Pi_{c}) < \frac{v_2}{n^2B}$.
		Therefore, we have: 
		$$E[r_{(2)}]< E_1[r_{(2)}] + \sum_{c=1}^{c_s} E_c[r_{(2)}]+\sum_{c=c_s}^{c_s+2\log_B n} E_c[r_{(2)}]$$
		
		\item $c_s > 2\log_B n$: similar to the above analysis, we have:
		$$\sum_{c > 2\log_B n}^{c_s} r_{(2)}(\Pi_{c}) + \sum_{c > c_s+2\log_B n} r_{(2)}(\Pi_{c}) < 2\frac{v_2}{n^2B}$$
		Furthermore, we have $$E[r_{(2)}]< E_1[r_{(2)}] + \sum_{c=1}^{2\log_B n} E_c[r_{(2)}]+\sum_{c=c_s}^{c_s+2\log_B n} E_c[r_{(2)}]$$
	\end{itemize}
		
	In total, we have $$\small{(4\log_B n+1) \cdot E_{c^*}[M_O] \ge \frac{1}{4Bn} E[r_{(2)}] }.$$
	That is $E_{c^*}[M_O]\ge 1/\Theta(n\log n) \cdot E[r_{(2)}] $ and $M_W$ is $\Theta(n\log n) $ competitive.
\end{proof}

\subsection{Discussion on Other Discount Functions $d(t)$}
\label{subsec:other-d}

In this subsection, we then study situations when the discount function $d(t)$ is not non-increasing.

\textbf{Case 1: $d(t)$ is an {arbitrary} known and non-decreasing}.
When the discount function $d(t)$ is non-decreasing, then the mechanism $M_R$ clearly is not truthful now.
In this situation, given the strategies of the other users, for a given user $i$, if it arrives in some class $c$ from the top $\hat{c}$
 classes originally, it will be better for user $i$ to delay its arrival to a later time-slot.
 Then its reported price will be increased. Thus, increasing the probability that the user $i$ will be chosen as the final winner.
 
 From the analysis of the case when $d(t)$ is non-decreasing, we can show that more than $1/2$ of the expected revenue $r_{(2)}(\Pi)$ is from 
  the last $\hat{c}$ discount-classes now.
  However, one challenge of designing a truthful mechanism with good competitive ratio is that most profitable  users are from the late arrived users.
  Thus, it is difficult to apply the observe-then-decide strategy adopted by the mechanism $M_R$.

\textbf{Case 2: $d(t)$ is an {arbitrary} and known function}.
In this case, when the $d(t)$ is public known to both the mechanism and the buyers, then all mechanisms based on partitioning discount-classes are not truthful here.
When the seller knows $d(t)$, while the buyers do not know what the exact $d(t)$ is, then it is expected that the buyers will not manipulate its arrival time.
Thus the mechanism becomes a single-parameter game. In this case, we can show that our mechanisms are truthful in  expectation.

In both cases, a mechanism that uses a fixed price $r_0$ as a reserved price to auction the item 
 (when a user $i$ whose reported price is larger than $r_0$, the item is sold to this user and charged a payment $r_0$) is always truthful.
Then we need to carefully compute the expected value of $r_0$ so that it will maximize the expected revenue.

%%%%%%%%%%%%%%%
\subsection{Mechanism For Known Optimum Expected Revenue}
\label{alg_known_optimum}
We then consider the case when we know the value of the optimum revenue.
Given a discounted function $d(t)$, the adversary chosen valuation, and adversary chosen arrival sequence, 
 let $OPT_1$ be the expected value of the maximum reported price, by assuming that all users are truthful.
Let $OPT_2$ be the expected value of the second reported price, \ie, the optimum revenue by the truthful Vickery mechanism.
Obviously, $OPT_2 \le OPT_1$ for any discount function and initial valuations.
 Assume that we have a good estimation of the value $OPT_1$, \ie, 
  we can compute a value $\hat{Z}$ satisfying that $  OPT_1 /C  \le \hat{Z} \le C \cdot OPT_1$ for some constant $C \ge 1$.
  Let $Z= \frac{\hat{Z}}{C}$, then  $  \frac{OPT_1} {C^2}  \le Z \le OPT_1$.
  
 Then similar to the mechanism proposed in~\cite{babaioff2009secretary}, we propose the following mechanism  with a reserved price $Z/2$.
 
 \textbf{Known-OPT Mechanism $M_Z$:}
\begin{compactenum}
\item 
Compute an estimation $Z$ of the optimum revenue $OPT_1$ when all users are assumed to be truthful.
Assume that value $Z$ is adjusted such that $  \frac{OPT_1} {C^2}  \le Z \le OPT_1$.
\item
Process each of the arrived users $i$.
If the reported price $r_i $ is at least $Z/2$, then we accept this user $i$ and charge this user a payment $Z/2$.
\end{compactenum}

 It is easy to show that this mechanism is clearly time-truthful and valuation-truthful.
Then we analyze the competitive ratio of this mechanism $M_{Z}$. 
Similar to the proof of Theorem 4.7 in~\cite{babaioff2009secretary}, we can prove the following lemma.

\begin{lemma}[Known-OPT$_1$]
If there is a good estimate $Z$ for the expected optimum revenue $OPT_1$, \ie, $c_0 \cdot OPT_1 \le Z \le OPT_1 $ for some constant $c_0 \in(0,1)$,
 then the expected revenue achieved by the mechanism $M_Z$ is at least $Z/4$.
\end{lemma}

Recall that $OPT_2 \le OPT_1$, and $E(M_Z) \le Z/4 \le c_0 \cdot OPT_1/4 \ge c_0 \cdot OPT_2/4$, then we have the following lemma for the competitive ratio of the
 mechanism $M_Z$.
 
\begin{lemma}[Known-OPT]
If there is a good estimate $Z$ for the expected optimum revenue $OPT_1$ (or $OPT_2$), \ie, $c_0 \cdot OPT_1 \le Z \le OPT_1 $ for some constant $c_0 \in(0,1)$,
 then the competitive ratio of the mechanism $M_Z$ is at most $\frac{4}{c_0}$.
\end{lemma}

\section{Valuations Drawn From A Given Distribution}
\label{alg_special_case}

We study the case that the buyers' initial valuations are drawn \textit{i.i.d.} from a given distribution.
Our first idea is to set a fixed take-it-or-leave-it price. 
Since both the valuation distribution and discount function are given, we try to compute a fixed reservation price to maximize the expected profit. 
Here, we can use the idea of Myerson auction to compute an optimal reservation price $x$, \ie, $\arg\max_x x(1-P_0(x))$ where $P_0(x)$ denotes the probability that all $v_j \in V$ are less than $x$. However, with the existence of $d(t)$, it is difficult to compute $P_0(x)$ even with a simple PDF $f$ (\eg, uniform distribution).
We take uniform distribution as an example to illustrate the key idea of our mechanism, w.l.o.g.,  the density function $f(y) = 1$ on $0 \leq y < 1$.  
We will extend our mechanism to general distribution in \S\ref{alg:extension}.
To further improve the performance, we also consider updating the reservation price dynamically. 
Accordingly, we design the following three different mechanisms to (approximately) maximize the broker's profit.

\subsection{\textbf{Fixed Reservation Price Mechanism} ($M_F$)}
\label{alg:fixed}
%\subsubsection{\textbf{Fixed reserve price} ($M_3$)}
%\label{alg_fixed}

\textbf{Mechanism $\bm{M_F}$:} let $x$ be the fixed reservation price, i.e., the decision price never changed over time.
That is, we will pick the first buyer who reports a price $r \ge x$, and charge him at the price $x$ as well.
The details of our mechanism $M_F$ is shown in Mechanism~\ref{alg:MF}.
%Our mechanism is shown in Mechanism~\ref{alg:M3} and our analysis on its performance is as follows.	
\vspace{-0.05 in}
\begin{algorithm}
	\small
	\caption{Fixed reservation-price mechanism $M_F$}
	\label{alg:MF}
	\KwIn{$t_e$: stopping time; $d(t)$: discount function; $\lambda$: arrival rate.}
	\KwOut{winner and payment.}
	$winner=0$, $payment=0$\;
	Compute the reservation price $x$ by solving~(\ref{qua:obj})\;
	$j\gets 1$; Wait for the $j$-th buyer\;
	\While {${t}<t_e$}
	{
		\eIf {$r_j > x $}{
			Choose this buyer as $winner$ and charge him at $x$ \;
		}
		{ $j\gets j+1 $\;
			Wait for the $j$-th buyer\;
		}
	}
	return $winner$, $payment$ \;
\end{algorithm}

We study the competitive ratio $\rho=\frac{E[M_Y]}{E[M_F]}$ where $M_Y$ is the Myerson optimal auction. 
Given the arriving rate $\lambda$ of buyers, we first compute the expected $j$-th arrived time ${t}_j = j/\lambda$ and its corresponding discount value ${d}_j = d({t}_j)$.
Denote the $j$-th arrived buyer's report price as $v_j *d_j$.

Given the discount function $d(t)$, and the arrival process, we compute the expected discount value sequence $\{d_1, d_2, \cdots, d_n\}$.
Then we have the following lemma for $E[M_Y]$.
Recall that we assume initial valuations drawn \iid from uniform distribution whose range is $[0,1].$
%We first introduce the following Lemma~\ref{lem-M_Y}, which shows that mechanism $M_Y$'s revenue is less than $d_2$. 
%Due to the space limitation, we omit the proof details of Lemma~\ref{lem-M_Y} here.
\begin{lemma}
\label{lem-M_Y}
	$E[M_Y]<d_2$.
\end{lemma}
\begin{proof}
	Note that $M_Y$ always selects the highest price $r_{(1)}$ and charges at the second largest price $r_{(2)}$.
	We study the value of $r_{(2)}$ in the following cases:
	\begin{enumerate}
	       \item $r_{(2)}$ appears in the first item of the sequence of the reported prices, i.e., $r_{(2)}=v_1d_1 $. 
	       In this case, we have $r_{(1)}=v_j*d_j$ with some $j\geq 2$, that is, $r_{(1)} \le 1*d_2$, as $v_j \le 1$ and $d_j \leq d_2$. Then we have $r_{(2)}\le r_{(1)} \le d_2$. 
	       %\item $r_{(2)}$ appears in the second item of the sequence of the reported prices, i.e., $r_{(2)}=v_2d_2 $. Since $v_2 \in [0,1)$, we have $r_{(2)}< d_2$.
	\item Otherwise, $r_{(2)}=v_jd_j$ with some $j\geq2$. Since $v_j \le 1$ and $d_{j} \leq d_2$ for $j\geq2$, we know that $r_{(2)} \le 1*d_2$.
	\end{enumerate}
	Overall,   the second largest reported price is always less than $d_2$.
We obtain Lemma~\ref{lem-M_Y}.
	\end{proof}

Next, we study the value of $E[M_F]$. 
Note that we assume that the valuation $a_i$ follows uniform distribution within $[0,1]$, therefore the probability that $a_id_j<x$ is $\min\{1, x/d_j\}$ for each $j$. Consequently, 
%for the obtained sequence of discount values $\{d_1, d_2, \cdots, d_n\}$, 
the probability that all the initial valuations  are less than $x$ is 
{\small{
		\begin{equation}
		P_0(x)=\min\{1,x/{d}_1\}*\min\{1,x/{d}_2\}*\dots*\min\{1,x/{d}_n\}.
		\end{equation}
	}}
	Here $\min\{1,x/{d}_i\}$ is the probability that the $i$-th arrived user has a reported price $r_i$ at most $x$.
%{\begin{eqnarray}
%\scriptsize
%{P_0(x)=\min\{1,x/{d}_1\}*\min\{1,x/{d}_2\}*\dots*\min\{1,x/{d}_n\}. \label{for-mb-1}}
%\end{eqnarray}}
Hence the expected profit of $M_F$ is $E[M_F]=(1-P_0(x))x$.
To maximize $E[M_F]$, we set the fixed reservation price as the solution to:
\begin{equation}
\label{qua:obj}
1-P_0(x) -P_0'(x)x=0
\end{equation}
In other words, $x$ is the solution where the derivative of $(1-P_0(x))x$ is $0$.
By setting the fixed reservation price as $\arg\max_x (1-P_0(x))x$, where we restrict $x\in \{d_1, d_2, \cdots, d_n\}$, we have the following 
 Lemma~\ref{lem-M_F} to bound the value  $E[M_F]$.
\begin{lemma}
	\label{lem-M_F}
	$E[M_F]\ge(1-d_2/d_1)*d_{2}$
\end{lemma}
The lemma follows by choosing a special reserved price $x= d_2$.
Recall we have $E[M_Y]\le d_2$.
Therefore, we have the following theorem.

\begin{theorem}
	$M_F$ has a competitive ratio $\frac{E(M_Y)} {E(M_F)}\le \frac{d_1}{d_1-d_2}$. 
	\end{theorem} 
Obviously $M_F$ is a bid independent auction.
It is easy to show that the mechanism $M_F$  is value truthful.
In addition, for any buyer $i$ arriving at time $t_j$, it is easy to know that for any time $t_q > t_j$, $a_i*d(t_j)-x \geq a_i*d(t_q)-x$; that is, $M_F$ is time truthful. Therefore, we have:
\begin{theorem} 
	Mechanism $M_F$ is value and time truthful. 
\end{theorem}

Notice when $d_2 = d_1$, the aforementioned theorem does not hold anymore.
In this case, assume that $d_1 = d_2 = \cdots = d_{i-1} >d_i$ for some $i \ge 3$.
Then it is easy to show that $E(M_Y) \le d_1$ and $E(M_F) \le (1-\frac{d_i}{d_1})d_i$.
Then 
\begin{lemma}
	$M_F$ has a competitive ratio $\frac{E(M_Y)} {E(M_F)}\le \frac{d_1^2}{d_i(d_1-d_i)}$. 
	\end{lemma} 
When all the discount valuations are the same, it degenerates to the traditional online-secretary problem which has a truthful mechanism with a constant competitive ratio.

\subsection{\textbf{Dynamic Reservation Price Mechanism} ($M_D$)} 
\label{alg:dynamic}

Intuitively, as time goes on, the probability of having  reported prices larger than a reserved price $x$ is decreasing as the discount function is non-increasing. 
Thus, we should reduce our reserved decision price accordingly. 
To further improve $M_F$ that has a fixed reserved decision price, we propose the following  mechanism $\bm{M_D}$ with a dynamically chosen decision price.

\textbf{Mechanism {$\bm{M_D}$}:} let $x(t)*d(t)$ be the decision price at time $t$. We pick the buyer at time $t$ if he reports a price $r> x(t)*d(t)$, and also charge him (the winner) $x(t)*d(t)$. The details of our mechanism $M_D$ is shown in Mechanism~\ref{alg:M_D}.

\begin{figure}[t!]
	\begin{center}
		{\includegraphics[width=2.8in,height=0.8in,trim={0.8cm 0.6cm 0.8cm 1.6cm}, clip]{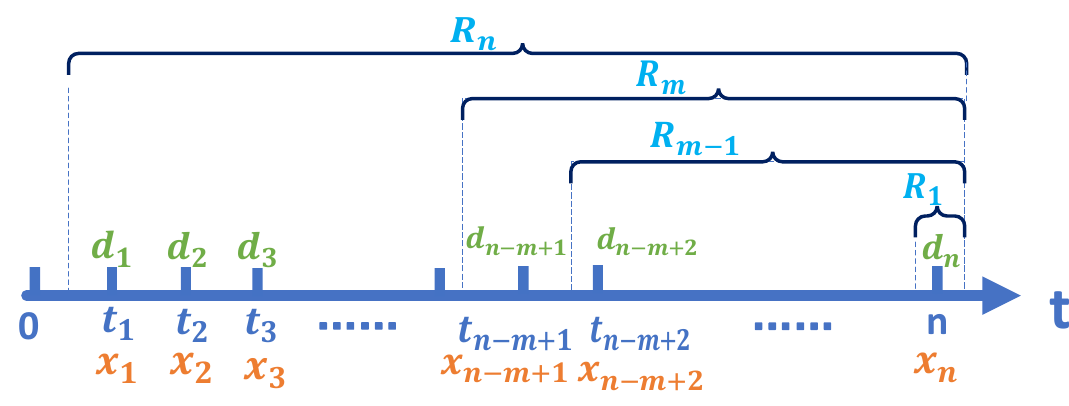}}%\linewidth 2.8 in 2.8 in %height=0.8in,
		\caption{An illustration of our mechanism $M_D$.} % left, bottom, right and top 
		\label{fig:keyidea}
	\end{center}
\end{figure}

\textbf{Key idea}:
To maximize the revenue, our key idea for designing $x(t)$ is as follows. %Intuitively, with time going on, the probability of we see high report price is decreasing. In other words, the expected profit is decreasing as time goes, so, we should decrease the decision price too.
Let $R_m$ ($1\leq m\leq n$) be the expected profit when there are $m$ buyers left. Our goal is to maximize $R_n$ by designing a series of decision prices where $n$ is the total number of possible buyers.
\iffalse
Let $\bar{t}$ denote the next buyer's expected arrival time. 
Set $x_{\bar{t}}*d(\bar{t})$ as the reservation price until time $\bar{t}$. 
We assume the buyer reports the price \textit{honestly}, then the probability that we accept the buyer and get the benefit $x_{\bar{t}}*d(\bar{t})$ is $(1-x_{\bar{t}})$ (recall that we assume that the buyers' valuations follow uniform distribution).
The probability that we reject the buyer and get the benefit $R_{m-1}$ is $x_{\bar{t}}$.
%So $R_m=(1-x_{\bar{t}})*x_{\bar{t}}*d(\bar{t})+x_{\bar{t}}*R_{m-1}$.
Let $\bar{t}_j=j/\lambda$ denote the expected arrival time of the $j$-th buyer, $d_j$ denote the corresponding discount value and $x_j$ be the calculated corresponding decision price. 
\fi
As shown in Fig.~\ref{fig:keyidea}, we divide the time domain into $n$ intervals. Similar to $M_F$, let $t_j=j/\lambda$ be the expected arriving time of the $j$-th buyer. 
Set $t_0=0$.
For each time interval $(t_{n-m}, t_{n-m+1}]$, we need to compute the decision price $x_{n-m+1}*d_{n-m+1}$. For this purpose, we conduct a backward analysis to determine the values. From $m=1$ to $n$, we recursively compute $x_{n-m+1}$ to maximize the expected profit $R_{m}$.
%When there are expected $m$ buyers left, we compute $x_{n-m+1}*d_{n-m+1}$, and set it as the decision price at time interval $(t_{n-m+1}, t_{n-m+1}]$.
%We first compute $x_n$ to maximize the expected profit $R_1$. 
%Then we compute the value of $x_{n-1}, \dots, x_{n-m+1},\dots,x_1$ to maximize $R_2, \dots, R_m, \dots, R_n$ recursively.
%In the following part, we will elaborate our strategy.
%\paragraph{Mechanism 1}  Assume there are remaining $i$ numbers. Let $P^{\underline{i}}$ be the probability of all $i$ numbers less than $x_t$, than $P^{\underline{i}}=\sum_{j=1}^{i} x_t*d_t/d_{t+j}$. If we pass $x_t$, then the probability we win is $P^{\overline{i}}=\sum_{j=1}^{i}(1-(x_t*d_t/d_{t+j}))* \prod_{k\not =j}^{i}(x_t*d_t/d_{t+k}) + 1/2*\sum_{j=1}^{i}\sum_{j'\not=j}^{i}(1-(x_t*d_t/d_{t+j}))(1-(x_t*d_t/d_{t+j'}))* \prod_{k\not =j,k\not =j' }^{i}(x_t*d_t/d_{t+k}) + 1/3* \dots$ to calculate $x_t$, let $P^{\underline{i}}=P^{\overline{i}}$.

\begin{algorithm}[ht!]
	\small
	\caption{Dynamic reservation price mechanism $M_D$}
	\label{alg:M_D}
	\KwIn{$t_e$: stopping time; $d(t)$: discount function; $\lambda$: arrival rate}
	\KwOut {$winner$ and $payment$.}
	${n}=\lambda*t_e$; $j\gets 1$;	$t_j=j/\lambda$\;	
	$winner=0$, $payment=0$\;
	\For{$j\in \{1,2, \dots,{n}\}$}
	{
		According to equation (\ref{recursive2}), compute $x_j$\;
	}	
	\While {${t}<t_e$} {	
		%$x_j=$ ComputeXT $(\lambda, d(t), j, n, t)$ \;			
		\While {$t<t_j$}
		{		
			Wait for the next buyer,  denoting his report price as $r$\;
			If {$r > x_j*d_j $}, choose this buyer as $winner$ and charge him at $payment=x_j*d_j$\;					
		}		
		$j \gets j+1$\;					
	}
	Return $winner$, $payment$\;
\end{algorithm}

%We use the idea of recursion 
%Note that we assume the auctioneer knows $n$, the total number of buyers, and he knows buyers' valuations' distribution. He must decide after each buyer reports her price whether to choose or reject it. The auctioneer's \textit{profit} is the price of winning buyer pays. If he has not chosen until the final buyer and the final buyer's report price is less than $x_{t_n}*d(t_n)$, the auctioneer will choose no one and its  profit is $0$, but we will show that in our mechanism, when $n$ is large enough, the probability of this circumstance occurs is very small. The object is to find a strategy that maximizes the expected price of the winning buyer pays.

{\textbf{Our strategy}:} we start from the base case $\bm{n=1}$. Since there is only one buyer, we just need to compute $x_1*d_1$ as the decision price.
 The probability that we accept the buyer is $1-x_1$. So the expected profit is $R_1=(1-x_1)x_1*d_1$. To maximize $R_1$, we set $x_1=1/2$ and therefore  $R_1=d_1/4$. 
%\iffalse
%If $n=2$, we must decide whether to keep the first price or to go on to the last. If the current one, $r_1>x_1*d_1$, we accept it, reject it if $r_1 \leq x_1*d_1$. And for the last price $r_2$, the decision price is $x_2*d_2$, so the expected profit is $(1-x_1)*x_1*d_1+x_1*(1-x_2)x_2*d_2$. Consider if we move on to time $t_2$, \ie, there is only one buyer left, we set $x_2=1/2$ to maximize our profit at time $t_2$. Then at time $t_1$, to maximize the expected profit: $(1-x_1)*x_1*d_1+x_1*d_2/4$, we set $x_1=1/2+d_2/8d_1$. The expected profit is $d_1/4+d_2^2/64d_1+d_2/8$.
%1) Therefore we keep the current one, $r$, if $r > d_2/2$, rejects it if $r < d_2/2$, and is indifferent if $r = d_2/2$.  The probability of accepting the first price is $1-d_2/2d_1$, condition on this, the expected first price value is $d_1*(1+d_2/2d_1)/2$. So for n = 2, the expected profit is $ (1-d_2/2d_1) *d_1*(1+d_2/2d_1)/2 + d_2/2d_1* 1/2*d_2=d_1/2+d_2^2/8d_1$. \linote{wrong!} The expected profit is $ (1-d_2/2d_1)*d_2/2 + d_2/2d_1*d_2/4$.
%%we use the method of induction to calculate each $x_j$.
%let $R_{m}$ be the expected profit when there are $m$ buyers left.
%%of a sequence of length $n$'s last $m$ prices under the optimum strategy. 
%Then let $x_{{n-m+1}}*d_{n-m+1}$ be the decision price for the ${(n-m+1)}$-th buyer in a sequence of length $n$. 
%We get a recursion relation between the expected profit of later length $m$ and the later $m-1$. 
%Let $v_{n-m+1}*d_{n-m+1}$ denote the $(n-m+1)$-th arrived buyer's report price. 
%\fi

Then we consider the case {$\bm{n>1}$}: for each $(t_{n-m},t_{n-m+1}]$, we need to determine the decision price $x_{n-m+1}*d_{n-m+1}$. 
%if the arrived buyer's report price exceeds $x_{n-m+1}*d_{n-m+1}$, we keep it and the buyer pays at $x_{n-m+1}*d_{n-m+1}$; otherwise, we reject it.
If we encounter a buyer reporting a price larger than the decision price, our profit is $x_{n-m+1}*d_{n-m+1}$;  %, conditional on its being larger than $x_{n-m+1}*d_{n-m+1}$. 
otherwise, our expected profit from the remaining $m-1$ buyers is $R_{m-1}$. 
%These two expectations must be weighted by their probabilities $1-x_{n-m+1}$ and $x_{n-m+1}$, respectively.
%\iffalse
%Let $q=n-m+1$, % $m=n-q+1, 1<n-q+1<=n, 1=<q<n$
%then we give the expectation for the last $n-q+1$ prices,
%\begin{equation}
%\label{}
%\begin{aligned}
%R_{}
%& R_{n-q+1}=(1-x_{q})*x_{q}*d_{{q}} + x_{q}*R_{n-q},  1 \leq q < n \\
%& R_1=(1-x_n)*x_n*d_n; \quad x_n=1/2, R_1=d_{n}/4. 
%\end{aligned}
%\end{equation}
%\fi
For convenience, let $R_{m}^{acc}= (1-x_{n-m+1})*x_{n-m+1}*d_{{n-m+1}}$ and $R_{m}^{rej}= x_{n-m+1}*R_{m-1}$ (the initial $R_0$ is set to be $0$). Then we have the following recursion
\begin{equation}
\label{for-recursive1}
\begin{aligned}
\begin{cases}
R_{m}^{acc} &= (1-x_{n-m+1})*x_{n-m+1}*d_{{n-m+1}};\\
R_{m}^{rej} &= x_{n-m+1}*R_{m-1};\\
R_{m} &= R_{m}^{acc} + R_{m}^{rej}, \text{ for } 1 \leq m \leq n. 
\end{cases}
\end{aligned}
\end{equation}
Given $R_{m-1}$ ($1 < m \leq n$), we can obtain the maximum $R_{m}$ and corresponding $x_{n-m+1}$ via (\ref{for-recursive1}), \ie, 
\begin{equation}
\begin{aligned}
\label{recursive2}
& x_{n-m+1}=1/2+R_{m-1}/(2d_{n-m+1}), \quad x_n=1/2 \\
R_m
& =(R_{m-1}+d_{n-m+1})^2/(4d_{n-m+1}), \quad R_1=d_{n}/4\\
\end{aligned}
\end{equation}

%\begin{equation}
%\begin{aligned}
%R_m
%& =(R_{m-1}+d_{n-m+1})^2/4d_{n-m+1}, \quad R_1=d_{n}/4\\
%%& =d_{n-m+1}/4+R_{m-1}/2+R_{m-1}^2/4d_{n-m+1}. 
%\end{aligned}
%\end{equation}
%$R_m=(R_{m-1}+d_{n-m+1})/2+(R_{m-1}^2-d_{n-m+1}^2)/(4*d_{n-m+1})$.

\iffalse
\begin{algorithm}
	\caption{ComputeXT($\lambda, d(t), j, n, t$)}
	\For{$k\in \{j, j+1, \dots,{n}\}$}
	{
		${t}_k \gets t+k/\lambda$; $d_k \gets d({t}_k)$\; 
	}
	According to equation (\ref{recursive2}), compute $x_j$\;
	Return {$x_j$}\;
\end{algorithm}
\fi

\begin{algorithm}[ht!]
	\small
	\caption{Semi-time truthful mechanism $M_T$}
	\label{alg:M_T}
	\KwIn{$t_e$: stopping time; $d(t)$: discount function; $\lambda$: arrival rate.}
	\KwOut {$winner$ and $payment$.}
	${n}=\lambda*t_e$; $j\gets 1$;	$t_j=j/\lambda$\;	
	$winner=0$, $payment=0$\;
	\For{$j\in \{1,2, \dots,{n}\}$}
	{
		According to equation (\ref{recursive2}), compute $x_j$\;
	}
	\While {${t}<t_e$} 
	{	
		%$x_j=$ ComputeXT $(\lambda, d(t), j, n, t)$\;	
		\While {$t<t_j$}
		{		
			Wait for the next buyer,  denoting his report price as $r$\;
			If {$r > x_j*d_j $}, choose this buyer as $winner$ and charge him at $payment=x_j*d_j$\;					
		}		
		$j \gets j+1$\;				
	}
	Return $winner$, $payment$\;
\end{algorithm}
%The details of our mechanism $M_D$ is shown in Mechanism~\ref{alg:M_D}. 

More importantly, we have the following theorem which is consistent with our intuition. We emphasize that we have the same result in Section~\ref{alg:extension} for general distribution of the valuations (see Theorem~\ref{generalMD}). 

\begin{theorem}
\label{the-mb-2}
Given a non-decreasing discount function $d(t)$,
 both the profit $R_m$ and decision price $x_{n-m+1}*d_{n-m+1}$ are strictly decreasing as time going on.
\end{theorem}

Theorem~\ref{the-mb-2} can be obtained by combining the following Lemmas~\ref{Rn} and \ref{xndn}.
We first introduce the following claim and prove it by induction on $m$. 
%{The proof is omitted due to the space limitation.}
\begin{claim}
	\label{claim:Rm}
	$\forall m,  1\leq m \leq n$, we have $R_m < d_{n-m+1}$.
\end{claim}
\begin{proof}
	We prove this claim by induction on $m$.
	
	\textbf{Base case:} {For ${m=1}$}, $R_1=d_n/4 <d_n$.
	
	\textbf{Induction step:} {For ${m>1}$}, we assume $R_{m-1}<d_{n-m+2}$. Since $x_{n-m+1}<1$, from the definition of $R_{m}^{acc}$ and $R_{m}^{rej}$, we have,
	\begin{equation*}
	\begin{aligned}
	&R_{m}^{acc} < (1-x_{n-m+1})*d_{n-m+1},\\
	&R_{m}^{rej}< x_{n-m+1}*d_{n-m+2}.\\
	\end{aligned}
	\end{equation*}
	Recall that we assume $d_1\geq d_2\geq \dots \geq d_n$, thus,
	$$ R_{m}^{rej} < x_{n-m+1}*d_{n-m+1}$$
	Thus, $ R_{m}= R_{m}^{acc} + R_{m}^{rej}< d_{n-m+1}$.
	The claim follows.
\end{proof}

\begin{lemma}
	\label{Rn}
	$\forall m,  2\leq m \leq n$, we have $R_m > R_{m-1}$.
\end{lemma}
\begin{proof}
	\iffalse
	We prove this lemma by induction on $m$.\\
	\textbf{Base case:} {For ${m=2}$}, since $R_2=d_{n-1}/4+d_{n}^2/64d_{n-1}+d_{n}/8$, $R_1=d_n/4$, obviously, $R_2>R_1$.\\	
	\textbf{Induction step:} {For ${m>2}$}, assume $R_m > R_{m-1}$, 
	\ie, 
	\begin{equation*}
	\begin{aligned}
	R_m-R_{m-1} & =\frac{(R_{m-1}+d_{n-m+1})^2}{4d_{n-m+1}}-R_{m-1}\\
	& =\frac{d_{n-m+1}^2-2*R_{m-1}*d_{n-m+1}+R_{m-1}^2}{4d_{n-m+1}}\\
	& = \frac{(R_{m-1}-d_{n-m+1})^2}{4d_{n-m+1}} > 0
	\end{aligned}
	\end{equation*}
	\fi
	From (\ref{recursive2}), proving $R_m > R_{m-1}$ is equivalent to proving $R_m-R_{m-1}  =\frac{(R_{m-1}+d_{n-m+1})^2}{4d_{n-m+1}}-R_{m-1}  = \frac{(R_{m-1}-d_{n-m+1})^2}{4d_{n-m+1}} > 0$.
	%(\ref{recursive2})
	%{\small{$$R_m-R_{m-1} =\frac{(R_{m-1}+d_{n-m+1})^2}{4d_{n-m+1}}-R_{m-1} > 0$$}}
%	\begin{equation*}
%	\small
%	\begin{aligned}
%	R_m-R_{m-1}  =\frac{(R_{m-1}+d_{n-m+1})^2}{4d_{n-m+1}}-R_{m-1}  = \frac{(R_{m-1}-d_{n-m+1})^2}{4d_{n-m+1}} > 0\\
%	%& =\frac{d_{n-m+1}^2-2*R_{m-1}*d_{n-m+1}+R_{m-1}^2}{4d_{n-m+1}}\\	
%	\end{aligned}
%	\end{equation*}
%	$$R_m-R_{m-1}=d_{n-m+1}/4-R_{m-1}/2+R_{m-1}^2/4d_{n-m+1}> 0.$$
%	so we just need to prove: $ d_{n-m+1}^2-2*R_{m-1}*d_{n-m+1}+R_{m-1}^2 > 0$.\\
%	Clearly,
%	\begin{equation*}
%	\begin{aligned}
%		& d_{n-m+1}^2-2*R_{m-1}*d_{n-m+1}+R_{m-1}^2\\
%		& = (R_{m-1}-d_{n-m+1})^2 \geq 0
%	\end{aligned}
%	\end{equation*}
	 According to Claim~\ref{claim:Rm}, $R_{m-1}<d_{n-m+2} \leq d_{n-m+1}$, we obtain $(R_{m-1}-d_{n-m+1})^2>0$. %Thus we have $\forall m, 2\leq m\leq n$, $R_m > R_{m-1}$.
\end{proof}
% $x_n, x_{n-1}, \dots, x_1$ does not satisfy monotonicity.

\begin{lemma}
	\label{xndn}
%	$x_1d_1> x_2d_2> \dots >x_nd_n $. That is,
	 $\forall m \in [1, n-1]$, $x_{n-m}d_{n-m} > x_{n-m+1}d_{n-m+1}$.	
\end{lemma}
\begin{proof}
%	If $n=2$, obviously $x_2 < x_1$, so $x_2d_2 < x_1d_1$.
%  Assume $n \geq 2$, $x_n=1/2, x_{n-1}=1/2 + d_{n-1}/8d_n$, clearly, $x_n d_n < x_{n-1} d_{n-1}$.
%	We need to prove when $1<m < n$, $x_{n-m+1}d_{n-m+1} < x_{n-m}d_{n-m}$. 
	From (\ref{recursive2}), we have
	\begin{equation}
	\label{for-xndn}
	\begin{aligned}
	\begin{cases}
	x_{n-m}*d_{n-m}	& =(1/2+\frac{R_{m}}{2d_{n-m}})*d_{n-m}\\
	x_{n-m+1}*d_{n-m+1}
	& =(1/2+\frac{R_{m-1}}{2d_{n-m+1}})*d_{n-m+1}
	\end{cases}	
	\end{aligned}
	\end{equation}
	%Note that we have $R_m>R_{m-1}$ and $d_{n-m} \geq d_{n-m+1}$ 
	According to Lemma~\ref{Rn}, we have $R_m>R_{m-1}$ and $d_{n-m} \geq d_{n-m+1}$, so $x_{n-m}d_{n-m} - x_{n-m+1}d_{n-m+1}>0$
	%we obtain Lemma~\ref{xndn} by (\ref{for-xndn}).
\end{proof}

%\paragraph{Zero profit}: If we do not select any buyer at last, our profit is $0$. Let $P_{fail}$ denote the expected probability of zero profit, $P_{fail}=x_1*x_2*\dots*x_n$. 
%%Since $\forall i, x_i<1$, 
%If $n$ is large enough, $P_{fail}$ can be very small. Thus, we say $M_D$ have little probability to achieve zero profit.
%%In Section~\ref{sec:eva}, the experiment results have verified the our statement.

Now we consider the  the competitive ratio $\rho=\frac{E[M_Y]}{E[M_D]}$.
According to the above analysis, we know that the expected revenue of $M_D$ is $R_n$.
For $n=1$, $\rho=\frac{E[M_D]}{E[M_Y]}=\frac{d_1/4}{d_1}=1/4$.
When $n>2$,
% ${R_n}/{E[Vickrey]}>{\sum_{i=1}^{n} d_{i}/2^{i+1}}/{\sum_{i=1}^{n}\sum_{j\not =i}^{n}E[P_{i,j}*d_j*v_j]}$\\
%$R_n=(R_{n-1}+1)^2/4$, $R_1=1/4$
we have %$x_{n-m+1}=1/2+R_{m-1}/2d_{n-m+1}$ and
$R_m=\frac{(R_{m-1}+d_{n-m+1})^2}{4d_{n-m+1}} > \frac{d_{n-m+1}}{4}+\frac{R_{m-1}}{2}$
%thus $R_n>\sum_{i=1}^{n} d_{i}/2^{i+1}$.
from (\ref{recursive2}). In addition, $E[M_Y]<1$. 
Then, $ \rho=\frac{E[M_Y]}{E[M_D]}<\frac{4}{d_1}.$
%> \frac{\sum_{i=1}^{n} d_{i}/2^{i+1}}{\sum_{i=1}^{n}\sum_{j\not =i}^{n}E[P_{i,j}*d_j*v_j]} 
%\begin{itemize}[]	 
%	\item $R_1=d_{n}/4$
%	%\item $R_2=d_{n-1}/4+d_n/8+ d_n^2/64d_{n-1}>d_{n-1}/4+d_n/8$
%	\item $R_2=d_{n-1}/4+d_n/8+ R_1^2/4d_{n-1}$
%	\item $R_3>d_{n-2}/4+d_{n-1}/8+d_n/16$
%	\item $R_4>d_{n-3}/4 + d_{n-2}/8+d_{n-1}/16+d_n/32$
%	\item \dots .
%	\item $R_m>\sum_{i=1}^{m} d_{n-i+1}/2^{m-i+2}$.
%	\item \dots .
%	\item So, $R_n>\sum_{i=1}^{n} d_{i}/2^{i+1}$
%\end{itemize}

%Let $P_{i,j}$ ($i \not = j$) denotes the probability of $d_i*v_i$ is the maximum and $d_j*v_j$ is the second largest one.
%Then:
%\begin{equation}
%\label{pij}
%\begin{aligned}
%P_{i,j}&=\max\{0,1-d_jv_j/d_i\}*\prod_{k\not=i,j}^{n} \min\{1,d_jv_j/d_k\} 
%%\\ & =(1-d_jv_j/d_i)[d_j <d_i/v_j]* \prod_{k\not=i,j}^{n} d_jv_j/d_k [d_j < d_k/vj]
%\end{aligned}
%\end{equation}
%Conditional on the probability $P_{i,j}$, the expected optimal revenue that \textit{Vickrey} auction achieve is:
%$$E[P_{i,j}*d_j*v_j]=\int_{0}^{\infty} P_{i,j}*d_j*v_j* f(v_j) dv_j$$
%So, $$E[Vickrey]=\sum_{i=1}^{n}\sum_{j\not =i}^{n}E[P_{i,j}*d_j*v_j]$$
%$\lambda$ is large enough and 
Next we study the change of $\rho$ when $n \to \infty$ (for instance, we can assume that there are sufficient buyers in a real scenario).
According to Lemma~\ref{Rn}, $\{R_1, \cdots, R_n\}$ is an increasing series. Moreover, we know that $R_n$ is always up-bounded by $d_1\leq 1$. 
Then there exists some $\alpha$ such that $\lim\limits_{n\to \infty}R_{n}=\alpha$.
%function with the increment of $m$.
%Assume $$\lim\limits_{n\to \infty}R_{n-1}=R_{n}=\alpha$$
From (\ref{recursive2}), we have $\alpha =(\alpha+d_{1})^2/4d_{1}$ which implies $\alpha=d_1$.
%$d_2 \to d_1$, $$\lim\limits_{n\to \infty}v_{(2)} = v_{(1)} \to 1$$
%
%Note that $E[M_Y]< 1$.
%\begin{equation*}
%\begin{aligned}
%\lim\limits_{n\to \infty} d(t_2) = d(t_1), \quad
%\lim\limits_{n\to \infty}E[P_{k,l}*d_l*v_l]=\lim\limits_{n\to \infty}E[r_{(1)}] = d_1
%\end{aligned}
%\end{equation*}
%Therefore, we have $$\lim\limits_{n\to \infty} E[P_{k,l}*d_l*v_l]=d_1$$
%In addition, $$\frac{R_m-R_{m-1}}{d_{n-m+1}-d_{n-m+2}}>\frac{1}{4}$$
Thus, $\lim\limits_{n\to \infty}\rho= \frac{E[M_Y]}{\alpha} = \frac{1}{d_1}.$

In summary, we have the following theorem.
\begin{theorem}
	Mechanism $M_D$ has a competitive ratio $<\frac{4}{d_1}$, and converging to a value $\frac{1}{d_1}$ when $n \to \infty$. 
\end{theorem} 
%We remark that the traditional offline Vikcrey auction is not the optimal mechanism under our problem model (considering $d(t)$), and thus it is not surprising that the ratio $\rho$ can converge to a value less than $1$.   
%In fact, the optimal offline mechanism under our model is an unexplored problem.
\begin{theorem}
	\label{M2:truthful}
	$M_D$ is value truthful but not semi-time truthful. 
\end{theorem}
%\begin{theorem}
%	\label{M2_timeTruthful}
%	$M_D$ is not time truthful.
%\end{theorem}
%\begin{proof}
	Obviously, $M_D$ is a bid independent auction and thus value truthful.
	To prove that it is not semi-time truthful, we can find a value $t_q$ yielding $a_i*d(t_j)-x_j*d(t_j) < a_i*d(t_q)-x_q*d(t_q)$ and $a_i*d(t_j) > x_j*d(t_j)$ by carefully selecting the valuations and discounts. 	
	We omit the details due to space limit.
%%%%%%%%%%%%%%	
%	We just need to provide an example that there exists some buyer $i$ that even he wins by reporting the true valuation at his arrival time $t_j$, he still can improve his utility by delaying his arrival. Formally, 
%	$ \exists t_q > t_j$, such that $$a_i*d(t_j)-x_j*d(t_j) < a_i*d(t_q)-x_q*d(t_q).$$
%	We design the example as follows. Without loss of generality, we study the case $n=2$ and let $j=1$ and $q=2$.
%	So we have $x_1*d_1=d_1/2+d_2/8$ and $x_2d_2=d_2/2$ by (\ref{recursive2}). We define the function 
%	$$g(a_i) = a_i*d_1-x_1*d_1-a_i*d_2+x_2*d_2,$$
%	and just need to find an $a_i>x_j$ such that $g(a_i) < 0$.
%	%If there exists $a_i$ such that $g(a_i) < 0$, then $M_D$ is not time truthful. 
%	Firstly, if $d_1=d_2$,  we know that $g(a_i)<0$ for any $a_i$ (because $x_1*d_1>x_2*d_2$ by Lemma~\ref{xndn}).
%	Second, if $d_1>d_2$, let $a_i=1/2+d_2/(8d_1)+ \delta $ where $\delta$ is any value between $0$ and ${d_2^2}/({8d_1(d_1-d_2)})$, and then $g(a_i)<0$. 
	
%\end{proof}

%To achieve time truthfulness, we further propose the following time truthful dynamic mechanism $M_T$.

\subsection{\textbf{Semi-time-Truthful Dynamic Reservation Mechanism} (${M_T}$)}
\label{alg:timeTruth}
%In the above proof, we have find $a_i>x_j$ such that $a_i*d_j-x_j*d_j < a_i*d_q-x_q*d_q$. if we slightly reduce the payment value $x_j*d_j$, such that $a_i*d_j-x_j*d_j \geq a_i*d_q-x_q*d_q$, \ie, the $j$-th arrived buyer cannot improve her utility by delaying her arrival to $t_q$. 
{\textbf{Key idea}}: Our new mechanism $M_T$ is
%Our new mechanism $M_T$ is 
almost identical to $M_D$. In particular, it uses the same decision price  $x(t)*d(t)$. 
 However, instead of charging the buyer the same value $x(t)*d(t)$, the payment is replaced by $\varrho(t)*d(t)$ where $\varrho(t)$ is a carefully crafted function $\varrho(t) \le x(t)$. 
The key point to achieve the truthfulness lies in the design of $\varrho(t)$ such that the function $v_j d(t) - \varrho(t)d(t)$ is non-increasing for the winning buyer with $v_j \ge x(t)$ after his arrival time.
%All the other parts of $M_T$ are same to those of $M_D$.

Also, we use the same backward analysis to maximize the revenue.
Let $R_m'$ denote the new expected profit (replacing the notations in Figure~\ref{fig:keyidea}) when there are $m$ buyers left. We determine the values of the variables through exactly the same manner of (\ref{recursive2}). Initially, $\varrho_n =x_n=1/2$ and $R_1=R_1'=1/2.$
%Let $R_{n-j+1}'$ denote the new expected profit when there are $(n-j+1)$ buyers left.
%Set $x_n=1/2, \varrho_n=x_n=1/2$, thus we have $R_1'=d_n/4$. 
For each $j=n-m+1$ ($1\leq j<n$), we have
%$\varrho_{n-1}*d_{n-1}=x_{n-1}'d_{n-1}-x_{n-1}'*d_{n}+\varrho_{n}*d_{n}$\\
%$R_2=(1-x_{n-1}')*\varrho_{n-1}*d_{n-1}+x_{n-1}'*R_1'$
%
%\begin{equation}
%\label{maximize2}
%\begin{aligned}
%\varrho_j&=(x_jd_j-x_j*d_{j+1}+\varrho_{j+1}*d_{j+1})/d_j\\
%R_{n-j+1}'& =  (1-x_j)*\varrho_j*d_j+x_j*R_{n-j}, \quad 1\leq j<n\\
%&\st \quad \varrho_jd_j < x_jd_j
%\end{aligned}
%\end{equation}
%\begin{equation}
%\begin{aligned}
%\label{}
%x_{n-m+1}' = & x_{n-m+1}  = 1/2+R_{m-1}/2d_{n-m+1}, \quad x_n=1/2 \\
%R_m = & (R_{m-1}+d_{n-m+1})^2/4d_{n-m+1}, \quad R_1=d_{n}/4\\
%\varrho_{n-m+1}= & \frac{(x_{n-m+1}(d_{n-m+1}-d_{n-m+2})+\varrho_{{n-m+2}}*d_{{n-m+2}})}{d_{n-m+1}}\\
%R_{n-j+1}'& =  (1-x_j)*\varrho_j*d_j+x_j*R_{n-j}', \quad 1\leq j<n\\
%\end{aligned}
%\end{equation}
\begin{equation}
\label{equ:M3}
\begin{aligned}
%R_{n-j+1} & =  (R_{n-j}+d_{j})^2/4d_{j}, \\
%x_j&=x_j = 1/2+R_{n-j}/2d_{j}, \\
\varrho_j&=(x_jd_j-x_j*d_{j+1}+\varrho_{j+1}*d_{j+1})/d_j, \\
R_{n-j+1}'& =  (1-x_j)*\varrho_j*d_j+x_j*R_{n-j}', \\ 
\end{aligned}
\end{equation}

Similar with $M_D$, we have the following theorem which is consistent with our intuition. 
%We omitted the proof details due to the space limitation.
%See the proof of extension to general distribution, \ie, Lemma~\ref{Rj}.

\begin{lemma}
	\label{MT}
	$\forall j,  1\leq j < n$ 
	%$\forall j \in [1,n)$, 
	$x_jd_j > x_{j+1}d_{j+1}$, $\varrho_jd_j\geq \varrho_{j+1}d_{j+1}$, $R_{n-j+1}' > R_{n-j}'$.
\end{lemma}

%[Why you need this theorem?]
%\begin{proof}
%	1) Since $x_j=x_j$, from Lemma~\ref{xndn}, we obtain $x_jd_j > x_{j+1}'d_{j+1}$.
%	2) From (\ref{equ:M3}), $\varrho_jd_j=(x_jd_j-x_j*d_{j+1}+\varrho_{j+1}*d_{j+1})$. $d_j \geq d_{j+1}$, thus, $\varrho_jd_j\geq \varrho_{j+1}*d_{j+1}$.
%	
%	3)$R_{n-j}'< \varrho_{j+1}*d_{j+1} \leq \varrho_jd_j$, 
%	$$R_{n-j+1}' =  (1-x_j)*\varrho_j*d_j+x_j*R_{n-j}'> (1-x_j+x_j)R_{n-j}'$$
%	Therefore, $R_{n-j+1} > R_{n-j}$.
%\end{proof}

Similar to the previous mechanisms, we study the \textbf{competitive ratio} $\rho=\frac{E[M_Y]}{E[M_T]}$.
We need the following claim first in our analysis.
{Due to the space limitations, we omit the proof.}
\begin{claim}
	\label{cj}
	Let $j^*$ denote the last one satisfying $d_{j^*} >  d_{j^*+1}$ in the sequence, \ie, 
	$d_1 \geq d_2 \geq \dots \geq d_{j^*}> d_{j^*+1}=\dots= d_n$.
	Then %$\varrho_{j^*+1}=\varrho_{j^*+2}=\dots =\varrho_n=1/2$.
	 $\forall j \in [1, j^*+1]$, $\varrho_j>1/2 $. %$\forall 1\leq j\leq j^*+1,$
\end{claim}
We focus on the change of $\rho$ when $n \to \infty$.
%According to Lemma~\ref{MT}, $\{R_1', \cdots, R_n'\}$ is an increasing series. Moreover, we know that $R_n'$ is always up bounded by $d_1\leq 1$. Therefore, we know that there exist some $\alpha$ such that
Similar to our previous analysis on $M_D$, we can assume that $\lim\limits_{n\to \infty}R'_{n}=\alpha$.
From (\ref{equ:M3}), we have $\alpha=(1-x_1)\varrho_1d_1+x_1*\alpha$,
 which implies $\alpha=\varrho_1d_1$. %{\color{red} [$x_1=1/2$? you need to show this]}.
Recall that $E[M_Y]<1$.
According to Claim~\ref{cj}, we have $\varrho_1 \geq  1/2$ and therefore
%The worst case for $\rho$ is when $d_1=d_2=\dots=d_n$, $\lim\limits_{n\to \infty}R_n=d_1/2$.
{{$\lim\limits_{n\to \infty}\rho=\frac{E[M_Y]}{E[M_T]}<\frac{2}{d_1}.$}}

We also consider the truthfulness of $M_T$. Obviously, $M_T$ is bid independent auction and thus value truthful.
	For any $j$-th arrived buyer $i$ who wins the auction at time $t_j$, from (\ref{equ:M3}), we have $u_i^j=a_i*d_j-\varrho_j*d_j=a_i*d_j-x_jd_j+x_j*d_{j+1}-\varrho_{j+1}*d_{j+1}$
	%{\color{red}[why? and what is $x_j$]} 
	and $u_i^{j+1}= a_i*d_{j+1}-\varrho_{j+1}*d_{j+1}$. So, if $a_i>x_j$, $u_i^j-u_i^{j+1}= a_i*(d_j-d_{j+1})-x_jd_j+x_j*d_{j+1}=(a_i-x_j)*(d_j-d_{j+1}) \ge 0$.
%	\begin{equation*}
%	\begin{aligned}
%%	u_i^j=&a_i*d_j-\varrho_j*d_j\\
%%	=&a_i*d_j-x_jd_j+x_j*d_{j+1}-\varrho_{j+1}*d_{j+1}\\
%%	u_i^{j+1}=&a_i*d_{j+1}-\varrho_{j+1}*d_{j+1}\\
%	u_i^j-u_i^{j+1}=& a_i*(d_j-d_{j+1})-x_jd_j+x_j*d_{j+1}\\
%	=&(a_i-x_j)*(d_j-d_{j+1}) \ge 0.
%	\end{aligned}
%	\end{equation*}
	% According to the definition of time truthfulness (see Definition~\ref{def:truthful}), we obtain that $M_F$ is time truthful.
	For the $j$-th arrived buyer $i$ who lose the auction at time $t_j$ may win at some time $t>t_j$. But even this event happens, the expected profit obtained from our mechanism will not decrease.
	Thus $M_T$ is semi-time truthful.
	
	Overall, we have the following theorem for $M_T$.

\begin{theorem}
	Mechanism $M_T$ is value and semi-time truthful. 
	$M_T$ has a competitive ratio $<\frac{2}{d_1}$ when $n \to \infty$.
\end{theorem}
Recall that $M_D$'s competitive ratio converges to a value $\frac{1}{d_1}$ when $n \to \infty$, which implying that $M_T$ has paid about a half profit of $M_D$ to achieve semi-time truthful.

\subsection{Extension to General Distribution}
\label{alg:extension}

Here, we extend $M_F$, $M_D$ and $M_T$ to general distributions. 
Consider the general probability density function $f(y)$. 
We use the same notations as that in Section~\ref{alg_special_case}.
From our above analysis, we can see that as long as the formula of the expected revenue can be obtained explicitly, we are able to compute the (approximate) maximum value by numerical optimization.
%computation cost

\subsubsection{Fixed Reservation Price Mechanism $M_F$}
The general maximization objective equation is:
\begin{equation}
\label{}
\begin{aligned}
E[M_F]=\left(1-\prod_{i=1}^{n} \int_{-\infty}^{\frac{x}{d_j}}f(y) \,dy\right )*x
\end{aligned}
\end{equation}
\iffalse
To calculate competetive ratio, rewrite $P_{i,j}$ in (\ref{pij}) as
\begin{equation}
\label{}
\begin{aligned}
P_{i,j}=\int_{\frac{d_jv_j}{d_i}}^\infty f(x)dx  *\prod_{k\not=i,j}^{n} \left (1-\int_{\frac{d_jv_j}{d_k}}^\infty f(x)dx \right )
\end{aligned}
\end{equation}
\fi
%$E[P_{i,j}*d_j*v_j]=\int_{0}^{\infty} P_{i,j}*d_j*v_j* f(v_j) dv_j$,
%$E[M_V]=\sum_{i=1}^{n}\sum_{j\not =i}^{n}E[P_{i,j}*d_j*v_j]$.
%So, $$\frac{R_n'}{E[M_V]}> \frac{\sum_{i=1}^{n} d_{i}/2^{i+1}}{\sum_{i=1}^{n}\sum_{j\not =i}^{n}E[P_{i,j}*d_j*v_j]}$$

%\linote{how maximize $E[ALG2]$ ?}

\subsubsection{Dynamic Reservation Price Mechanism $M_D$}
%\input{MD_extension.tex}
%we first calculate $x_n$, by maximizing $R_1$ in (\ref{R1}). 
%Thus we can get the corresponding $ R_1$ and $x_n$.
By maximizing (\ref{Rm}), we can recursively get the value of $x_{n-m+1}$ and $R_m$ ($1\leq m\leq n$), where $R_0=0$.
%\begin{equation}
%\label{R1}
%\begin{aligned}
%& R_1=\int_{x_n}^{\infty}f(y)dy * x_n*d_n \quad\quad\quad\quad\quad\quad\quad\quad\quad
%\end{aligned}
%\end{equation}

\begin{equation}
\small
\label{Rm}
\begin{aligned}
& R_{m}=\int_{x_{n-m+1}}^{\infty}f(y)dy *x_{n-m+1}*d_{{n-m+1}}\\
& \quad\quad + \left (1-\int_{x_{n-m+1}}^{\infty}f(y)dy \right)*R_{m-1}, \quad 1 \leq m \leq n\\
%& R_1=\int_{x_n}^{\infty}f(y)dy * x_n*d_n
\end{aligned}
\end{equation}
Similar to $M_D$, both Lemma~\ref{Rn} and Lemma~\ref{xndn} hold for general distribution, which implies that the expected profit and decision prices are decreasing with time going on.
{Due to space limitation, we omit the proof of following Lemma~\ref{generalMD}.}
\begin{lemma}
	\label{generalMD}	
	$\forall m, \quad 2\leq m \leq n$, $R_m > R_{m-1}$, $x_{n-m+1}d_{n-m+1} > x_{n-m+2}d_{n-m+2}$.	
\end{lemma}

\subsubsection{Semi-Time Truthful Mechanism $M_T$}
\label{MT_extension}

%\input{MT_extension.tex}
%We first compute $x_j (1 \leq j\leq n)$ by maximizing (\ref{R1}) and (\ref{Rm}).
According to (\ref{Rmmm}), for any $1 \leq j < n$, we can recursively obtain $\varrho_j$ and $R_{n-j+1}'$, where $R_1'=R_1$, $\varrho_n=x_n'$.
\begin{equation}
\small
\label{Rmmm}
\begin{aligned}
\varrho_j  & = ( x_jd_j -x_j*d_{j+1}+\varrho_{j+1}*d_{j+1})/d_j\\
%R_{n-j+1}^{acc'} & = \int_{x_{j}}^{\infty}f(y)dy *\varrho_j*d_j \\
%R_{n-j+1}^{rej'}& =\left(1-\int_{x_{j}}^{\infty}f(y)dy \right) *R_{n-j}' \\
R_{n-j+1}'& =  \int_{x_{j}}^{\infty}f(y)dy *\varrho_j*d_j + \left(1-\int_{x_{j}}^{\infty}f(y)dy \right) *R_{n-j}' %j \in[1,n) %1\leq j<n
\end{aligned}
\end{equation}
\iffalse
From the definition of $\varrho_j$, we easily obtain the following claim.
\begin{claim}
	$\forall j,  1\leq j < n$, $\varrho_jd_j\geq \varrho_{j+1}d_{j+1}$,
\end{claim}
\fi
Similar to lemma~\ref{MT}, we have the following lemma for the general case, which is consistent with our intuition, \ie, as time goes on, the expected profit is decreasing.
Proof details are omitted due to the space limitation.
\begin{lemma}
	\label{Rj}
	$\forall j,  1\leq j < n$, $R_{n-j+1}' > R_{n-j}'$.
\end{lemma}

\subsection{Mechanism with Learning Phase}
\label{alg_learning}
%\subsection{Algorithms with learning phase}
%\label{alg_learning}
The mechanisms studied in Section~\ref{alg_special_case} are all based on the assumption that we already know all parameters about the probability density function $f(y)$. 
However, it is very likely that we do not have such prior knowledge. 
%need the observation process to learn the distribution function.  
Here we propose a learning-based mechanism $M_L(n_s,n)$ which consists two phases: learning and decision (w.l.o.g, the number of the buyers $n$ is large enough to support our learning phase).\\
1) \textbf{Learning phase}: firstly, we keep observing and collecting the first $n_s$ buyers' reported prices as samples without choosing any one, where $n_s$ is set to be much smaller than $n$. 
%(we assume $n>>n_s$.) %the total number of buyers $n$ is large enough, 
The purpose of the learning phase is to estimate the parameters of $f(y)$; obviously, for different distributions, we may apply different methods (such as maximum likelihood estimate). \\
%Take the normal distribution $N(\mu,\sigma)$ as an example.
%%we have {\small {$f(v(\mathbb{D})|\mu,\sigma^2)=\frac{1}{\sigma\sqrt{2\pi}}e^{-\frac{(v(\mathbb{D})-\mu)^{2}}{2\sigma^2}}.$}}
%With $1-\delta$ confidence, we have the estimated value of $\mu$ and $\sigma$.
%%$\mu_s \in ( \bar{v}_m-\frac{S_m}{\sqrt{m}}t_{\frac{\delta_1}{2}}(m-1),\bar{v}_m+\frac{S_m}{\sqrt{m}}t_{\frac{\delta_1}{2}}(m-1))$, 
%%$\sigma_s^2 \in (\frac{(m-1)S_m^2}{\chi^2_{\delta_1/2}(m-1)},\frac{(m-1)S_m^2}{\chi^2_{1-\delta_1/2}(m-1)})$ respectively, where $\bar{v}_n_s$ denotes the sample mean and $S_m^2$ denotes the sample variance.
%We set $\delta$ to be a desirable big number by adjusting $n_s$.
%We fixing the width of confidence interval, by looking up the \textit{student's t-distribution} table and \textit{chi-square distribution} table, we set
%$t_{0.05}(m-1) \leq 1.658$ and $\chi^2_{0.05}(m-1)\leq 146.567$, $\chi^2_{0.95}(m-1)\leq 95.705$.
%Then we set $m \geq 121$, the larger $n_s$ but also causes larger value loss.	
%For example uniform distribution ($X \sim U(a,b)$), given a series of observed reported prices $\{v_1*d_1, v_2*d_2, \dots, v_m*d_m\}$ and the function $d(t)$, it is easy to get the estimated parameters $a$ and $b$ by solving the following equations: 1) $1/n*\sum_{i=1}^{m}v_i=(a+b)/2$; 2) $1/n*\sum_{i=1}^{m}v_i^2=(a+b)^2/4+(b-a)^2/12 $.
2) \textbf{Decision phase}: given $f(y)$, run $M_T$ (or $M_F$) to choose the winner. In this phase, we also update the parameters of $f(y)$ by the new observed knowledge and we dynamically update the stopping time $t_e$, thus the value of $n$.
%Since we run $M_T$ (or $M_F$) in the decision phase,  
%we have $E[M_L(n_s,n)]=(1-\delta)*m/n *E[M_T]$. Thus, with carefully selecting $n_s$, $M_L(n_s,n)$ can achieve an approximate revenue maximization.{\color{red}[ Why?]} 

%$M_L$'s performance is closely related to the effectiveness of the learning phase. Since we run $M_T$ (or $M_F$) in the decision phase, we can get constant competitive.

%It is worth to mention the truthfulness of $M_L(n_s,n)$.
%Firstly, if delaying arrival time is not an option for buyers, 
We can prove that $M_L$ is value truthful but not time truthful.
A buyer appeared in the learning phase can delay his arrival to the decision phase, possibly coupled with bid manipulation, he can get a positive profit, instead of zero profit in the learning phase.
To get  \textit{time and value truthful}, we propose to give some \textit{compensation} to  buyers  in the learning phase.

%{The choice of the compensation value is left for our future analysis.}
 
\paragraph{Compensation design} (when run $M_T$ in the decision phase): here we consider buyers are risk-neutral who aim to maximize their expected profit.
%Let $S$ denote the set of buyers appear at learning phase.
Let $P_{win}^{j}$ denote the probability of winning for each buyer $i$ who appeared at learning phase and delayed his arrival to  time instance $t_{j \geq n_s+1}$. Let $\varepsilon_i^j$ denote the compensation value for $i$ at time $t_j$ (when $M_T$ chooses someone as winner).
%For each $j \geq n_s+1$: $P_{win}^{j}=\frac{(x_{j-1}-x_{j})}{j} $, we have
\begin{equation}
\label{compensation2}
 \varepsilon_i^j = E[u^j_i]= (\frac{x_{j-1}+x_j}{2}* d_j-\varrho_jd_j)*\frac{(x_{j-1}-x_{j})}{j} \\
\end{equation}	
% We have the following theorem for $E[M_L(n_s,n)]$. Due to the space limitation, we omit the proof details of here.
\begin{theorem}
	\label{claim:compensation2}
	$M_L(n_s,n)$ has $\Theta(1/d(\frac{1}{\lambda}))$ competitive ratio when $M_T$ is the decision mechanism at its decision phase and the compensation is calculated by Equation (\ref{compensation2}).
\end{theorem}	
It's worth to mention that the value of compensation  depends on the final revenue of the broker. We can prove that
%Thus, buyers in learning-phase have same objective function with the broker. Therefore, 
\begin{theorem}
	$M_L(n_s,n)$ is value and semi-time truthful in expectation.
\end{theorem}
\iffalse
%In other words, a buyer appeared in the learning phase can return to the decision phase and have a positive profit rather than the zero profit in the learning phase. 
%We propose this challenge that how to ensure both \textit{time and value truthful} for a learning based mechanism as an open problem in future work. 

%To ensure both \textit{time and value truthful}, we propose to give some \textbf{compensation} to those buyers appeared in the learning phase.
%The choice of the compensation value is left for our future analysis.
\input{compensation.tex}
\fi

\section{Performance Evaluation}
\label{sec:eva}
\begin{figure*}[ht]
\centering
\begin{minipage}[t]{0.32\linewidth}
	\centering
	\includegraphics[width = \linewidth]{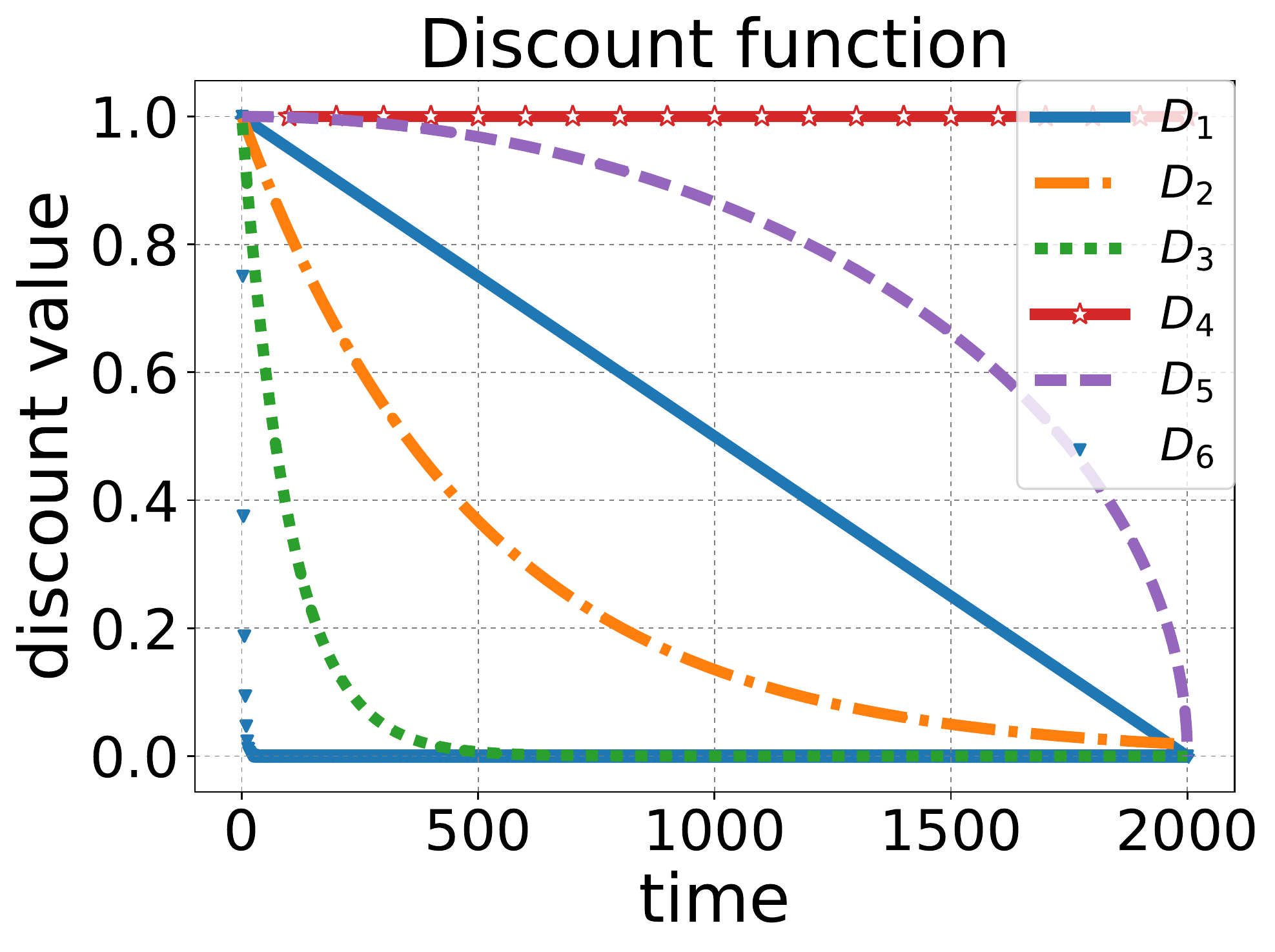}
	\caption{Different discount functions used.}
	\label{Discount_function}
\end{minipage}\quad
\begin{minipage}[t]{0.32\linewidth}
	\centering
	\includegraphics[width =\linewidth]{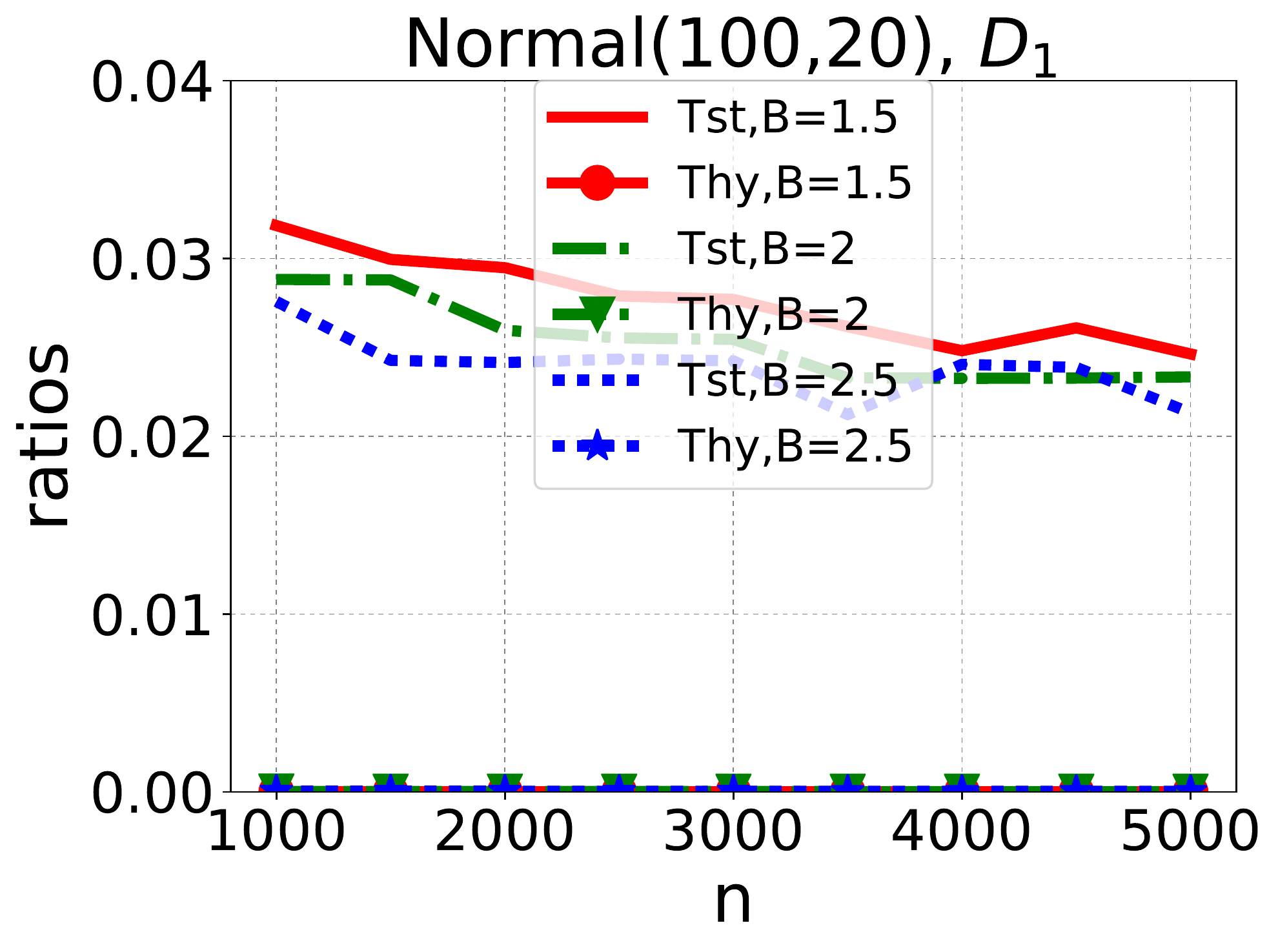}
	\caption{$\mathcal{M}_R$'s performance with different bases.}
	\label{drawBasesRatios}
\end{minipage}\quad
\begin{minipage}[t]{0.32\linewidth}
	\centering
	\includegraphics[width = \linewidth]{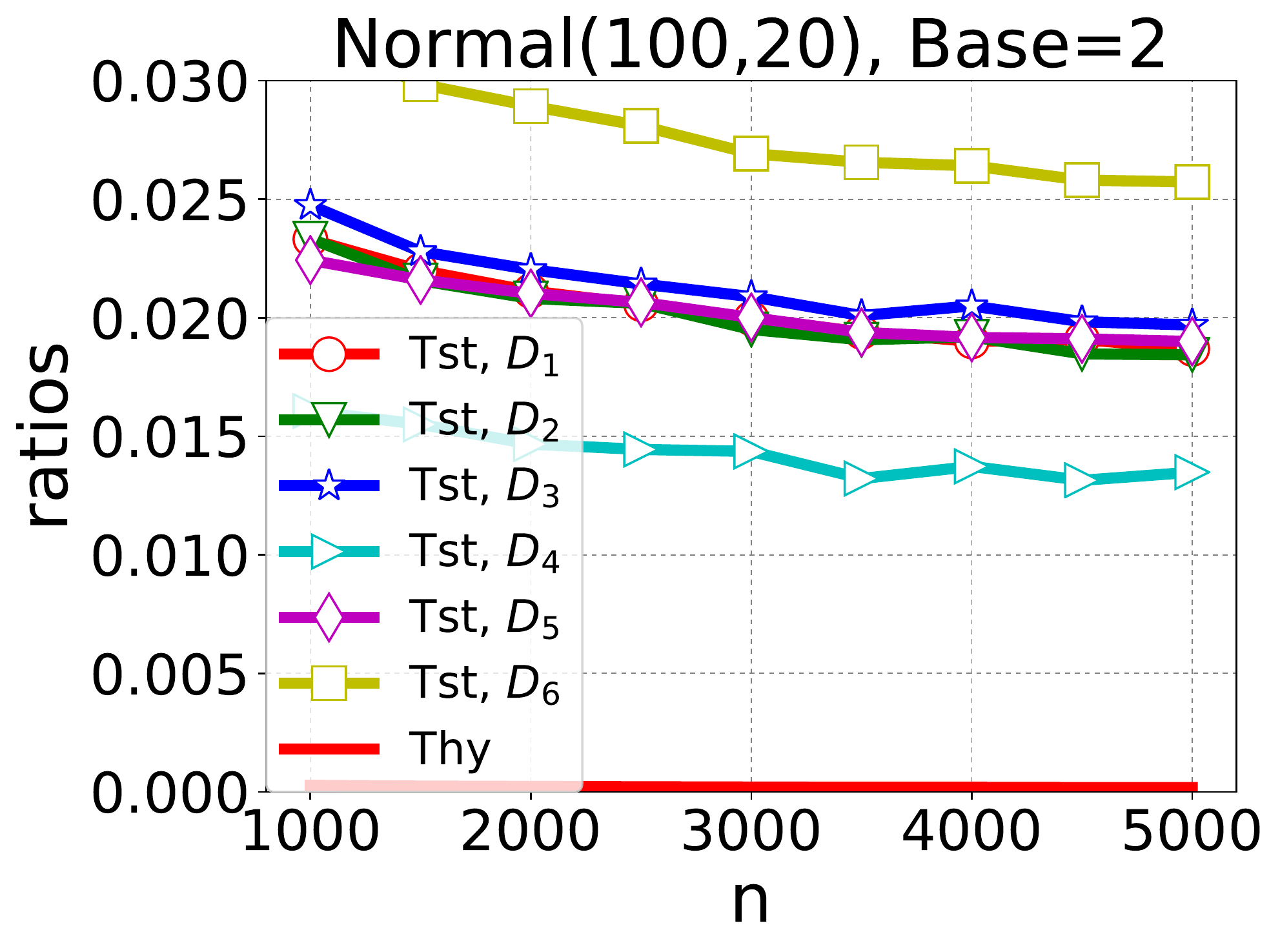}
	\caption{$\mathcal{M}_R$'s performance with different discount functions.}
	\label{drawDisRatios}
\end{minipage}
\quad
\begin{minipage}[t]{0.32\linewidth}
	\centering
	\includegraphics[width =\linewidth]{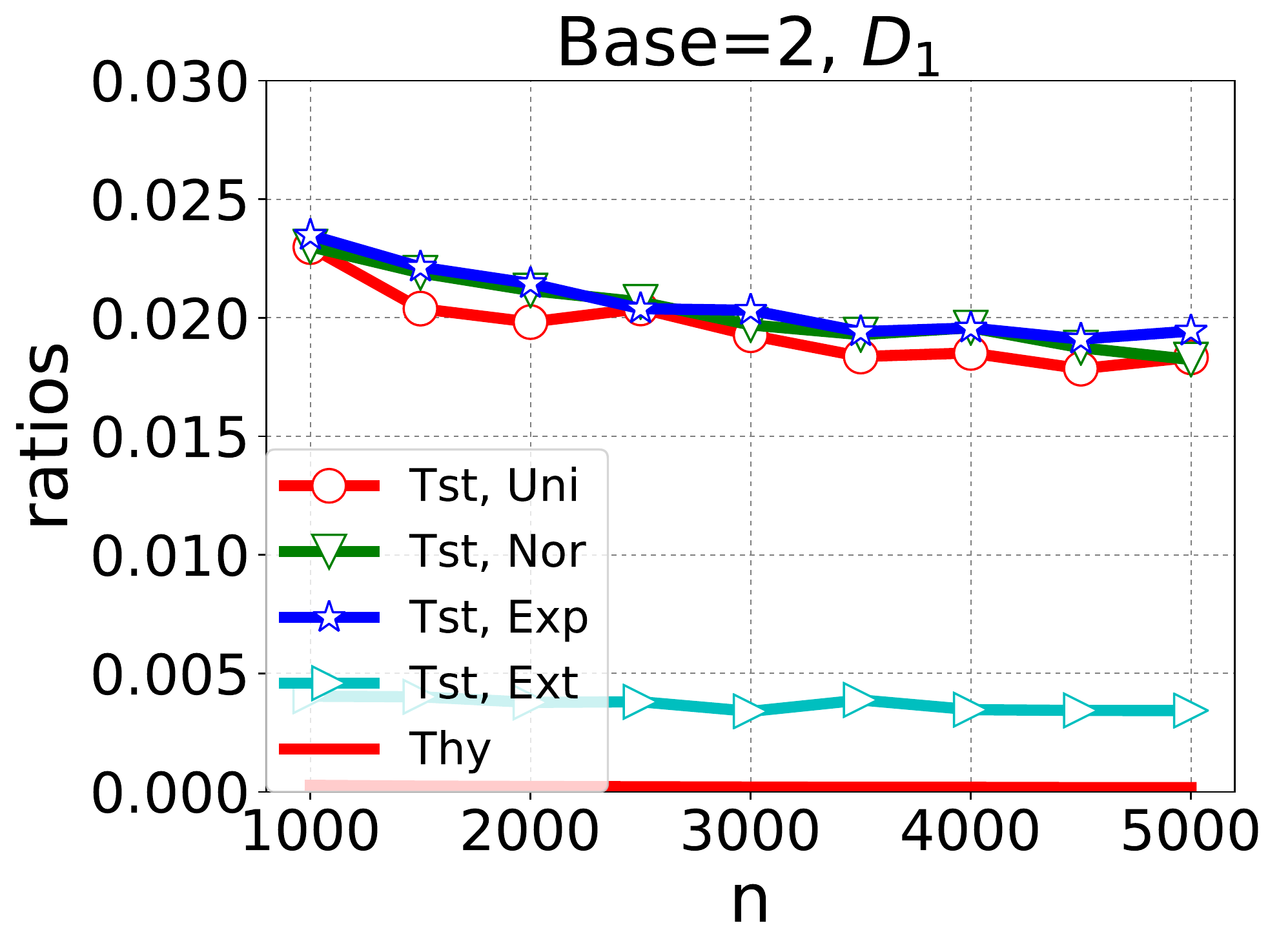}
	\caption{$\mathcal{M}_R$'s performance with different valuation distributions.}
	\label{drawDistributionRatios}
\end{minipage}\quad
\begin{minipage}[t]{0.32\linewidth}
	\centering
	\includegraphics[width = \linewidth]{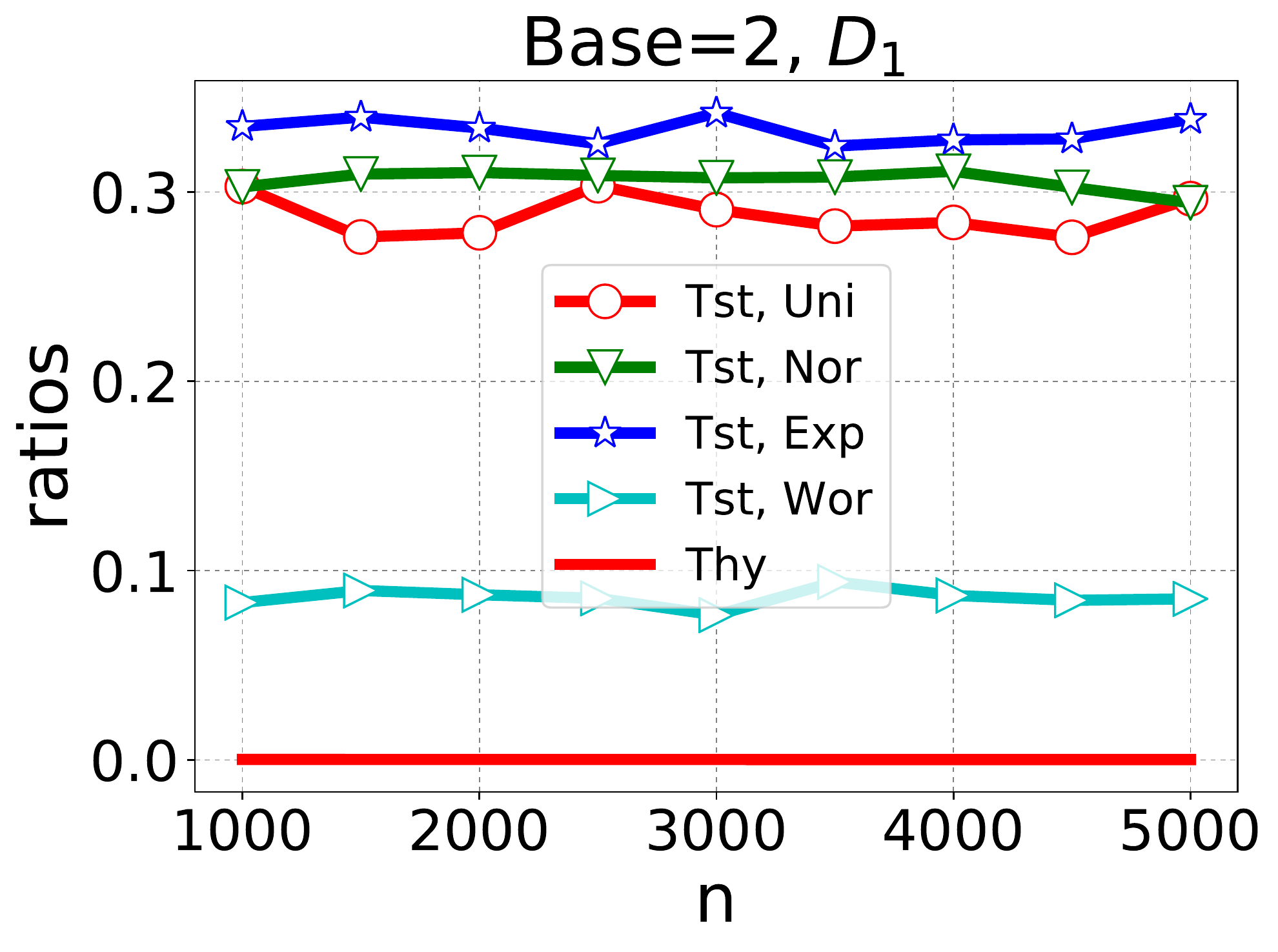}
	\caption{$\mathcal{M}_W$'s performance with different valuation distributions.}
	\label{drawDistributionRatiosMU2}
\end{minipage}\quad
\begin{minipage}[t]{0.32\linewidth}
	\centering
	\includegraphics[width = \linewidth]{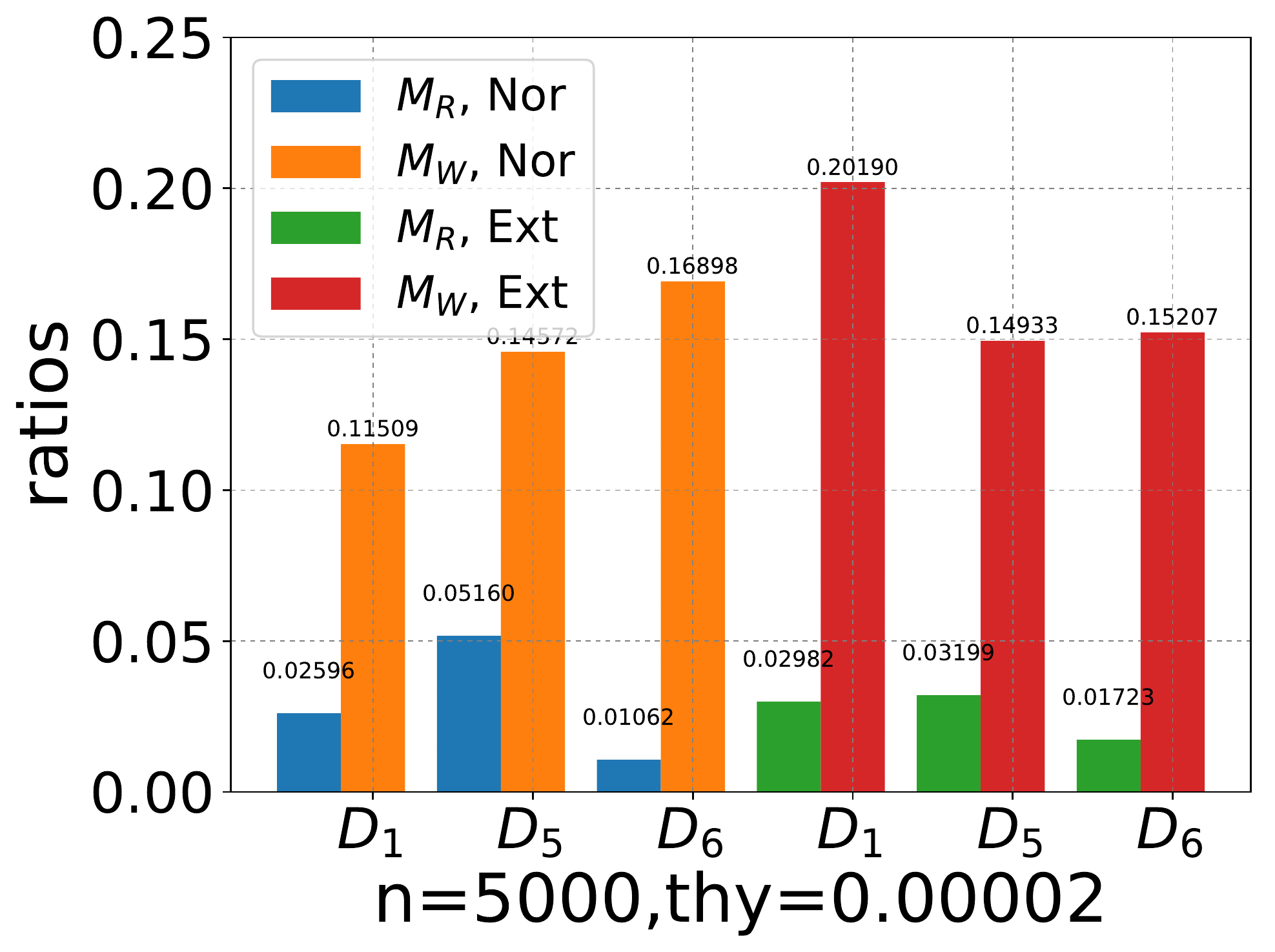}
	\caption{Comparison of $\mathcal{M}_R$ and $\mathcal{M}_W$.}
	\label{drawBarUniNorExpWor_smallN}
\end{minipage}
\end{figure*}

In this section, we evaluate our mechanisms with simulation study. We use the offline Vickrey auction's revenue $E[\mathcal{M}_V]$ as comparison baseline for all of our mechanisms.
For the sake of understanding intuition,  we use $1/\rho=\frac{E[\mathcal{M}]}{E[\mathcal{M}_V]}$ as the tested ratio to present our experimental performance measurement results.

\subsection{Experiment Settings}
Our experiments are conducted on a server equipped with a 12-core i7 Intel CPU, 64G of RAM.
We set $T=2000$,  and test our mechanisms with different $\lambda$, \ie, the expected total number of buyers $n$.
and vary the number of buyers $n$ from $1000$ to $5000$ with a step of $500$.
All the expected revenue results are averaged over $10*n$ runs and our mechanisms finished in $1$ minutes when $n=2000$. 
We test mechanism $\mathcal{M}_R$ with different combinations of valuation distribution and discount function. % $D_1(t)$. 
Let ``Tst" and ``Thy" denote our experiment and theoretical results respectively.
Specifically, we use the following six discount functions with different decreasing speeds (as shown in Fig.~\ref{Discount_function}). Abbreviated as $D_1$, $D_2$, $D_3$, $D_4$, $D_5$ and $D_6$ respectively, which captures different decreasing speed of data valuations in reality.
Especially, we have $D_4$ as the case that there are no discount factors.
\begin{equation*}
\small
\begin{aligned}
\begin{cases}
\text{Linear Discounting: }  D_1(t)=(1-t/T) \\
\text{Exponential Discounting: }  D_2(t)={\exp^{-1}\frac{4*(t-1)}{(T-1)}} \\
\text{Fast Exponential Discounting: }  D_3(t)={0.99^t} \\
\text{No Discounting: }  D_4(t)=1 \\
\text{Slow Discounting: } D_5(t)=\sqrt{((\lambda*T)^2-t^2)/(\lambda*T)^2} \\  %y5=[math.sqrt((a*a*b*b-b*b*x*x)/math.pow(a,2)) for x in n] a=n,b=1
\text{Extreme case: } D_6(t), \text{\ie the discount function in~(\ref{worst_case})}.%$2^{-(c-1)}, \quad t \in [t_{2*c-1},t_{2*c}]$
\end{cases}
\end{aligned}
\end{equation*}

\begin{figure}[th!]
\centering{
\includegraphics[width = 0.8\linewidth]{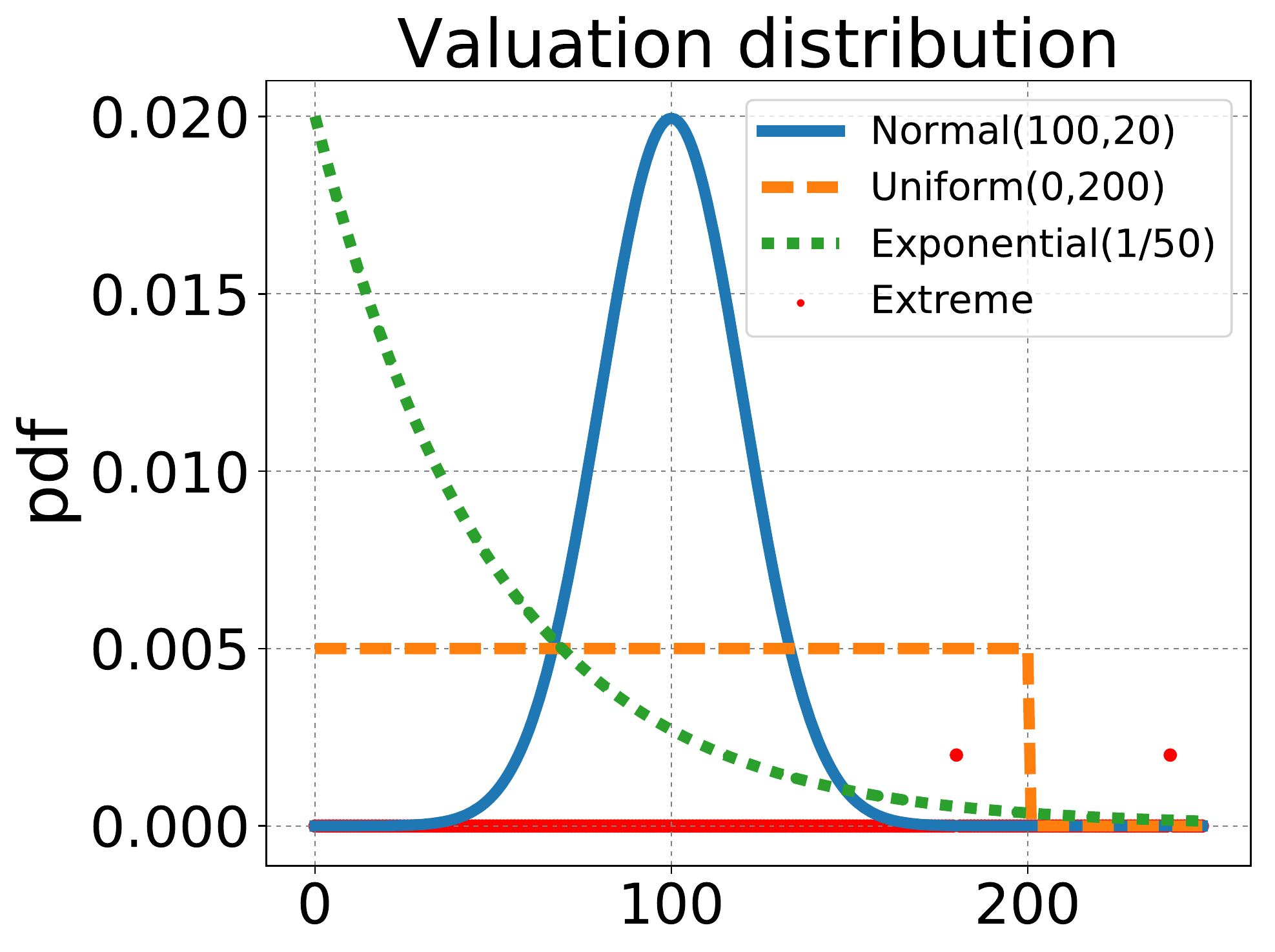}
\caption{Four different valuation's distributions used in our evaluation} 	
\label{Valuation_distribution}	}
\end{figure}
For valuation distribution, we consider the following ones (see Figure~\ref{Valuation_distribution}):
\begin{compactenum}
\item uniform distribution $Uni(0,200)$; 
\item normal distribution $Nor(100,20)$; 
\item exponential distribution $Exp(\frac{1}{50})$; 
\item the extreme case with only two positive valuations (as in Eq.~\eqref{worst_case}), other valuations are $0$. 
\end{compactenum}
Below, we use  ``Uni'', ``Nor'', ``Exp'' and ``Ext'' for short representation of these valuation sets.

\subsection{Numerical Results for $M_R$ and $M_W$}
\label{eva}
We mainly investigate the impact of various factors on the mechanisms' performance and show the differences between $\mathcal{M}_R$ and $\mathcal{M}_W$
with different discount functions.
We show mechanism $\mathcal{M}_R$'s performance with different valuations' distributions in Figure~\ref{drawDistributionRatios} 
 and we can see that even with the extreme case, the practical competitive ratios are much better than the theoretical results. 

\textbf{Impact of Dividing Base.}
To show the generalizability of our mechanism, 
 we consider different values of the base $B$  for dividing the discount value domain in $\mathcal{M}_R$.
As shown in Fig.~\ref{drawBasesRatios}, the competitive ratios of $\mathcal{M}_R$ in our evaluations are close with different $B$s. 
The main reason here is that 
 when we divide the values of $D_1$ into multiple classes with $B\in\{1.5, 2, 2.5\}$,
 all buyers' expected arrivals fall in the reserved classes. 
 That is, the globally second-largest expected report price $r_{(2)}$ gets all of its value, instead of $E[r_{(2)}]/B$ by our theoretical analysis, from the top $\hat{c}$ classes.

\textbf{Impact of Discount Function.}
Fig.~\ref{drawDisRatios} demonstrates the results of  $\mathcal{M}_R$ with different discount functions. 
All the competitive ratios by our evaluation results are much higher than the theoretical results.
Intuitively, the faster decreasing speed, the lower competitive ratios. 
However, the competitive ratio with a discount function $D_2$ with faster decreasing speed is much better than that for discount function $D_1$. 
According to our analysis in \S\ref{analysis_M_R}, not only the decreasing speed but also the number of reserved classes and the number of buyers in each reserved class 
 will affect the performance of $M_R$. 
Moreover, when without discount factors, i.e., $D_4$, $M_R$'s revenue ratio is much smaller than the theoretical results, i.e., $1/4$ in ~\cite{hajiaghayi2004adaptive}. 
Note that we add a random winner selection in the ``Observation" phase in Algorithm~\ref{alg:M_o} to guarantee the time truthfulness and consumer sovereignty.
This verifies  our mechanisms' effectiveness on time-sensitive valuations.

However, as plotted in Fig.~\ref{drawDisRatios}, the competitive ratios of $\mathcal{M}_R$ on the extreme case (with only two large positive initial valuations $v_{(1)}$ and $v_{(2)}$) are decreasing when the decreasing speed of $d(t)$ is increasing.
This phenomena only occurs on the extreme case due to the following facts:
 the first and second-largest reported prices are always the corresponding report price of buyers whose initial valuations are $v_{(1)}$ or $v_{(2)}$.
 1) $\mathcal{M}_R$ only obtains positive revenue of when $v_{(1)}$ and $v_{(2)}$ appear at the same class and $v_{(2)}$ arrives before $v_{(1)}$; 
 2) the $d(t)$ with faster decreasing speed has more number of discount classes with small discount value.
Thus, $\mathcal{M}_R$ obtains a larger portion of revenue from discount classes with small discount value when compared with $\mathcal{M}_R$ on the normal distribution.
Moreover, the competitive ratios of $\mathcal{M}_R$ on the worst case (\ie, $Ext$ with $D_6$) are very close to our theoretical results, which verifies our theoretical analysis.
Thus, $\mathcal{M}_R$ works poorly on the extreme valuation case with fast decreasing speed of $d(t)$.
$M_R$ performs better on discount functions with fast decreasing speeds than those decrease slowly, 
As plotted in Fig.~\ref{drawDisRatios} and Fig.~\ref{drawDisRatios}, different discount functions combine with different valuation distributions result in different ratios. 
Fig.~\ref{drawDistributionRatios} demonstrates the evaluation results of  $\mathcal{M}_R$ with different valuation distributions and we can see that even in the extreme case, the competitive ratio is much better than the theoretical result. 
Fig.~\ref{drawDistributionRatios} shows that the tested ratios of $\mathcal{M}_R$ are very close with different valuation distributions except the extreme case, \ie, $Ext$.

\textbf{Impact of Valuation Distribution.}
Fig.~\ref{drawDistributionRatios} presents how different valuation distributions affect the performance of $M_R$. 
Interestingly, the competitive ratios of $\mathcal{M}_R$ by our evaluations are very close with different valuation distributions except the extreme case,
  \ie, $Ext$ has only two large positive initial valuations.
The ratios are decreasing slowly when the total number of buyers $n$ is increasing.
The reason is that when $n$ increases, the capability gap of obtaining large reported prices between different discount classes is becoming larger.
However $M_R$ chooses each class equally.
With the increase of $n$, 
 the expected number of buyers in each candidate class increases, 
 thus the chosen probability of the class with larger contribution to revenue decreases.
To further improve $\mathcal{M}_R$'s performance, we hope to design some more sophisticated strategy to select each class in the future work.
Without discount factors, \ie, $D_4$, $\mathcal{M}_R$'s revenue competitive ratio is much smaller than the optimal mechanism's theoretical results (\ie, 1/4 in ~\cite{hajiaghayi2004adaptive}).

\textbf{Comparison of $\mathcal{M}_R$ and $\mathcal{M}_W$.}
Different from $M_R$,
 the competitive ratios of $M_W$ are almost the same when the total number of buyer $n$ is increasing as shown in 
Fig.~\ref{drawDistributionRatiosMU2}.
Recall that $\mathcal{M}_W$ selects different reserved classes with carefully designed weights. 
This verifies the effectiveness of $M_W$'s ``Weighted-Select"  when compared with $M_R$'s ``Random-Select" strategy.
From Fig.~\ref{drawBarUniNorExpWor_smallN}, it is obvious that $\mathcal{M}_W$ performs about ten times better than $\mathcal{M}_R$ under different discount function settings.
Even in the worst case (\ie, $D_6$ and $Ext$), $\mathcal{M}_W$ achieves much better tested results than the theoretical results.

%%\subsection{The simulation study of $M_U$}
%%\label{eva:MU}
%%\input{eva_MU.tex}
%
%%\subsection{The performance of $M_U^+$}
%%\label{eva:MU+}
%%\input{eva_MU+.tex}
%
%\subsection{The performance of $M_F,M_D,M_T$}
%\label{eva:MFDT}
%\input{eva_M_FDT.tex}
%
%\subsection{The performance of $M_L(n_s,n)$}
%\label{eva:ML}
%\input{eva_ML.tex}

\section{Discussion and Conclusion}
\label{sec:conclusion}

In this work, we  studied online truthful data auction mechanisms for trading time-sensitive valued data product.
We assumed that the discount function $d(t)$ is known to both the adversaries and the mechanism designer.
By partitioning the discount function into different classes where the discount values in each class is within a factor $B>1$ of each other, 
 we designed several revenue-competitive truthful online auction mechanisms with time-sensitive valuations.
We mainly presented an $\Theta(\log^2 n)$-competitive truthful online auction mechanism $M_R$, 
 an $\Theta(\log n)$-competitive truthful online auction mechanism $M_1$, when the number of users $n_c$ is within a constant $\eta$ factor of each other
  for all discount class $ c \le \hat{c}$.
It is worth to mention that our results still hold when we know a constant approximation of the discounting function $d(t)$,
 \ie, when we are given $d'(t)$ instead of the actual discount function $d(t)$ with $\alpha_1 d(t) \le d'(t) \le \alpha_2 d(t)$ for some positive constant 
 $0< \alpha_1 < \alpha_2$ and any $t>0$.
Further, when different buyers have different discounting functions $d(t)$, our results still hold as long as they are within a constant factor of each other.
Our mechanisms could perfectly be applied to other kinds of goods whose value decreases with time, like diamonds, flowers, vegetables and so on.

There are a number of interesting and challenging unsolved questions left for future research.
First, we need design truthful mechanisms with a good competitive ratio when the discount function $d(t)$ is arbitrary unknown,
 or different buyers have heterogeneous discount functions.
Second, in this work, we assume that the number of users $n$ is large enough, 
so one interesting question is to design a good mechanism when there is only a constant number of users to arrive.
Third, for most upper-bounds and lower-bounds on the competitive ratios of the truthful mechanisms, we assumed that the
 adversary is adaptive-online.
 It will be interesting to analyze some lower-bounds of the truthful mechanisms for other adversary models, such as 
  oblivious adversary, adaptive offline adversary, and valuation adaptive-online adversary; and then design truthful mechanisms with 
   competitive ratio that can (closely and asymptotically) match the lower bounds.
Finally, we need design online truthful and competitive mechanisms for selling data to $k \ge 2$ buyers where the value of $k$ could be some large constant
 or some function $f(n)$.

{
\bibliographystyle{IEEEtran}
\bibliography{reference-3}
}

%\section{Appendix}
%\label{sec:appendix}
%\input{appendix-2021.tex}

\end{document}